\pdfoutput=1

\documentclass[11pt]{article}
\usepackage{jheppub}

\usepackage{amsmath}
\usepackage{amssymb}
\usepackage{graphicx,xcolor,slashed}
\usepackage{enumerate}
\usepackage{ifpdf}
\usepackage[english]{babel}
\usepackage{url}
\usepackage{booktabs}
\usepackage{xspace}
\usepackage{xinttools}
\usepackage{mathrsfs}
\usepackage{blkarray}
\notoc

\renewcommand{\baselinestretch}{1.18}

\definecolor{darkgreen}{rgb}{0.0, 0.26, 0.15}
\definecolor{darkred}{rgb}{0.65,0.15,0}

\usepackage{float}

\hypersetup{bookmarksnumbered,bookmarksdepth=3}

\allowdisplaybreaks[1]

\DeclareFontFamily{U}{mathx}{\hyphenchar\font45}
\DeclareFontShape{U}{mathx}{m}{n}{
	<5> <6> <7> <8> <9> <10>
	<10.95> <12> <14.4> <17.28> <20.74> <24.88>
	mathx10
}{}
\DeclareSymbolFont{mathx}{U}{mathx}{m}{n}
\DeclareFontSubstitution{U}{mathx}{m}{n}
\DeclareMathAccent{\widecheck}{0}{mathx}{"71}

\font\tenshuffle=shuffle10 \font\sevenshuffle=shuffle7 \font\fiveshuffle=shuffle7 at 5pt
\def\shuffle{{%
		\def\Dshuffle{\mathbin{\hbox{\tenshuffle\char'001}}}%
		\def\Sshuffle{\mathbin{\hbox{\sevenshuffle\char'001}}}%
		\def\SSshuffle{\mathbin{\hbox{\fiveshuffle\char'001}}}%
		\mathchoice{\Dshuffle}{\Dshuffle}{\Sshuffle}{\SSshuffle}}}

\restylefloat{figure}
\definecolor{dgreen}{rgb}{0,0.70,0.30}
\definecolor{gold}{rgb}{0.85,.66,0}
\definecolor{purple}{rgb}{1.0,0.3,0.6}

\usepackage{tikz}
\usepackage[compat=1.1.0]{tikz-feynman}
\usetikzlibrary{calc}
\usetikzlibrary{patterns}
\usetikzlibrary{decorations.pathreplacing}
\usetikzlibrary{decorations.markings}
\usetikzlibrary{decorations.pathmorphing}
\usetikzlibrary{positioning}
\usetikzlibrary{arrows.meta}

\newcommand{\bea}{\begin{eqnarray}}
\newcommand{\eea}{\end{eqnarray}}
\def\beq{\begin{equation}}
\def\eeq{\end{equation}}

\newcommand{\Tr}{{\rm Tr}}

\newcommand{\PT}{{\rm PT}}

\newcommand{\ep}{\epsilon}

\newcommand{\dd}{\mathrm{d}}
\newcommand{\te}{\textrm}

\newcommand{\vecb}{\left(\begin{array}{c}}
	\newcommand{\vece}{\end{array}\right)}
\newcommand{\ccb}{\left(\begin{array}{cc}}
	\newcommand{\cce}{\end{array}\right)}
\newcommand{\cccb}{\left(\begin{array}{ccc}}
	\newcommand{\ccce}{\end{array}\right)}
\newcommand{\ccccb}{\left(\begin{array}{cccc}}
	\newcommand{\cccce}{\end{array}\right)}
\newcommand{\cccccb}{\left(\begin{array}{ccccc}}
	\newcommand{\ccccce}{\end{array}\right)}

\newcommand{\NN}{\mathbb N}

\DeclareMathOperator{\Pf}{Pf}

\usepackage[format=plain,
labelfont=bf,
textfont=it]{caption}

\title{One-loop amplitudes in Einstein-Yang-Mills from forward limits}

\author[a,b]{Franziska Porkert,}
\author[c]{Oliver Schlotterer}

\affiliation[a]{Institut f\"ur Physik und IRIS Adlershof, Humboldt-Universit\"at zu Berlin, 
	Zum Gro\ss{}en Windkanal 6, D-12489 Berlin, Germany}
\affiliation[b]{Bethe Center for Theoretical Physics, Wegelerstra\ss{}e 10, 
Universit\"at Bonn, D-53115, Germany}
\affiliation[c]{Department of Physics and Astronomy, Uppsala University, L\"agerhyddsv\"agen 1, SE-752 37 Uppsala, Sweden}

\emailAdd{fporkert@uni-bonn.de}
\emailAdd{oliver.schlotterer@physics.uu.se}

\date{\today}

\abstract{We present a method to compute the integrands of one-loop 
Einstein-Yang-Mills amplitudes for any number of external gauge and gravity 
multiplets. Our construction relies on the double-copy structure of Einstein-Yang-Mills 
as (super-)Yang-Mills with the so-called YM+$\phi^3$ theory -- pure Yang-Mills 
coupled to bi-adjoint scalars -- which we
implement via one-loop Cachazo-He-Yuan formulae. The YM+$\phi^3$ building blocks are
obtained from forward limits of tree-level input in external gluons and scalars, and we give
the composition rules for any number of traces and orders in the couplings $g$ and $\kappa$.
On the one hand, we spell out supersymmetry- and dimension-agnostic relations that reduce
loop integrands of Einstein-Yang-Mills to those of pure gauge theories. On the other hand,
we present four-point results for maximal and half-maximal supersymmetry where all
supersymmetry cancellations are exposed. In the half-maximal case, we determine 
six-dimensional anomalies due to chiral hypermultiplets in the loop.}

\preprint{BONN-TH-2022-01 \\  \phantom{~} \hfill UUITP-04/22}

\begin{document}

\maketitle{}

\newpage

\setcounter{page}{1}
\pagenumbering{roman}

\setcounter{tocdepth}{2}

\renewcommand{\baselinestretch}{1.12}\normalsize
\tableofcontents

\renewcommand{\baselinestretch}{1.18}\normalsize

\numberwithin{equation}{section}

 \newpage


\section{Introduction}
\label{sec:intro}

\setcounter{page}{1}
\pagenumbering{arabic}

Recent studies of scattering amplitudes revealed a variety of symmetries of and
connections between different quantum field theories that are hidden in traditional Lagrangian formulations.
The most famous example concerns the double-copy structure of (super-)gravity:
The loop integrands of supergravity amplitudes can often be assembled from suitably chosen
squares of gauge-theory building blocks. This phenomenon relies on the Bern-Carrasco-Johansson
(BCJ) duality between color and kinematics in gauge theories \cite{BCJ, loopBCJ, Bern:2010yg}, see
\cite{Bern:2019prr} for a comprehensive review. In many cases, the gravitational double
copy can be naturally understood from string theory, e.g.\ from the famous Kawai-Lewellen-Tye (KLT) relations at tree level \cite{Kawai:1985xq}, and from chiral splitting \cite{DHoker:1988pdl, DHoker:1989cxq} at the level of loop integrands.

The field-theoretic double-copy structure applies to a growing list of theories including
Born-Infeld, special Galileons \cite{Cachazo:2014xea, Cachazo:2016njl} and even open-string 
theories \cite{Mafra:2011nv, Zfunctions, Azevedo:2018dgo}. In particular, amplitudes
of Einstein-Yang-Mills (EYM) theories can be obtained from the double copy of (super-)Yang-Mills 
with the so-called YM$+\phi^3$ theory \cite{Chiodaroli:2014xia}. 
EYM refers to Einstein gravity, extended by a dilaton \& B-field\footnote{The couplings of the dilaton \& B-field contribute to the one-loop EYM amplitudes in this work and are natural from the viewpoints of the double copy, supersymmetric extensions and string-theory realizations.}
and minimally coupled to Yang-Mills, including supersymmetric extensions with up to 16 supercharges. 
The non-supersymmetric ingredient YM$+\phi^3$ of its double copy augments pure Yang-Mills by 
a minimal coupling to bi-adjoint scalars with a cubic self interaction. Similar double-copy descriptions
have been found for variants of EYM with spontaneous symmetry breaking \cite{Chiodaroli:2015rdg, Chiodaroli:2017ehv}.

The double-copy structure of EYM implies relations between EYM amplitudes and those of
(super-)Yang-Mills. At tree level, such EYM amplitude relations have been analyzed from
a multitude of perspectives including open strings interacting with closed strings 
\cite{Stieberger:2016lng, Mazloumi:2022lga},
the Cachazo-He-Yuan (CHY) formalism\footnote{Representations of EYM tree amplitudes
in the CHY formalism \cite{Cachazo:2013gna, Cachazo:2013hca, Cachazo:2013iea} were given
in \cite{Cachazo:2014nsa, Cachazo:2014xea, Roehrig:2017wvh}, also see \cite{Casali:2015vta} for 
an underpinning via ambitwistor strings.} \cite{Nandan:2016pya, delaCruz:2016gnm, Du:2016wkt, Teng:2017tbo, Du:2017gnh}, heterotic strings \cite{Schlotterer:2016cxa}, gauge invariance \cite{Fu:2017uzt, Feng:2020jck} and the color-kinematics duality of YM$+\phi^3$ \cite{Chiodaroli:2017ngp}. 
Similar double-copy structures
and resulting tree-amplitude relations apply to conformal supergravity coupled to 
gauge theories \cite{Johansson:2017srf, Azevedo:2018dgo}.

There is considerably less literature on {\it loop-level} amplitudes of EYM:
Specific four-point one-loop amplitudes with half-maximal supersymmetry and external gluons
have been determined in \cite{Chiodaroli:2014xia} whereas rational one-loop EYM amplitudes can
be found in \cite{Nandan:2018ody} at leading order in the gravitational coupling $\kappa$ and
in \cite{Faller:2018vdz} at general orders. Moreover, all-loop results for one-graviton-$n$-gluon 
amplitudes have been obtained in~\cite{Chiodaroli:2017ngp}.
At leading order in $\kappa$, relations between 
gauge-invariant building blocks of one-loop EYM with higher numbers of gravitons 
and (super-)Yang Mills amplitudes 
have been pioneered in \cite{He:2016mzd}. The relations among ``partial integrands'' in 
the reference take a universal form for EYM theories with four to sixteen supercharges,
but they only capture the contributions from gauge multiplets in the loop.

In this work, we describe a method to extend the one-loop EYM amplitude relations of
\cite{He:2016mzd} to arbitrary combinations of gauge and gravity multiplets in the
loop and the external legs, i.e.\ to all orders in $\kappa$. Our results again reduce
loop integrands of EYM to dimension-agnostic partial integrands
of super-Yang-Mills without any coupling to gravity. In particular, the relations
we will derive take a universal form for any non-zero number of supercharges - they are 
expressible in terms of the gauge-invariant partial integrands
of \cite{He:2016mzd} that carry the variable amount of supersymmetry in the double copy. 
The universality of the relations stems from the non-supersymmetric YM$+\phi^3$ constituent
that appears in the double-copy construction of EYM theories with {\it any} number of supercharges. 
The dependence of
our relations on color factors can be straightforwardly adapted to gauge groups $U(N)$ 
and $SU(N)$.

The amplitude relations in this work are derived from forward limits of tree-level building blocks,
where divergences in intermediate expressions are bypassed via ambitwistor-string methods
\cite{Geyer:2015bja, He:2015yua, Geyer:2015jch, Cachazo:2015aol}.\footnote{As detailed
for instance in \cite{Catani:2008xa} and references therein, 
the significance of forward limits for one-loop amplitudes can 
be anticipated from the Feynman tree theorem \cite{Feynman:1963ax}.} This way of implementing
forward limits has been applied to explicitly construct loop integrands of gauge theories 
\cite{He:2017spx, Geyer:2017ela, Edison:2020uzf} and one-loop matrix elements of higher-mass-dimension
operators in the low-energy effective action of superstrings \cite{Edison:2021ebi}. However,
the ambitwistor methods at work lead to an unconventional form of the Feynman propagators
in the loop integrand: The inverse propagators of the partial integrands are
{\it linearized} in the loop momentum $\ell$, i.e.\ given by $2\ell \cdot K + K^2$ instead
of $(\ell + K)^2$ for (combinations of) external momenta $K$ \cite{Geyer:2015bja, He:2015yua, 
Geyer:2015jch, He:2016mzd}.

We provide the detailed form of the loop integrands in four-point EYM amplitudes with 
sixteen and eight supercharges in terms of traditional quadratic propagators $(\ell + K)^{-2}$.
In particular, we address various configurations of
external gauge and gravity multiplets as well as different orders in $\kappa$, i.e.\
all admissible contributions from gauge and/or gravity multiplets in the loop.
The conversion between linearized and quadratic propagators in the loop
is performed via elementary partial-fraction manipulations in the examples
of this work, see \cite{Gomez:2016cqb, Gomez:2017lhy, Gomez:2017cpe,
Ahmadiniaz:2018nvr, Agerskov:2019ryp, Farrow:2020voh, Edison:2021ebi} for more general  
recent discussions of this conversion. However, the supersymmetry-agnostic 
amplitude relations of this work still feature linearized propagators in intermediate
steps. We leave it as an open problem to preserve the universal form of the relations
for EYM loop integrands while manifesting the quadratic propagators $(\ell + K)^{-2}$
of the super-Yang-Mills building blocks.

\subsection*{Outline}
\label{sec:outline}

This paper is organized as follows: In section \ref{sec:2}, we review the EYM double copy including
the CHY methods relevant to this work and state the main formulae for our construction of one-loop 
EYM amplitudes. Their central building blocks are so-called half integrands of YM+$\phi^3$ theory; we spell out the detailed form of their color decomposition in terms of tree-level data for any 
number of external states in
section \ref{sec:hi}. Based on the four-point examples of YM+$\phi^3$ half integrands in 
section \ref{sec:nexthi}, we proceed to constructing four-point one-loop EYM amplitudes
at all orders in the coupling: maximally supersymmetric loop integrands in $D\leq 10$ spacetime
dimensions in section \ref{sec:maxsusy} and half-maximally supersymmetric ones in $D\leq 6$ in
section \ref{sec:halfmax}. In both cases, we expose all supersymmetry cancellations and convert
the output of the CHY double copy to quadratic Feynman propagators. Moreover, the chiral
fermions in six-dimensional EYM theories with 8 supercharges give rise to gauge- and diffeomorphism
anomalies whose integrated results can be found in section \ref{sec:anomaly}.

This work is supplemented by three appendices, starting with a review of the general form 
of CHY integrands for EYM tree amplitudes in appendix \ref{app:ymphi3tree}.  Moreover,
we have gathered background information on kinematic factors with half-maximal supersymmetry
and rational Feynman integrals in the six-dimensional anomalies in appendices \ref{app:cs}
and \ref{app:gaugemeth}, respectively. Finally, some of our 
results are available in machine-readable form in an ancillary file 
of the arXiv submission of this work.


\section{Review and basics}
\label{sec:2}

In this section, we review the basics of the EYM double copy, the CHY formulation of its tree-level amplitudes and the construction of one-loop amplitudes from forward limits of trees in the ambitwistor framework.


\subsection{Einstein-Yang-Mills as a double copy}
\label{sec:2.1}

The oldest incarnation of the double-copy structure of perturbative gravity is the
KLT formula for its $n$-point tree-level amplitudes \cite{Kawai:1985xq}
\begin{align}
M^{\rm tree}_{n, {\rm GR}} = \sum_{\rho,\tau \in S_{n-3}} A^{\rm tree}_{\rm YM}(1,\rho,n{-}1,n) S(\rho|\tau)_1
\bar A^{\rm tree}_{\rm YM}(1,\tau,n,n{-}1)\, .
\label{review.3}
\end{align}
On the right hand side, $\rho=\rho(2,3,\ldots,n{-}2)$ and $\tau$ are permutations of $n{-}3$ legs in
the color-ordered gauge-theory amplitudes $A^{\rm tree}_{\rm YM}(1,2,\ldots,n)$ referring to the coefficient
of ${\rm Tr}(t^{a_1} t^{a_2}\ldots t^{a_n})$ in the color 
decomposition. The entries
of the $(n{-}3)! \times (n{-}3)!$ KLT matrix $S(\rho|\tau)_1$ \cite{Bern:1998sv, momentumKernel} 
are degree-$(n{-}3)$ polynomials in Mandelstam invariants 
\beq
s_{ij} = k_i\cdot k_j = \frac{1}{2}(k_i{+}k_j)^2 \, , \ \ \ \ \ \ s_{ ij \ldots p} = \frac{1}{2}(k_{i}{+}k_j{+}\ldots{+}k_p)^2 \, ,
\label{review.4}
\eeq
(such as $S(2|2)_1 = -s_{12}$ at four points), where all the external momenta $k_i$ are 
lightlike throughout this work. The $(n{-}3)!$ permutations of 
$A^{\rm tree}_{\rm YM},\bar A^{\rm tree}_{\rm YM}$
form a basis of color-ordered amplitudes via BCJ relations \cite{BCJ} and are
therefore sufficient to generate a permutation invariant gravity amplitude via (\ref{review.3}).

The KLT formula calculates the tree-level amplitudes in a variety of further theories 
with double-copy structure including EYM \cite{Chiodaroli:2014xia}, Born-Infeld and special
Galileons \cite{Cachazo:2014xea, Cachazo:2016njl} as well as even open-string 
theories \cite{Mafra:2011nv, Zfunctions, Azevedo:2018dgo}. In general, (\ref{review.3}) yields the tree amplitudes $M^{\rm tree}_{n, B\otimes C}$ in the double-copy
theory $B\otimes C$ such as general relativity (GR) from the color-ordered amplitudes 
$A^{\rm tree}_{B},\bar A^{\rm tree}_{C}$ in theories $B$ and $C$, say Yang-Mills (YM). 
Since (\ref{review.3}) features the outer products of the polarizations of theories
$B$ and $C$, the spins of the external legs add up under double copy. The color degrees of
freedom common to theories $B$ and $C$ in turn are stripped off.

In the case of $B\otimes C = {\rm EYM}$, the constituent theories are $B= {\rm YM}$ and
an extended gauge theory $C= {\rm YM}+\phi^3$ \cite{Chiodaroli:2014xia} whose Lagrangian
($\mu,\nu=0,1,\ldots,D{-}1$ in $D$ spacetime dimensions)
\begin{align}
{\cal L}_{{\rm YM}+\phi^3} &= - \frac{1}{4} F^a_{\mu \nu} F^{a \mu \nu} + \frac{1}{2}(D_\mu \phi^A)^a (D^\mu \phi^A)^a 
- \frac{g^2}{4} f^{abe} f^{ecd} \phi^{aA} \phi^{bB} \phi^{cA} \phi^{dB}  \notag \\
&\quad+ \frac{ \lambda g}{3!} f^{abc} \hat f^{ABC} 
\phi^{aA} \phi^{bB} \phi^{cC} 
\label{review.1}
\end{align}
with coupling constants $g,\lambda$ involves
gluons $A_\mu = A_\mu^a t^a$ coupled to bi-adjoint scalars $\Phi= \phi^{aA} t^a \otimes T^A$
with two species of gauge-group generators $t^a, T^A$ and respective structure constants
$f^{abc},\hat f^{ABC}$. Our conventions for the non-linear gluon field-strengths $F_{\mu \nu}^a$
and gauge-covariant derivatives $D_\mu$ are
\begin{align}
F_{\mu \nu}^a &= \partial_\mu A_\nu^a - \partial_\nu A_\mu^a + g f^{abc} A_\mu^b A_\nu^c \notag \\
(D_\mu \phi^A)^a &= \partial_\mu \phi^{aA} + g f^{abc} A_\mu^b \phi^{cA} \, .
\label{review.2}
\end{align}
In adapting the KLT formula (\ref{review.3}) to EYM, the color-ordering of the ${\rm YM}+\phi^3$
amplitudes is performed w.r.t.\ the generators $t^a$ common to the scalars and the gauge bosons
(rather than the $T^A$ exclusive to scalars).
In other words, the color-dressed $n$-point amplitudes\footnote{For gauge-group generators $t^{a}$ 
and $T^A$, we will often abbreviate the adjoint indices
$a_i \rightarrow i$ and $A_i \rightarrow i$ referring to the color degrees of freedom of 
the $i^{\rm th}$ external leg and write $t^i$ and $T^i$ in the place of $t^{a_i}$ and $T^{A_i}$, respectively.}
\begin{align}
M^{\rm tree}_{n, {\rm YM}+\phi^3} = \sum_{\rho \in S_{n-1}} {\rm Tr}(t^1 t^{\rho(2)} t^{\rho(3)} \ldots t^{\rho(n)} ) A^{\rm tree}_{{\rm YM}+\phi^3}(1,\rho(2,3,\ldots,n)) 
\label{review.6}
\end{align}
decompose into color-ordered amplitudes $A_{{\rm YM}+\phi^3}$ entering (\ref{review.3})
that still depend on the $T^A$ of the external scalars. In case of scalars in the first legs $1,2,\ldots,r$,
further color decomposition w.r.t.\ $T^A$ gives rise to doubly-partial amplitudes $m^{\rm tree}_{{\rm YM}+\phi^3}$,
\beq
A^{\rm tree}_{{\rm YM}+\phi^3}(1,\rho)  = \sum_{\gamma \in S_{r-1}} 
{\rm Tr}(T^1 T^{\gamma(2)} T^{\gamma(3)} \ldots T^{\gamma(r)} )
m^{\rm tree}_{{\rm YM}+\phi^3}(1,\rho|1,\gamma) + {\rm multitrace} \, .
\label{review.7} 
\eeq
The multitrace terms receive contributions from the contractions $\phi^{aA} \phi^{bB} \phi^{cC} \phi^{dD} \delta_{AC} \delta_{BD}$ in (\ref{review.1}) and gluon propagators. They realize all 
cyclically inequivalent partitions of the 
$r$ scalars into up to $\lfloor \frac{r}{2} \rfloor$ traces, e.g.\ three permutations 
of ${\rm Tr}(T^1 T^2) {\rm Tr}(T^3 T^4)$ at 
four points. In the double copy to EYM, the $T^A$ are re-interpreted as the gauge-group generators of the 
YM states. For $r$ external gluons in the first legs $1,2,\ldots,r$ and $n{-}r$ external gravitons, the color-decomposition (\ref{review.7}) as well as the KLT formula (\ref{review.3}) carry over to
\begin{align}
M^{\rm tree}_{n,r, {\rm EYM}} &=\sum_{\gamma \in S_{r-1}} 
{\rm Tr}(T^1 T^{\gamma(2)}  \ldots T^{\gamma(r)} )
A^{\rm tree}_{\rm EYM}(1,\gamma(2,\ldots,r)) + {\rm multitrace}
\label{review.8} \\
A^{\rm tree}_{\rm EYM}(1,\gamma(2,\ldots,r)) &= \sum_{\rho,\tau \in S_{n-3}} A^{\rm tree}_{\rm YM}(1,\rho,n{-}1,n) S(\rho|\tau)_1
m^{\rm tree}_{{\rm YM}+\phi^3}(1,\tau,n,n{-}1|1,\gamma)\, .
\label{review.9} 
\end{align}
The results of this work concern the explicit form of the analogous one-loop
amplitudes. More specifically, we present a general method to determine $n$-point loop integrands
of EYM from one-loop building blocks of (super-)Yang-Mills and spell out the detailed
form of four-point examples. Our relations between loop integrands apply to
supersymmetric EYM theories in any number of 
spacetime dimensions $D$ compatible with the
variable amount of supersymmetry, and the
four-point examples in sections \ref{sec:maxsusy} and \ref{sec:halfmax} 
preserve 8 or 16 supercharges.

The double copy for supersymmetric
EYM has all the supersymmetries on the YM side while taking ${\rm YM}+\phi^3$ as a purely bosonic
theory with color structure $t^a$ for each gluon and $t^a \otimes T^A$ for each scalar.
Furthermore, in the double copy to EYM, the coupling constants $g$ and $\lambda$ from the constituent theories are mapped onto the gauge coupling $g$ and gravitational coupling $\kappa$ of EYM theory according to
\begin{align}\label{review.10}
(g^2,\lambda)\rightarrow \left(\frac{\kappa}{4},~ 4~ \frac{g}{\kappa}\right)\, .
\end{align}
In four-point one-loop amplitudes of EYM, for instance, (\ref{review.10})
leads to the powers of $\kappa$ and $g$ listed in table \ref{tab:couplings}. 
In the examples of sections \ref{sec:maxsusy} and \ref{sec:halfmax}, we will determine the 
contributions at the orders of $g^m \kappa^{4-m}$ ($0\leq m \leq 4$)
for different numbers of external gluons and gravitons.

\begin{table}
	\begin{center}
		\begin{tabular}{|c|c || c  |}\hline 
			YM+$\phi^3$& YM& EYM \\\hline \hline
			$g^4\lambda^4$ & $g^4$ & $g^4$\\ \hline
			$g^4\lambda^3$ & $g^4$ & $\frac{1}{4}\kappa g^3$\\ \hline
			$g^4\lambda^2$ & $g^4$ & $\frac{1}{16} \kappa^2 g^2$\\ \hline
			$g^4\lambda$ & $g^4$ &$\frac{1}{64} \kappa^3 g$ \\ \hline
			$g^4$ & $g^4$ &$\frac{1}{256}\kappa^4$\\ \hline
		\end{tabular}
	\end{center}
\caption{In four-point one-loop amplitudes of ${\rm YM}\otimes ({\rm YM}+\phi^3)={\rm EYM}$, 
the couplings $g$ and $\lambda$ on the left-hand side of the double copy are mapped to
the following powers of the couplings $\kappa$ and $g$ in EYM via (\ref{review.10}).}
\label{tab:couplings}
\end{table} 

\subsection{The Einstein-Yang-Mills double copy in the CHY formalism}
\label{sec:2.2}

There are several equivalent formulations of the double-copy structure of tree-level amplitudes
including the KLT formula (\ref{review.3}), the BCJ double copy based on cubic-vertex 
diagrams \cite{BCJ, loopBCJ, Bern:2010yg} and the CHY formalism \cite{Cachazo:2013gna, Cachazo:2013hca, Cachazo:2013iea}. The backbone of the CHY formulae for $n$-point tree-level amplitudes 
are moduli-space integrals over punctures $\sigma_1,\sigma_2,\ldots, \sigma_n$ on 
the Riemann sphere $\mathbb C \cup \{\infty\}$ which are completely localized by the 
scattering equations. The latter are imposed by the delta functions in the measure
\beq
\dd \mu^{\rm tree}_n = \frac{ \dd \sigma_1 \, \dd \sigma_2 \, \ldots \, \dd \sigma_n }{{\rm vol} \, {\rm SL}_2(\mathbb C)} \prod_{i=1}^n{}' \delta \bigg(\sum_{j=1 \atop{j\neq i}}^n \frac{ s_{ij} }{\sigma_{ij}} \bigg) \, , \ \ \ \ \ \ \sigma_{ij} = \sigma_i - \sigma_j \, .
\label{chysec.1}
\eeq
The inverse ${\rm vol} \, {\rm SL}_2(\mathbb C)$ and the prime along with the product
instruct to drop any three $\dd \sigma_i \, \dd \sigma_j \, \dd \sigma_k$ and 
the associated delta functions, and the respective punctures
can be fixed to $(\sigma_i,\sigma_j,\sigma_k) \rightarrow (0,1,\infty)$ after inserting
the Jacobian $|\sigma_{ij} \sigma_{ik} \sigma_{jk}|^2$. The CHY reformulation of 
the KLT double copy then reads
\beq
M_{n,B\otimes C}^{\rm tree} = \int \dd \mu^{\rm tree}_n \, I_B^{\rm tree}(\{1,2,\ldots,n\})  \,
 I_C^{\rm tree}(\{1,2,\ldots,n\}) \, ,
\label{chysec.2}
\eeq
where the integral is over the moduli space ${\cal M}_{0,n}$ of $n$ marked points on the Riemann sphere.
The so-called half integrands $I^{\rm tree}_B,I^{\rm tree}_C$ are functions of the $\sigma_j$
that both transform with weight two under M\"obius transformations $\sigma_j \rightarrow \frac{ a\sigma_j + b }{c \sigma_j + d}$
with $( \smallmatrix a &b \\ c &d \endsmallmatrix) \in {\rm SL}_2(\mathbb C)$. Moreover, the half integrands
$I^{\rm tree}_B,I^{\rm tree}_C$ depend on momenta and polarization or color degrees of freedom of the particles enclosed in $\{\ldots\}$.

\subsubsection{Basic half integrands for color and kinematics}
\label{sec:2.2.0}

The CHY formulae for bi-adjoint scalars, YM  and gravity are based on two types of half integrands:
\begin{itemize}
\item Color degrees of freedom are encoded in
\beq
I_{\phi^3}^{\rm tree}(\{1,2,\ldots,n\}) = \sum_{\rho \in S_{n-1}} {\rm Tr}(t^1 t^{\rho(2)} t^{\rho(3)}\ldots t^{\rho(n)}) {\rm PT}(1,\rho(2,3,\ldots,n))
\label{chysec.3}
\eeq
with Parke-Taylor factor
\beq
{\rm PT}(1,2,\ldots,n) = \frac{1}{\sigma_{12} \sigma_{23}\ldots \sigma_{n-1,n} \sigma_{n1}}  \, .
\label{chysec.4}
\eeq
Upon color ordering w.r.t.\ two species of gauge-group generators, Parke-Taylor integrals
\beq
m_{\phi^3}(1,2,\ldots,n|\rho(1,2,\ldots,n)) = \int  \dd \mu^{\rm tree}_n  \, {\rm PT}(1,2,\ldots,n){\rm PT}(\rho(1,2,\ldots,n))
\label{chysec.5}
\eeq
yield the doubly-partial amplitudes of biadjoint scalars. We will frequently apply this formula 
to determine Parke-Taylor integrals from the straightforward Feynman-diagram computation 
of $m_{\phi^3}$,
for instance using the Berends-Giele recursion of \cite{Mafra:2016ltu}.
\item The dependence on polarization vectors $\epsilon_j$ (subject to transversality $\epsilon_{j}\cdot k_j=0$) in YM and gravity is carried by the reduced Pfaffian
\beq
I^{\rm tree}_{\rm YM}(\{1,2,\ldots,n\}) = {\rm Pf}' \Psi_n( \{1,2,\ldots,n\}) = \frac{ (-1)^{i+j} }{\sigma_{ij}} {\rm Pf}\big[  \Psi_n(\{1,2,\ldots,n\})\big]^{ij}_{ij} \, ,
\label{chysec.6}
\eeq
where $^{ij}_{ij}$ instruct to remove the $i^{\rm th}$ and $j^{\rm th}$ rows and columns from
the $2n\times 2n$ matrix~$\Psi_n$,
\begin{align}
\Psi_n(\{1,2,\ldots,n\})&= \ccb A &-C^t \\ C &B \cce &B_{ij} &= \left\{ \begin{array}{cl} \frac{ \epsilon_i\cdot \epsilon_j }{\sigma_{ij}} &: \ i\neq j \\ 0 &: \ i=j \end{array} \right. \label{chysec.7} \\
A_{ij} &= \left\{ \begin{array}{cl} \frac{ s_{ij} }{\sigma_{ij}} &: \ i\neq j \\ 0 &: \ i=j \end{array} \right. 
&C_{ij} &= \left\{ \begin{array}{cl} \frac{ \epsilon_i\cdot k_j }{\sigma_{ij}} &: \ i\neq j \\ -\sum^n_{m \neq i} \frac{ \epsilon_i \cdot k_m}{\sigma_{im}} &: \ i=j \end{array} \right. \, .
\notag
\end{align}
The reduced Pfaffian is linear in all of $\epsilon_{1},\epsilon_{2},\ldots,\epsilon_{n}$ and, on the support
of the scattering equations in (\ref{chysec.1}), independent on the choice of $i,j\in \{1,2,\ldots, n\}$ 
in (\ref{chysec.6}) and invariant under linearized gauge transformations $\epsilon_{m} \rightarrow k_m$.
\end{itemize}
%

\subsubsection{Integrands of EYM from half integrands of YM$+\phi^3$}
\label{sec:2.2.1}

The CHY formula (\ref{chysec.2}) for EYM tree-level amplitudes with $r$ gauge bosons and $n{-}r$ gravitons
reads \cite{Cachazo:2014nsa, Cachazo:2014xea}
\beq
M_{n,{\rm EYM}}^{\rm tree} = \int  \dd \mu^{\rm tree}_n \, 
 {\rm Pf}' \Psi_n( \{1,2,\ldots,n\}) \, I_{{\rm YM}+\phi^3}^{\rm tree}(\{1,2,\ldots,r\}; \{r{+}1,\ldots,n\}) \, ,
\label{chysec.0}
\eeq
and supersymmetric EYM amplitudes can be obtained by replacing the Pfaffian (\ref{chysec.6}) 
by its fermionic completion\footnote{For ten-dimensional SYM, the supersymmetrization of
the Pfaffian may be imported from the open pure-spinor superstring \cite{Gomez:2013wza},
based on the correlation functions of
massless vertex operators in \cite{Mafra:2011kj, Mafra:2011nv, Mafra:2015vca}. Alternatively, 
the simplified spin-field correlation functions of \cite{Edison:2020uzf} yield an analogue of
the Pfaffian (\ref{chysec.7}) with two and four fermions and arbitrary numbers of bosons
among the external states, also see \cite{Frost:2017} for two fermions.}.
The non-supersymmetric half integrand $I_{{\rm YM}+\phi^3}^{\rm tree}$ refers to
YM$+\phi^3$ theory with Lagrangian (\ref{review.1}). It depends on the gauge-group 
generators $T^j$ of the external scalars $j=1,2,\ldots,r$ in the first set of labels and the polarizations
$\epsilon_j$ of the external gauge bosons $j=r{+}1,\ldots,n$ in the second set.
By analogy with the color decomposition (\ref{review.7}), it will be convenient to
separately analyze color-ordered half integrands $J_{{\rm YM}+\phi^3}^{\rm tree}$ 
for single and double traces,
\begin{align}
J_{{\rm YM}+\phi^3}^{\rm tree}(1,2,\ldots,r;\{r{+}1,\ldots,n\}) &= I^{\rm tree}_{{\rm YM}+\phi^3} \, \big|_{{\rm Tr}(T^1 T^2\ldots T^r)} \label{chysec.8} \\
J_{{\rm YM}+\phi^3}^{\rm tree}(1,2,\ldots,p|p{+}1,\ldots,r;\{r{+}1,\ldots,n\}) &= I^{\rm tree}_{{\rm YM}+\phi^3} \, \big|_{{\rm Tr}(T^1 \ldots T^p)
{\rm Tr}(T^{p+1} \ldots T^r)}\, , \notag 
\end{align}
where vertical bars are used to separate multiple traces (e.g.\ $J_{{\rm YM}+\phi^3}^{\rm tree}(\ldots|\ldots|\ldots;\{r{+}1,\ldots,n\})$ for triple traces). The curly-bracket notation $\{\ldots\}$ refers to permutation-invariant functions of the data of the enclosed particles, e.g.\
$J_{{\rm YM}+\phi^3}^{\rm tree}(\ldots;\{i,j,\ldots\})=J_{{\rm YM}+\phi^3}^{\rm tree}(\ldots;\{j,i,\ldots\})$.
By contrast, the $J_{{\rm YM}+\phi^3}^{\rm tree}$ in (\ref{chysec.8}) are 
only cyclically invariant in each of the slots associated with trace structures, e.g.\ 
\beq
J_{{\rm YM}+\phi^3}^{\rm tree}( \ldots | i_1,i_2,\ldots,i_p|\ldots;\{\ldots\}) 
=J_{{\rm YM}+\phi^3}^{\rm tree}( \ldots |i_2,i_3,\ldots,i_p,i_1|\ldots;\{\ldots\}) \, .
\eeq
The single-trace instances of the half integrands are given by \cite{Cachazo:2014nsa}
\beq
J_{{\rm YM}+\phi^3}^{\rm tree}(1,2,\ldots,r;\{r{+}1,\ldots,n\})  = {\rm PT}(1,2,\ldots,r) {\rm Pf} \, \Psi_n(\{r{+}1,\ldots,n\})\, ,
\ \ \ \ r \geq 2
 \label{chysec.11}
\eeq
and combine the Parke-Taylor factor (\ref{chysec.4}) from  $I_{\phi^3}^{\rm tree}$ with the
Pfaffian from $I_{\rm YM}^{\rm tree}$. The $2(n{-}r)\times 2(n{-}r)$ matrix $\Psi_n(\{r{+}1,\ldots,n\})$ 
slightly generalizes the definition (\ref{chysec.7}) of the matrix $\Psi_n(\{1,\ldots,n\})$:
The sum over $m$ in any $C_{ii} = - \sum_{m=1}^n \epsilon_i \cdot k_m/\sigma_{im}$ entering (\ref{chysec.11}) 
runs over all of $\{1,2,\ldots,n\}$ instead of the shorter list $\{r{+}1,\ldots,n\}$ of gluon labels. Note that (\ref{chysec.11}) only applies to $r\geq 2$ scalars. For $n$ gluons, the results of appendix \ref{app:ymphi3tree} lead to the half integrand 
 \begin{align}
 \label{chysec.14}
 J_{{\rm YM}+\phi^3}^{\rm tree}(\emptyset;\{1,2,\ldots,n\})=  {\rm Pf}' \Psi_n( \{1,2,\ldots,n\}) = I^{\rm tree}_{\rm YM}(\{1,2,\ldots,n\})\, ,
 \end{align}
 and tree amplitudes with a single external scalar ($r=1$) vanish. For single-trace amplitudes of
 $r=n$ scalars in turn, we recover the result  of the pure $\phi^3$ theory in (\ref{chysec.3}),
 \beq
 J_{{\rm YM}+\phi^3}^{\rm tree}(1,2,\ldots,n;\emptyset)=
  J_{\phi^3}^{\rm tree}(1,2,\ldots,n)=
 {\rm PT}(1,2,\ldots,n)\, .
 \label{oldphicube}
 \eeq 
In the double-trace situation, the simplest cases with zero and one gluon are \cite{Cachazo:2014nsa, Cachazo:2014xea}
\begin{align}
J_{{\rm YM}+\phi^3}^{\rm tree}(1,2,\ldots,p|p{+}1,\ldots n;\emptyset)  &= 
s_{12\ldots p}  {\rm PT}(1,2,\ldots,p)   {\rm PT}(p{+}1,\ldots,n) 
 \label{chysec.12} \\
 J_{{\rm YM}+\phi^3}^{\rm tree}(1,2,\ldots,p|p{+}1,\ldots n{-}1;\{n\})  &= 
  {\rm PT}(1,2,\ldots,p)   {\rm PT}(p{+}1,\ldots,n{-}1) 
  \label{chysec.13} \\
&\hspace{-3.5cm}  \times \bigg[
  \sum_{i=1}^{p-1} \sum_{j=i+1}^p \frac{ 
 (s_{j,n} \epsilon_{n}\cdot k_i - s_{i,n} \epsilon_{n}\cdot k_j)
    \sigma_{ij} }{\sigma_{i,n} \sigma_{n,j}}
  +s_{12\ldots p} \sum_{j=1}^{n-1} \frac{ \epsilon_{n} \cdot k_j }{\sigma_{j,n}}
  \bigg]\, ,
   \notag
\end{align}
and more general half integrands of ${\rm YM}+\phi^3$ can be found in appendix \ref{app:ymphi3tree}.

\subsubsection{Kleiss-Kuijf relations}
\label{sec:2.2.2}

We note for future reference that partial-fraction relations between Parke-Taylor factors
imply so-called Kleiss-Kuijf relations \cite{Kleiss:1988ne} for $J_{{\rm YM}+\phi^3}^{\rm tree}$: 
Within each trace, different cyclic orderings are related by 
\beq
J_{{\rm YM}+\phi^3}^{\rm tree}(\ldots | i,P,j,Q |\ldots ;\{r{+}1,\ldots,n\}) = (-1) ^{|Q|} \! \! \! \!
\sum_{R \in P\shuffle \tilde Q} \! \! \! \! J_{{\rm YM}+\phi^3}^{\rm tree}(\ldots | i,R,j |\ldots ;\{r{+}1,\ldots,n\})\, .
\label{KKrels}
\eeq
The shuffle product $\shuffle$ is defined recursively by
\begin{align}
Pa \shuffle Qb = (P\shuffle Qb) a + (Pa\shuffle Q) b  \label{hiexample.9}
\end{align}  
for any two words $P=(p_1, p_2,\ldots,p_{|P|})$ and $Q=(q_1,q_2,\ldots,q_{|Q|})$ 
concatenated with words $a$ and $b$ of length one and
$P \shuffle \emptyset = \emptyset \shuffle P = P $ in case of the empty word $\emptyset$. 
The length of the word $Q$ is denoted by $|Q|$, and we use a tilde-notation for the reversal
$\tilde Q=(q_{|Q|},\ldots,q_2,q_1)$. Intuitively, the shuffle product $P \shuffle Q$ collects all
possibilities to interleave the letters in $P$ and $Q$ while preserving the order 
among the $p_i$ and $q_j$.


\subsection{One-loop CHY formulae and forward limits}
\label{sec:2.3}

The CHY formulae for tree-level amplitudes were underpinned by ambitwistor-string theories
in the Ramond-Neveu-Schwarz \cite{Mason:2013sva, Adamo:2013tsa} and
pure-spinor formulations \cite{Berkovits:2013xba, Adamo:2015hoa}. The ambitwistor-string 
prescription for one-loop amplitudes is centered on moduli-space integrals over punctured
genus-one surfaces or tori, and all integrations are again localized by
scattering equations involving a $D$-dimensional loop momentum $\ell$. 
In particular, the manipulations of \cite{Geyer:2015bja, Geyer:2015jch} 
localize the integrand at the cusp $\tau \rightarrow i\infty$, where the torus 
with modular parameter $\tau$ degenerates to a nodal Riemann
sphere. This limit reduces the genus-one scattering equations for the punctures to ($i=1,2,\ldots,n$)
\beq
\frac{ \ell \cdot k_i }{\sigma_i} + \sum_{j=1\atop{j\neq i}}^n \frac{k_i \cdot k_j}{\sigma_{ij}} =0
\label{looprev.1}
\eeq
which are the forward limits $k_{\pm} \rightarrow \pm \ell$ of the genus-zero scattering 
equations in (\ref{chysec.1}) with two additional legs $+,-$. Accordingly, $D$-dimensional
$n$-point one-loop amplitudes in theories $B\otimes C$ with double-copy structure are
given by \cite{Geyer:2015bja, Geyer:2015jch}
\beq
M^{\te{1-loop}}_{n, B \otimes C} = \int \frac{ \dd^D \ell}{\ell^2} \lim_{k_{\pm} \rightarrow \pm \ell} \int \dd \mu_{n+2}^{\te{tree}} \,I^{\te{1-loop}}_B(\{1,2,\ldots,n\};\ell) \,I^{\te{1-loop}}_C(\{1,2,\ldots,n\};\ell)
\label{looprev.2}
\eeq
in terms of forward limits of $(n{+}2)$-point CHY integrals at tree level. For theories with an ambitwistor-string
description, the half integrands $I^{\te{1-loop}}_B,I^{\te{1-loop}}_C$ can be obtained from correlation functions on
a torus in its degeneration limit to a nodal Riemann sphere. For instance, the maximally supersymmetric four-point
correlators known from type-I and type-II superstrings \cite{Green:1982sw} give rise to
\beq
I^{\te{1-loop}}_{{\rm YM,max}}=
I^{\te{1-loop}}_{{\rm YM,max}}(\{1,2,3,4\};\ell)  = t_8(1,2,3,4) \sum_{\rho \in S_4} {\rm PT}(+,\rho(1,2,3,4),-)\,,
\label{looprev.3}
\eeq
where the external polarizations conspire to the permutation-symmetric $t_8$-tensor
contracting the linearized field strengths $f_j^{\mu \nu}$ (not to be confused with
structure constants)
\begin{align}
t_8(1,2,3,4) &= {\rm tr}(f_1f_2f_3f_4) - \frac{1}{4}  {\rm tr}(f_1f_2) {\rm tr}(f_3f_4) + {\rm cyc}(2,3,4) \notag \\
&= s_{12} s_{23} A_{\rm YM}^{\rm tree}(1,2,3,4)
\label{looprev.4} \\
f_i^{\mu \nu} &= k_i^\mu \epsilon_i^\nu - k_i^\nu \epsilon_i^\mu \, .
\notag
\end{align}
Similar to (\ref{looprev.3}), the half integrands $I^{\te{1-loop}}_{{\rm YM}}(\{1,2,\ldots,n\};\ell)$ 
of gauge theories with 0 to 16 supercharges can be expressed in terms of
$(n{+}2)$-point Parke-Taylor factors (\ref{chysec.4}) involving the double points $(\sigma_{+},\sigma_-)
\rightarrow (0,\infty)$ of the nodal Riemann sphere. This can for instance be seen from the 
conformal-field-theory origin of these
correlators\footnote{One can again import simplified correlators from the chiral-splitting formulation
\cite{DHoker:1988pdl, DHoker:1989cxq} of superstrings to the ambitwistor setup such as the four-point expression (\ref{looprev.3}) from \cite{Green:1982sw}
and the multiparticle correlators of \cite{Tsuchiya:1988va} for external bosons and of \cite{Mafra:2018qqe}
for the entire gauge multiplet.} \cite{Adamo:2013tsa} and the manipulations of Ramond-Neveu-Schwarz spin sums in \cite{He:2017spx} based on \cite{Tsuchiya:1988va}.

Alternatively, (\ref{looprev.3}) as well as generalizations to higher multiplicity and reduced
supersymmetry \cite{Geyer:2015jch, Frost:2017, He:2017spx} can be obtained from forward limits of genus-zero half integrands.
For one-loop amplitudes in pure YM, the forward limit of the Pfaffian of (\ref{chysec.7}) was shown
to match the appropriate correlation functions on the torus \cite{Geyer:2015jch} and to yield combinations
of Parke-Taylor factors that reproduce known amplitude representations \cite{Geyer:2017ela}.
Apart from aligning the momenta $k_{\pm} \rightarrow \pm \ell$, the forward limit in pairs of
external gluons $+,-$ amounts to the replacement (see section \ref{sec:3.1.1} for the analogous color
replacement rules for scalars)
\begin{align}
\sum_{+,-} \epsilon_+^\mu \epsilon_-^\nu &= \Delta^{\mu \nu} \, , \ \ \ \ \ \ \eta_{\mu \nu} \Delta^{\mu \nu} = D{-}2 
\label{looprev.5}  \\
V_\mu W_\nu \Delta^{\mu \nu} &= V\cdot W \, , \ \ \ \ \ \  V,W \in \{ k_i,\epsilon_i : \ i=1,2,\ldots,n\}\, .
\notag
\end{align}
Similarly, by combining forward limits (\ref{looprev.5}) in gluon polarizations with those in 
fermions and scalars yields $n$-point correlators with supersymmetric multiplets
in the loop such as (\ref{looprev.3}) 
\cite{Edison:2020uzf}.

The key idea in this work is to extend the construction of one-loop half integrands
from forward limits of tree-level ones to the YM$+\phi^3$ theory. As will be detailed in section \ref{sec:hifromfw},
forward limits of the tree-level half integrands of YM$+\phi^3$ in the gluons and scalars
yield the non-supersymmetric half integrands 
\begin{align}
I^{\te{1-loop}}_{{\rm YM}+\phi^3}(\{1,\ldots,r\};\{r{+}1,\ldots,n\};\ell)
&=  \lim_{k_{\pm} \rightarrow \pm \ell}   \sum_{+,-}  \big[  I^{\te{tree}}_{{\rm YM}+\phi^3}(\{1,2,\ldots,r,+,-\};\{r{+}1,\ldots,n\})
\notag \\
&\quad \quad +I^{\te{tree}}_{{\rm YM}+\phi^3}(\{1,2,\ldots,r\};\{r{+}1,\ldots,n,+,-\}) \big]
\label{fwlimits}
\end{align}
in one-loop amplitudes of EYM with $r$ external 
gluons and $n{-}r$ external gravitons:
\beq
M^{\te{1-loop}}_{n, {\rm EYM}, \alpha} = \int \frac{ \dd^D \ell}{\ell^2} \lim_{k_{\pm} \rightarrow \pm \ell} \int \dd \mu_{n+2}^{\te{tree}} \,I^{\te{1-loop}}_{\rm YM,\alpha}(\{1,\ldots,n\};\ell) \,I^{\te{1-loop}}_{{\rm YM}+\phi^3}(\{1,\ldots,r\};\{r{+}1,\ldots,n\};\ell)
\label{looprev.6}
\eeq
The amount of supersymmetry will be indicated by the parameter $\alpha$ on the left-hand side
and in the subscript of the YM half integrand on the right-hand side (e.g.\ $\alpha =$max or $\frac{1}{2}$-max).
In the same way as $I^{\rm tree}_{{\rm YM}+\phi^3}$ are available in Parke-Taylor form \cite{Nandan:2016pya, Teng:2017tbo, Du:2017gnh, Edison:2020ehu},\footnote{It follows from the work of Aomoto in the 
mathematics literature \cite{Aomoto87} that an arbitrary half integrand in the tree-level 
formula (\ref{chysec.2}) with ${\rm SL}_2(\mathbb C)$-weight two in all of 
$\sigma_1,\sigma_2,\ldots,\sigma_n$ can be 
decomposed in terms of Parke-Taylor factors.} 
the forward limits in (\ref{fwlimits}) can be brought into the $n!$-term form
\begin{align}
I^{\te{1-loop}}_{{\rm YM}+\phi^3}(\ell)&= \sum_{\rho \in S_n} 
N_{+|\rho(12\ldots n)| -}(\ell)  {\rm PT}(+,\rho(1,2,\ldots,n),-) \, ,
\label{looprev.7}
\end{align}
where $N_{+|\rho(12\ldots n)| -}(\ell)$ are local combinations of the color degrees of freedom, momenta
and polarizations of the YM$+\phi^3$ states. Together with the Parke-Taylor representations of
$I^{\te{1-loop}}_{\rm YM,\alpha}$ \cite{He:2017spx, Geyer:2017ela, Edison:2020uzf}, all the
$\dd \mu_{n+2}^{\te{tree}}$-integrals in (\ref{looprev.6}) can be straightforwardly performed in
terms of doubly-partial amplitudes of biadjoint $\phi^3$ via (\ref{chysec.5}).  In this procedure, 
the forward limit $k_{\pm} \rightarrow \pm \ell$ of the integrals over $\sigma_j$ should be 
performed {\it after} summing the permutations to avoid divergences. More specifically, forward-limit 
divergences occur in non-supersymmetric theories and can be addressed in the ambitwistor framework 
using the methods of \cite{Geyer:2017ela}.\footnote{In general non-supersymmetric
one-loop CHY formulae, the regularization of forward-limit divergences can be traced
back to the dropout of singular solutions of the scattering equations \cite{He:2015yua, Cachazo:2015aol}.} 
The supersymmetric examples in sections \ref{sec:maxsusy} and \ref{sec:halfmax} will be unaffected by 
subtleties related to forward-limit divergences, and the discussion of half integrands of YM$+\phi^3$
in sections \ref{sec:hi} and \ref{sec:nexthi} are completely supersymmetry-agnostic.

\subsection{Linearized versus quadratic propagators}
\label{sec:2.4}

\begin{figure}
\begin{center}
\begin{tikzpicture} [scale=0.75, line width=0.30mm]
\begin{scope}[xshift=-0.8cm]
\draw (0.5,0)--(-0.5,0);
\draw (-0.5,0)--(-0.85,-0.35);
\draw [dashed](-0.85,-0.35)--(-1.2,-0.7);
\draw (0.5,0)--(1.2,-0.7);
\draw[dashed] (-1.2,-1.7)--(-1.2,-0.7);
\draw (1.2,-1.7)--(1.2,-0.7);
\draw (1.2,-1.7)--(0.85,-2.05);
\draw[dashed] (0.85,-2.05)--(0.5,-2.4);
\draw[dashed] (-0.5,-2.4)--(0.5,-2.4);
\draw[dashed] (-0.5,-2.4)--(-1.2,-1.7);
\draw (-0.5,0)--(-0.7,0.4)node[left]{$n$};
\draw (0.5,0)--(0.7,0.4)node[right]{$1$};
\draw (1.2,-0.7)--(1.6,-0.5)node[right]{$2$};
\draw (1.2,-1.7)--(1.6,-1.9)node[right]{$3$};
\draw (0,0) node{$| \! |$};
\draw (-0.25,0.2)node{$-$};
\draw (0.25,-0.2)node{$+$};
\end{scope}
\draw[-> ](1.7,-1.2)  -- (3.2,-1.2);
\begin{scope}[xshift=1.7cm, yshift=0.5cm]
\draw(11.4,-2)node{$+ \ {\rm cyclic}(1,2,\ldots,n)$};
\draw (2.9,-2)node[left]{$+\ell$} -- (5.8,-2);
\draw (7.2,-2) -- (8.1,-2)node[right]{$-\ell$};
\draw (3.5,-2) -- (3.5,-1.5)node[above]{$1$};
\draw (4.5,-2) -- (4.5,-1.5)node[above]{$2$};
\draw (5.5,-2) -- (5.5,-1.5)node[above]{$3$};
\draw[dashed] (5.8,-2) -- (7.2,-2);
\draw (7.5,-2) -- (7.5,-1.5)node[above]{$n$};
\end{scope}
\end{tikzpicture}
\caption{Terms in the partial-fraction representation of $n$-point loop integrals
in (\ref{looprev.11}) can be interpreted 
as $(n{+}2)$-point tree-level diagrams.}
\label{linprops}
\end{center}
\end{figure}
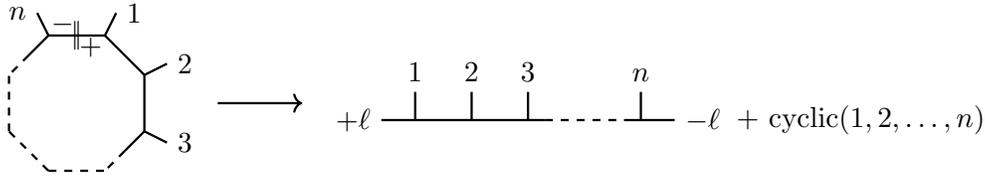

Given that doubly-partial amplitudes (\ref{chysec.5}) at tree level are functions of $k_i\cdot k_j$
with all the $k_j^2$ set to zero, the forward limit in (\ref{looprev.2}) and (\ref{looprev.6})
can only involve the loop momentum via $k_i\cdot \ell$ and not via $\ell^2$.
Hence, the square of $\ell$ only enters the loop integrand of $M^{\te{1-loop}}_{n, B \otimes C}$ as 
a global prefactor $\ell^{-2}$ outside the
$\dd \mu_{n+2}^{\te{tree}}$-integral, see \cite{Geyer:2015bja, Geyer:2015jch} for its origin.
This can be reconciled with the Feynman propagators $(\ell +K)^{-2}$ (with combinations
$K$ of external momenta) expected for loop
diagrams in massless field theories by the following rewriting of an $n$-gon integral,
\begin{align}
\int {2^{n-1} \ {\rm d}^D \ell \over \ell^2 \ell_1^2 \ell_{12}^2 \ldots \ell_{12\ldots n-1}^2}
&= \sum_{i=0}^{n-1} \int { 2^{n-1} \ {\rm d}^D \ell \over \ell_{12\ldots i}^2 } \prod^n_{j \neq i} {1\over \ell_{12\ldots j}^2 - \ell_{12\ldots i}^2 } \label{looprev.11} \\
&= \sum_{i=0}^{n-1} \int { {\rm d}^D \ell \over \ell^2}   \prod_{j=0}^{i-1} {1\over s_{j+1,j+2,\ldots,i,-\ell}} \prod_{j=i+1}^{n-1} {1\over s_{i+1,i+2,\ldots,j,\ell} } \, , \notag
\end{align}
where our notation for composite momenta and $\ell$-dependent Mandelstam invariants is
\beq
k_{12\ldots p} = \sum_{j=1}^p k_j \, , \ \ \ \ \ \ \ell_{12\ldots p} = \ell+ k_{12\ldots p} \, , \ \ \ \ \ \ s_{12\ldots p,\pm \ell} = s_{12\ldots p} \pm \ell \cdot k_{12\ldots p} \, .
\label{looprev.12}
\eeq
The first step of (\ref{looprev.11}) is based on partial-fraction manipulations, and we have 
shifted the loop momentum by external momenta in passing to the second line such that the 
only quadratic propagator is $\ell^{-2}$ rather than $(\ell + K)^{-2}$. The transition to linearized 
propagators in (\ref{looprev.11}) can be straightforwardly extended to
massive corners. As visualized in figure \ref{linprops}, each of the $n$ terms in the partial-fraction
decomposition (\ref{looprev.11}) of the $n$-gon can be thought of as the $(n{+}2)$-point tree-level diagram
obtained from cutting one of the $n$-gon propagators.

With the $n$ possibilities to cut a given $n$-gon and its $(n{-}1)!$ cyclically inequivalent orderings, 
the linearized $n$-gon propagators can be associated with $n!$ tree-level diagrams of half-ladder
topology. These are the master diagrams in the sense of the BCJ color-kinematics 
duality \cite{BCJ, loopBCJ}, i.e.\ their kinematic
numerators w.r.t.\ linearized propagators generate all the lower-gon numerators by kinematic
Jacobi identities. The $n!$ Parke-Taylor coefficients $N_{+|\rho(12\ldots n)| -}(\ell)$ in (\ref{looprev.7}) 
with $\rho \in S_n$ are the master numerators associated with the ladder diagram in figure~\ref{linprops} 
and its permutations in $1,2,\ldots,n$. The kinematic Jacobi identities that determine the remaining 
cubic-diagram numerators (of $(n{-}1)$-gons, $(n{-}2)$-gons etc.) can be traced back to properties of the 
$\dd \mu_{n+2}^{\te{tree}}$-integrals over Parke-Taylor factors in (\ref{looprev.6}).

However, the forward limit of individual doubly-partial amplitudes (\ref{chysec.5}) will usually not 
recombine to quadratic propagators. It requires the sum over $n$ terms in (\ref{looprev.11}) to 
recover the $n$-gon integral with Feynman propagators $(\ell + k_{12\ldots p})^{-2}$ by reversing 
the shifts of loop momentum and the partial-fraction manipulations. It will be an important cross-check 
for the $I^{\te{1-loop}}_{{\rm YM}+\phi^3}$ to be derived from forward limits (\ref{fwlimits}) that 
their $\dd \mu_{n+2}^{\te{tree}}$-integrals against $I^{\te{1-loop}}_{{\rm YM},\alpha}$
admit a recombination to quadratic propagators.  We will do so at the level of color-ordered 
EYM amplitudes where, in analogy with the tree-level notation (\ref{chysec.8}),
\begin{align}
&J^{\te{1-loop}}_{{\rm YM}+\phi^3}(1,2,\ldots,j | j{+}1,\ldots,p | \ldots ;\{r{+}1,\ldots,n\};\ell)  \label{looprev.15} \\
& \ \ = I^{\te{1-loop}}_{{\rm YM}+\phi^3}(\{1,2,\ldots,r\};\{r{+}1,\ldots,n\};\ell) \big|_{ {\rm Tr}(T^1 T^2 \ldots T^j) 
	{\rm Tr}(T^{j+1} \ldots T^p)  \ldots  }\, .
\notag
\end{align} 
We can express the loop integrand of color-ordered EYM amplitudes in terms of so-called partial 
integrands in YM theory
\begin{align}
a^{\te{1-loop}}_{ \text{YM},\alpha }(+,\rho(1,\dots,n),-) = \lim_{k_{\pm} \rightarrow \pm \ell} \int \dd \mu_{n+2}^{\te{tree}} {\rm PT}(+,\rho(1,\ldots,n),-) I^{\te{1-loop}}_{\rm YM,\alpha}(\{1,\ldots,n\};\ell)
\label{defpartint}
\end{align}
which were introduced in \cite{He:2016mzd} as gauge-invariant building blocks of loop integrands
in linearized-propagator representations. The idea is to calculate the $\dd \mu_{n+2}^{\te{tree}}$-integral separately for each Parke-Taylor factor in the decomposition (\ref{looprev.7}) of the YM+$\phi^3$
half integrand (\ref{looprev.15}) and to thereby obtain the decomposition
\begin{align}
&A^{\te{1-loop}}_{ {\rm EYM}, \alpha}(1,2,\ldots,j | j{+}1,\ldots,p | \ldots ;\{r{+}1,\ldots,n\} ) \label{looprev.14}\\
&=  \int \frac{ \dd^D \ell}{\ell^2} \! \! \lim_{k_{\pm} \rightarrow \pm \ell} \int \dd \mu_{n+2}^{\te{tree}} \,
J^{\te{1-loop}}_{{\rm YM}+\phi^3}(1,\ldots,j | j{+}1,\ldots,p | \ldots ;\{r{+}1,\ldots,n\};\ell)
I^{\te{1-loop}}_{\rm YM,\alpha}(\{1,\ldots,n\};\ell)  \notag\\
& = \int \frac{ \dd^D \ell}{\ell^2}   \sum_{\rho \in S_n} 
{N}_{+|\rho(12\ldots n)| -}(\ell)\big|_{ {\rm Tr}(T^1 T^2 \ldots T^j) 
{\rm Tr}(T^{j+1} \ldots T^p)  \ldots  } a^{\te{1-loop}}_{ \text{YM},\alpha }(+,\rho(1,2,\dots,n),-)\, .\notag
\end{align}
The partial integrands $a^{\te{1-loop}}_{ \text{YM},\alpha }$ with maximal and half-maximal supersymmetry are available from forward limits of doubly-partial amplitudes \cite{He:2016mzd, He:2017spx} and will be reviewed in sections \ref{sec:pimax} and \ref{sec:pihalfmax}, respectively. While the $a^{\te{1-loop}}_{ \text{YM},\alpha }$ are still given in terms of linearized propagators, the single- and multi-trace amplitudes $A^{\te{1-loop}}_{ {\rm EYM}, \alpha}$ admit quadratic-propagator representations. By combining the contributions of
all the $n!$ permutations $\rho$ in (\ref{looprev.14}), we will find the expected
quadratic-propagator expressions for EYM loop-integrands at $n=4$ for different amounts
$\alpha$ of supersymmetry, and separately at each order in the couplings $g$ and $\kappa$ (see
table \ref{tab:couplings}).


\section{YM+$\phi^3$ half integrands at one loop: all-multiplicity results}
\label{sec:hi}

The first step in the construction of one-loop amplitudes in EYM theories is to obtain the non-supersymmetric 
half integrands $I^{\te{1-loop}}_{{\rm YM}+\phi^3}$ in their CHY representation (\ref{looprev.6}). 
In this section we investigate the color decomposition of these one-loop half integrands in 
YM+$\phi^3$ theory, with single- and multi-trace coefficients $J^{\te{1-loop}}_{{\rm YM}+\phi^3}$
entering the color-ordered EYM amplitudes in (\ref{looprev.14}). 
 In particular, we explain how to choose the forward limits of tree-level half integrands in (\ref{fwlimits})
that contribute to a given color-ordered one-loop half integrand $J^{\te{1-loop}}_{{\rm YM}+\phi^3}$. 
Four-point examples can be found in the next section \ref{sec:nexthi}.

\subsection{First look at one-loop YM+$\phi^3$ half integrands from forward limits}
\label{sec:hifromfw}

 In YM+$\phi^3$ theory, various forward limits of tree-level half integrands in both scalars and gluons contribute to the same color-ordered one-loop half integrand. The basic premise for the choice of these contributions is to include all forward limits of $(n{+}2)$-point tree-level half integrands that have the external states, color structure and powers of the couplings $\lambda,g$ compatible with the desired $n$-point one-loop half integrand. 
 
\subsubsection{Color management in forward limits} 
 \label{sec:3.1.1}
 
In EYM theories with gauge groups $SU(N)$ and $U(N)$, forward limits
in color degrees of freedom may change the number of traces. At the level of
the double-copy constituents YM+$\phi^3$, this concerns the trace structure of the
generators $T^A$ specific to the bi-adjoint scalars which changes upon forward
limits in the scalars. The sum $\sum_{+,-}$ over adjoint degrees of freedom $T^+ = T^{A_+}$ 
and $T^- = T^{A_-}$ of scalar legs $+$ and $-$ in (\ref{fwlimits}) is 
implemented through the completeness relations
\begin{align}
\sum_{+,-} (T^+)_i{}^j (T^-)_k{}^l  &=  \delta^l_i \delta^j_k &&: \ U(N)
\label{extend.1} \\
\sum_{+,-} (T^+)_i{}^j (T^-)_k{}^l  &= \delta^l_i \delta^j_k - \frac{1}{N} \delta^j_i \delta^l_k  &&:\ SU(N) \, ,
\notag
\end{align}
with fundamental indices $i,j,k,l = 1,2,\ldots,N$.
Depending on the relative positions of legs $+$ and $-$, 
the forward limits of traces follow from one of
\begin{align}
\sum_{+,-}\Tr(P,+,Q,-) =\, &\Tr(P)\Tr(Q)+c_1 \Tr(PQ) \notag\\
 \sum_{+,-} \Tr(P,+,-) =\, &c_2\Tr(P) \label{colsums1}\\
 \sum_{+,-} \Tr(+,-)=\, &N c_2 \notag\\ 
\sum_{+,-} \Tr(P,+) \Tr(Q,-)  =\, &\Tr(PQ)+c_1\Tr(P)\Tr(Q)\, . \notag
  \end{align}
We employ the shorthand notation  
 \begin{align}
 \label{hiformfw.2}
 \Tr(i_1,i_2,\ldots,i_n)= \Tr(T^{A_{i_1}}T^{A_{i_2}}\dots T^{A_{i_n}})
 \end{align}  
and use capital letters from the second half of the alphabet for words 
$P=(i_1 ,i_2,\ldots,i_{|P|})$ of length $|P|$ as
in section \ref{sec:2.2.2}. Moreover, the factors $c_1$ and $c_2$ subject to 
$c_2-c_1=N$ differ for gauge groups $U(N),SU(N)$ and are listed in table \ref{tab:colorfactors}. 
The first and last line of (\ref{colsums1}) illustrate that the number of traces may be raised
or lowered under forward limits in color factors.
 
 	\begin{table}
 		\begin{center}
 		 \begin{tabular}{|c||c|c|}\hline 
 		 	factor & $U(N)$ & $SU(N)$\\\hline \hline &&
 		 	\\[-1em]
 		 	$c_1$&$0$&$-\frac{1}{N}$ \\\hline &&\\[-1em]
 		 	$c_2$&$N$&$\frac{N^2-1}{N}$ \\\hline
 \end{tabular}
\end{center}
\caption{The color factors $c_1$ and $c_2$ subject to $c_2-c_1=N$ arise in the forward limits of traces
over generators of the gauge groups $U(N)$ and $SU(N)$, see (\ref{colsums1}).} 
\label{tab:colorfactors}
 \end{table} 
 
\subsubsection{Coupling dependence in forward limits} 
 \label{sec:3.1.2}
 
\begin{figure}
\begin{center}
\begin{tikzpicture} [scale=1.25, line width=0.30mm]
\draw (-1,0)node[left]{$\ldots$} -- (0,0);
\draw (0,0) -- (0,1)node[above]{$\phi$};
\draw (1,0) -- (1,1)node[above]{$\phi$};
\draw (1,0) -- (2,0)node[right]{$\phi$};
\draw(0,0)node[below]{$g\lambda$} -- (1,0) node[below]{$g\lambda$};
\draw[->](-1,0.5) .. controls (-2.5,0.5) .. (-2.5,-0.5);
\draw(-2.7,0.6)node{$\times\lambda^{-1}$};
\draw(3.7,0.6)node{$\times\lambda^{-2}$};
\draw[->](2,0.5) .. controls (3.5,0.5) .. (3.5,-0.5);
\scope[xshift=-3cm, yshift=-2cm]
\draw (-1,0)node[left]{$\ldots$} -- (0,0);
\draw (0,0) -- (0,1)node[above]{$\phi$};
\draw (1,0) -- (1,1)node[above]{$\phi$};
\draw(2,0)--(2,0)node[right]{$A$};
\draw(0,0)node[below]{$g\lambda$} -- (1,0) node[below]{$g$};
\begin{feynman}
            \vertex (g1)  at (0.9,0) {};
            \vertex (g2) at (2.15,0)  {};
            \diagram* {
                (g1) -- [gluon]  (g2),
            };
\end{feynman}
\endscope
\scope[xshift=3cm, yshift=-2cm]
\draw (-1,0)node[left]{$\ldots$} -- (0,0);
\draw (0,0) -- (0,1)node[above]{$\phi$};
\draw (1,0) -- (1,1)node[above]{$\phi$};
\draw (1,0) -- (2,0)node[right]{$\phi$};
\draw(0,0) -- (0,0) node[below]{$g$};
\draw(1,0) -- (1,0) node[below]{$g$};
\begin{feynman}
            \vertex (g1)  at (-0.1,0) {};
            \vertex (g2) at (1.15,0)  {};
            \diagram* {
                (g1) -- [gluon]  (g2),
            };
\end{feynman}
\endscope
\end{tikzpicture}
\caption{Trading an external scalar for a gluon effectively removes a power of 
$\lambda$ (left panel). Similarly, increasing the number of traces 
effectively removes two powers of $\lambda$ (right panel).}
\label{countgs}
\end{center}
\end{figure}
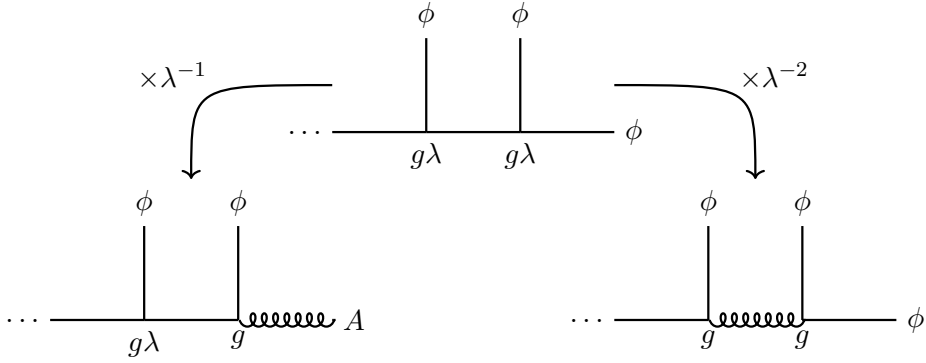

Even though we defined the half integrands $J^{\te{tree}}_{{\rm YM}+\phi^3}$ to 
exclude the couplings in the ${\rm YM}+\phi^3$ Lagrangian (\ref{review.1}), one 
can straightforwardly associate powers of $g$ and $\lambda$ to each trace- and
external-state configuration of $J^{\te{tree}}_{{\rm YM}+\phi^3}$: Single-trace
integrands of $n$ external scalars generate diagrams with $n{-}2$ cubic vertices $\sim \phi^3$,
so they are associated with the powers $(g \lambda)^{n-2}$. 
Cubic vertices involving one or three gluons in turn involve a factor of
$g$ rather than $g \lambda$. As exemplified in figure \ref{countgs}, 
this reduces the power-counting of $\lambda$ by one whenever an external 
scalar is replace by an external gluon and by two for each additional trace.
Hence, we associate
\begin{align}
\te{$m$-trace} \ J^{\te{tree}}_{{\rm YM}+\phi^3} \ \te{@ $r\geq 2$ scalars \& $n{-}r$ gluons} \ &\leftrightarrow \ g^{n-2} \lambda^{r-2m} \, ,\label{extend.2}
\end{align}
for instance $g^{n-2} \lambda^{r-2}$ to single-trace examples and $g^{n-2} \lambda^{r-4}$ to
double-trace examples with $r\geq 2$ scalars and $n{-}r$ gluons. 

By the color-factor identities (\ref{colsums1}), forward limits in a scalar can introduce 
or eliminate one trace. As a result, the tree-level power counting of (\ref{extend.2})
yields a bandwidth of three different powers of $\lambda$
in each $J^{\te{1-loop}}_{{\rm YM}+\phi^3}$ with $m\geq 2$ traces, where the lowest power 
$\lambda^{r-2m}$ also arises from forward limits in a pair of gluons,
\begin{align}
&\te{$m$-trace} \ J^{\te{1-loop}}_{{\rm YM}+\phi^3} \ \te{@ $r\geq 2$ scalars \& $n{-}r$ 
gluons with $m \geq 2$} 
\label{extend.3}
\\
 &\leftrightarrow \ \left\{ \begin{array}{rl}g^{n} \lambda^{r-2m+4} \ \& \ g^{n} \lambda^{r-2m+2} \ \& \ g^{n} \lambda^{r-2m} &: \te{scalar forward limit} \\
 g^{n} \lambda^{r-2m} &: \te{gluon forward limit} \, . \end{array} \right.  \notag
\end{align}
For single-trace half integrands at one loop, the option with the highest power of $\lambda$
in (\ref{extend.3}) is absent since there is no underlying forward limit with fewer traces.
Hence, one arrives at only two dependences on the couplings for single traces,
where the lower power $\lambda^{r-2}$ also arises from forward limits in gluons. 
\begin{align}
&\te{single-trace} \ J^{\te{1-loop}}_{{\rm YM}+\phi^3} \ \te{@ $r\geq 2$ scalars \& $n{-}r$ gluons} \label{extend.4} \\ 
&\leftrightarrow \  \left\{ \begin{array}{rl}  g^{n} \lambda^{r} \ \& \ g^{n} \lambda^{r-2} &: \te{scalar forward limit} \\
g^{n} \lambda^{r-2} &: \te{gluon forward limit}\, . \end{array} \right. \notag
\end{align}
Upon double copy through the CHY-representation (\ref{looprev.14}) of EYM amplitudes,
the dictionary (\ref{review.10}) for the couplings leads to the following power-counting on $g$ and $\kappa$,
\begin{align}
&\te{single-trace} \ A^{\te{1-loop}}_{ {\rm EYM}, \alpha} \ \te{@ $r\geq 2$ gluons \& $n{-}r$ gravitons}\label{extend.4a} \\
&\leftrightarrow \ \left\{ \begin{array}{rl}  g^{r} \kappa^{n-r} \ \& \ g^{r-2} \kappa^{n+2-r}  &: \te{gluon forward limit} \\
g^{r-2} \kappa^{n+2-r} &: \te{graviton forward limit} \end{array} \right. \notag
\\
&\te{$m$-trace} \ A^{\te{1-loop}}_{ {\rm EYM}, \alpha} \ \te{@ $r\geq 2$ gluons \& $n{-}r$ gravitons
with $m\geq 2$}  \notag \\ 
&\leftrightarrow \ \left\{ \begin{array}{rl} g^{r-2m+4} \kappa^{n+2m-4-r}  \ \& \ 
g^{r-2m+2} \kappa^{n+2m-2-r} \ \& \ g^{r-2m} \kappa^{n+2m-r} &: \te{gluon forward limit} \\ 
g^{r-2m} \kappa^{n+2m-r}  &: \te{graviton forward limit}\ .\end{array} \right. \notag
\end{align}


\subsection{Explicit single-trace YM+$\phi^3$ half integrands at one loop}
\label{sec:1trace}

We shall now spell out the detailed decomposition
of single-trace YM+$\phi^3$ half integrands at one loop
into forward limits of their color-ordered tree-level counterparts $J^{\te{tree}}_{{\rm YM}+\phi^3}$.
This will be done separately for the two orders in the coupling noted in (\ref{extend.4}), i.e.\
for both half integrands on the right-hand side of
\begin{align}
J^{\te{1-loop}}_{{\rm YM}+\phi^3}(1,2,\ldots,r; \{r{+}1,\ldots,n\} ;\ell) 
&= J^{\te{1-loop}}_{{\rm YM}+\phi^3}(1,2,\ldots,r; \{r{+}1,\ldots,n\} ;\ell)\big|_{g^n\lambda^r}
\label{extend.5} \\
&\quad
+J^{\te{1-loop}}_{{\rm YM}+\phi^3}(1,2,\ldots,r; \{r{+}1,\ldots,n\} ;\ell)\big|_{g^n\lambda^{r-2}} \, . \notag
\end{align}

\subsubsection{No external gluons}

In order to simplify the bookkeeping, we shall focus on the case with $r=n$ scalars first.
The contribution $\sim \lambda^n$ to (\ref{extend.5}) can then be obtained by adding up
all the forward limits that yield the single-trace $\Tr(1,2,\ldots,n)$ on the right-hand side of (\ref{colsums1}).
As we shall see, both $U(N)$ and $SU(N)$ gauge groups give rise to the same $\lambda^{n}$ contributions
\begin{align}
 J^{\te{1-loop}}_{{\rm YM}+\phi^3}(1,2,\ldots,n; \emptyset ;\ell)\big|_{g^n\lambda^n}
 &= N \big[
 J^{\te{tree}}_{{\rm YM}+\phi^3}(1,2,\ldots,n,+,-;\emptyset) 
 +  J^{\te{tree}}_{{\rm YM}+\phi^3}(1,2,\ldots,n,-,+;\emptyset)  \notag \\
 &\ \ \ \ \ \
 +{\rm cyc}(1,2,\ldots,n) \big]\, . \label{extend.7}
\end{align}
With the Parke--Taylor form (\ref{oldphicube}) of the single-trace half integrands of
the pure $\phi^3$-theory, this reproduces the color
factors in the ambitwistor-string formulae for planar one-loop super-Yang-Mills amplitudes
\cite{Geyer:2015bja, Geyer:2015jch}. Still, it is instructive to see how it arises from the forward-limit
computations and a careful tracking of all single traces at the $\lambda^n$ order in
(\ref{colsums1}): Intermediate steps towards (\ref{extend.7}) give
\begin{align}
 J^{\te{1-loop}}_{{\rm YM}+\phi^3}(1,2,\ldots,n; \emptyset ;\ell)\big|_{g^n\lambda^n} &= c_2 \big[
 J^{\te{tree}}_{{\rm YM}+\phi^3}(1,2,\ldots,n,+,-;\emptyset) 
 +  J^{\te{tree}}_{{\rm YM}+\phi^3}(1,2,\ldots,n,-,+;\emptyset)\big]  \notag \\
 &\! \! \! \! \! \! \! \! \! \! \! \! \! \! \! \! \! \! \! \!  \! \! \! \! \! 
  + c_1 \sum_{j=1}^{n-1} J^{\te{tree}}_{{\rm YM}+\phi^3}(1,2,\ldots,j,+,j{+}1,\ldots,n,-;\emptyset) 
 +{\rm cyc}(1,2,\ldots,n) \, ,
  \label{extend.9}
 \end{align}
where the special case $J^{\te{tree}}_{{\rm YM}+\phi^3}(1,2,\ldots,n,-,+;\emptyset) 
+{\rm cyc}(1,2,\ldots,n,-)=0$ of the Kleiss-Kuijf relations (\ref{KKrels}) identifies the
coefficient of $c_1$ to be minus the coefficient of $c_2$. By virtue of the relation
$c_2-c_1=N$ universal to $SU(N)$ and $U(N)$, one arrives at the simplified expression~(\ref{extend.7}).

At the subleading order of $\lambda^{n-2}$ in turn, the $N$-dependence varies between $U(N)$ and $SU(N)$
through the color factor $c_2$ in table \ref{tab:colorfactors} (with $Nc_2=N^2$ for $U(N)$ and
$Nc_2=N^2-1$ for $SU(N)$),
\begin{align}
 J^{\te{1-loop}}_{{\rm YM}+\phi^3}(1,2,\ldots,n; \emptyset ;\ell)\big|_{g^n\lambda^{n-2}}
 &= N c_2   J^{\te{tree}}_{{\rm YM}+\phi^3}(1,\ldots,n|+,-;\emptyset) 
 + J^{\te{tree}}_{{\rm YM}+\phi^3}(1,\ldots,n;\{+,- \})  \notag \\
 & \! \! \! \! \! \! \! \! \! \! \! \! \! \! \! \! \! \! \! \!  \! \! \! \! \! \! \! \! \! 
 +\sum_{j=1}^{n-1} \big[ J^{\te{tree}}_{{\rm YM}+\phi^3}(1,2,\ldots,j,+|-,j{+}1,\ldots,n;\emptyset) 
 + {\rm cyc}(1,2,\ldots,n) \big]\, . \label{extend.8}
\end{align}
The forward limit in the gluon legs $\{+,-\}$ is understood to incorporate the sum 
over $D{-}2$ physical polarization vectors $\epsilon_+,\epsilon_-$ as in (\ref{looprev.5}).

\begin{figure}
\begin{center}
\begin{tikzpicture} [scale=0.75, line width=0.30mm]
\begin{scope}[xshift=-0.8cm]
 \begin{feynman}
            \vertex (g1)  at (0.7,0) {};
            \vertex (g2) at (-0.7,0)  {};
            \diagram* {
                (g1) -- [gluon]  (g2),
            };
        \end{feynman}
\draw (-0.5,0)--(-0.85,-0.35);
\draw [dashed](-0.85,-0.35)--(-1.2,-0.7);
\draw (0.5,0)--(1.2,-0.7);
\draw[dashed] (-1.2,-1.7)--(-1.2,-0.7);
\draw (1.2,-1.7)--(1.2,-0.7);
\draw (1.2,-1.7)--(0.85,-2.05);
\draw[dashed] (0.85,-2.05)--(0.5,-2.4);
\draw[dashed] (-0.5,-2.4)--(0.5,-2.4);
\draw[dashed] (-0.5,-2.4)--(-1.2,-1.7);
\draw (-0.5,0)--(-0.7,0.4)node[left]{$n$};
\draw (0.5,0)--(0.7,0.4)node[right]{$1$};
\draw (1.2,-0.7)--(1.6,-0.5)node[right]{$2$};
\draw (1.2,-1.7)--(1.6,-1.9)node[right]{$3$};
\end{scope}
\draw[-> ](1.7,-1.2)  -- (3.2,-1.2);
\begin{scope}[xshift=1.7cm, yshift=1.5cm]
 \begin{feynman}
            \vertex (g1)  at (2.6,-2) {};
            \vertex (g2) at (3.8,-2)  {};
            \diagram* {
                (g1) -- [gluon]  (g2),
            };
        \end{feynman}
\draw (3.5,-2)  -- (5.8,-2);
 \draw (2.9,-2)node[left]{$+\ell$};
\draw (7.2,-2) -- (7.5,-2);
\draw (8.1,-2)node[right]{$-\ell$};
\draw (3.5,-2) -- (3.5,-1.5)node[above]{$1$};
\draw (4.5,-2) -- (4.5,-1.5)node[above]{$2$};
\draw (5.5,-2) -- (5.5,-1.5)node[above]{$3$};
\draw[dashed] (5.8,-2) -- (7.2,-2);
\draw (7.5,-2) -- (7.5,-1.5)node[above]{$n$};
 \begin{feynman}
            \vertex (g1)  at (7.3,-2) {};
            \vertex (g2) at (8.4,-2)  {};
            \diagram* {
                (g1) -- [gluon]  (g2),
            };
        \end{feynman}
%
\end{scope}
\begin{scope}[xshift=1.7cm, yshift=-0.5cm]
\draw (2.9,-2)node[left]{$+\ell$} -- (3.5,-2);
\draw (4.5,-2) -- (5.8,-2);
\begin{feynman}
            \vertex (g1)  at (3.3,-2) {};
            \vertex (g2) at (4.8,-2)  {};
            \diagram* {
                (g1) -- [gluon]  (g2),
            };
\end{feynman}
\draw (7.2,-2) -- (8.1,-2)node[right]{$-\ell$};
\draw (3.5,-2) -- (3.5,-1.5)node[above]{$n$};
\draw (4.5,-2) -- (4.5,-1.5)node[above]{$1$};
\draw (5.5,-2) -- (5.5,-1.5)node[above]{$2$};
\draw[dashed] (5.8,-2) -- (7.2,-2);
\draw (7.5,-2) -- (7.5,-1.5)node[above]{$n{-}1$};
%
%
\draw(11.0,-2)node{$+\ (n{-}2 \ {\rm others})$};
\end{scope}
\end{tikzpicture}
\caption{In the partial-fraction decomposition of an $n$-gon diagram with one gluon propagator
and otherwise scalar propagators, the associated tree-level diagrams describe one gluonic
forward limit with internal scalars and $n{-}1$ scalar forward limits with one internal gluon line.}
\label{linprops2}
\end{center}
\end{figure}
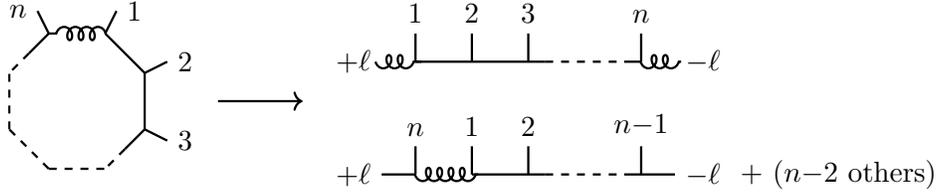

In the expression (\ref{extend.8}) for the $\lambda^{n-2}$-order, the gluonic forward
limits $J^{\te{tree}}_{{\rm YM}+\phi^3}(\ldots;\{+,- \})$ and the scalar forward limits 
$ J^{\te{tree}}_{{\rm YM}+\phi^3}(1,\ldots,j,+|-,j{+}1,\ldots,n;\emptyset) $ capture different
terms in the partial-fraction decomposition of various Feynman integrals. For instance,
the $n$-gon diagram in figure \ref{linprops2} with one gluon propagator and otherwise scalar propagators
yields one tree diagram corresponding to a gluonic forward limit and $n{-}1$ tree diagrams
corresponding to scalar forward limits under partial-fraction decomposition. As a result, 
the recombination of the loop integrand of (\ref{looprev.14}) to quadratic propagators
relies on having the correct relative normalization of the terms 
$J^{\te{tree}}_{{\rm YM}+\phi^3}(\ldots;\{+,- \})$ and 
$ J^{\te{tree}}_{{\rm YM}+\phi^3}(1,\ldots,j,+|-,j{+}1,\ldots,n;\emptyset) $ at the same
order of $g,\lambda$ in (\ref{extend.8}).
Since this recombination has to occur for any value of $N$, its
first term with coefficient $N c_2 $ will separately yield quadratic propagators
after performing the $\dd \mu_{n+2}^{\rm tree}$ integral in (\ref{looprev.14}).

One can also confirm from one-loop Feynman-diagram computations that the only admissible
dependence of color-ordered EYM amplitudes on the group-theory data can occur 
via Kronecker-deltas in the fundamental indices $\delta_i^i = N$ and in the adjoint ones 
$\delta_a^a = N c_2$. This can be manifested in the forward-limit approach of this work
by applying Kleiss-Kuijf relations, both at the single-trace and at the multitrace level.

\subsubsection{Adjoining external gluons}

The single-trace expressions (\ref{extend.7}) and (\ref{extend.8}) can be straightforwardly
generalized to external gluons. The two contributions in (\ref{extend.5}) at different orders
in the couplings then read (for $r\geq 2$ scalars)
\begin{align}
 J^{\te{1-loop}}_{{\rm YM}+\phi^3}(1,2,\ldots,r; \{r{+}1,\ldots,n\} ;\ell)\big|_{g^n\lambda^r}
 &= N \big[
 J^{\te{tree}}_{{\rm YM}+\phi^3}(1,2,\ldots,r,+,-; \{r{+}1,\ldots,n\}) 
  \notag \\
 &\hspace{-3cm} +  J^{\te{tree}}_{{\rm YM}+\phi^3}(1,2,\ldots,r,-,+; \{r{+}1,\ldots,n\})
 +{\rm cyc}(1,2,\ldots,r) \big]  \label{extend.11}
\end{align}
as well as
\begin{align}
 J^{\te{1-loop}}_{{\rm YM}+\phi^3}(1,2,\ldots,r; \{r{+}1,\ldots,n\} ;\ell)\big|_{g^n\lambda^{r-2}}
 &= N c_2   J^{\te{tree}}_{{\rm YM}+\phi^3}(1,2,\ldots,r|+,-; \{r{+}1,\ldots,n\})  \notag \\
 &\! \! \! \! \! \! \! \! \! \! \! \! \! \! \! \! \! \! \! \!  \! \! \! \! \! \! \! \! \!  \! \! \! \! \! \! \! \! \! 
  \! \! \! \! \! \! \! \! \!  \! \! \! \! \! \! \! \! \!  \! \! \! \! \! \! \! \! \! 
 + J^{\te{tree}}_{{\rm YM}+\phi^3}(1,2,\ldots,r; \{r{+}1,\ldots,n,+,-\})  \label{extend.12} \\
 & \! \! \! \! \! \! \! \! \! \! \! \! \! \! \! \! \! \! \! \!  \! \! \! \! \! \! \! \! \!  \! \! \! \! \! \! \! \! \! 
  \! \! \! \! \! \! \! \! \!  \! \! \! \! \! \! \! \! \!  \! \! \! \! \! \! \! \! \! 
 +\sum_{j=1}^{r-1} \big[ J^{\te{tree}}_{{\rm YM}+\phi^3}(1,2,\ldots,j,+|-,j{+}1,\ldots,r; \{r{+}1,\ldots,n\}) 
 + {\rm cyc}(1,2,\ldots,r) \big]\, . \notag
\end{align}
In particular, the simplification of the leading order in $\lambda$ literally follows the discussion
around (\ref{extend.9}) since the tree-level half integrands (\ref{chysec.11}) with external gluons obey
the same Kleiss-Kuijf relations as in the case without gluons.

While ${\rm YM}+\phi^3$ amplitudes
and half integrands with a single scalar vanish, the purely gluonic cases can be obtained by
truncating (\ref{extend.12}) to its first two lines,
\begin{align}
 J^{\te{1-loop}}_{{\rm YM}+\phi^3}(\emptyset; \{1,2,\ldots,n\} ;\ell) &=
  J^{\te{1-loop}}_{{\rm YM}+\phi^3}(\emptyset; \{1,2,\ldots,n\} ;\ell)\big|_{g^n}  \label{trunc.12} \\
 &= N c_2   J^{\te{tree}}_{{\rm YM}+\phi^3}(+,-; \{1,2,\ldots,n\}) 
  + J^{\te{tree}}_{{\rm YM}+\phi^3}(\emptyset; \{1,2,\ldots,n,+,-\})  \, , \notag
\end{align}
see (\ref{chysec.11}) and (\ref{chysec.14}) for the respective tree-level building blocks.
The first and second term on the right-hand side of (\ref{trunc.12}) describe
diagrams with scalars and gluons in the loop, respectively.
 

\subsection{Explicit multi-trace YM+$\phi^3$ half integrands at one loop}
\label{sec:2trace}

The single-trace results of the previous section will now be generalized to multiple traces.
By (\ref{extend.3}), there are three possible dependences on the couplings
\begin{align}
&J^{\te{1-loop}}_{{\rm YM}+\phi^3}(\Tr_1|\Tr_2|\ldots |\Tr_m; P ;\ell)
= J^{\te{1-loop}}_{{\rm YM}+\phi^3}(\Tr_1|\Tr_2|\ldots |\Tr_m; P;\ell) \,\big|_{g^n \lambda^{r-2m+4}}
 \label{extend.15}  \\
& \ \ + J^{\te{1-loop}}_{{\rm YM}+\phi^3}(\Tr_1|\Tr_2|\ldots |\Tr_m; P;\ell) \,\big|_{g^n \lambda^{r-2m+2}}
+ J^{\te{1-loop}}_{{\rm YM}+\phi^3}(\Tr_1|\Tr_2|\ldots |\Tr_m; P ;\ell) \,\big|_{g^n \lambda^{r-2m}}
\notag
\end{align}
for a total of $n{-}r$ gluons in $P=  \{r{+}1,\ldots,n\} $ and $r$ scalars in the union of the cyclically 
ordered sets ${\rm Tr}_1,{\rm Tr}_2,\ldots,{\rm Tr}_m$ with $m\geq 2$. In the same way as the 
combinatorial structure of the single-trace formulae 
(\ref{extend.7}) and (\ref{extend.8}) does not change upon addition of gluons in (\ref{extend.11}) 
and (\ref{extend.12}), also the multitrace results can be presented in a unified way for any choice of $P$.

\subsubsection{Double trace}

In the double-trace example, i.e.\ (\ref{extend.15}) at $m=2$, the three different orders in
couplings contribute with
\begin{align}
J^{\te{1-loop}}_{{\rm YM}+\phi^3}(1,2,\ldots,s|s{+}1,\ldots,r; P ;\ell) \,\big|_{g^n  \lambda^r} &=
\! \! \sum_{Q \in {\rm cyc}(1,2,\ldots,s) \atop{ R \in {\rm cyc}(s+1,\ldots,r)} }  \! \! 
\big[ J^{\te{tree}}_{{\rm YM}+\phi^3}(Q,+,R,-; P)
+(+\leftrightarrow -)\big]  \label{extend.16}  \\
J^{\te{1-loop}}_{{\rm YM}+\phi^3}(1,2,\ldots,s|s{+}1,\ldots,r; P ;\ell) \,\big|_{g^n  \lambda^{r-2}} &=
N \bigg\{  \sum_{Q \in {\rm cyc}(1,2,\ldots,s)  }   
 J^{\te{tree}}_{{\rm YM}+\phi^3}(Q,+,-|s{+}1,\ldots,r; P) \notag \\
&\hspace{-4cm}+ \sum_{R \in {\rm cyc}(s+1,\ldots,r)}
 J^{\te{tree}}_{{\rm YM}+\phi^3}(1,2,\ldots,s | R,+,-; P)+ (+\leftrightarrow -) \bigg\} 
   \label{extend.17} \\
J^{\te{1-loop}}_{{\rm YM}+\phi^3}(1,2,\ldots,s|s{+}1,\ldots,r; P ;\ell) \,\big|_{g^n  \lambda^{r-4}} &= 
N c_2  J^{\te{tree}}_{{\rm YM}+\phi^3}(1,2,\ldots,s|s{+}1,\ldots,r|+,-; P ) \notag \\
&\hspace{-5cm}+ J^{\te{tree}}_{{\rm YM}+\phi^3}(1,2,\ldots,s|s{+}1,\ldots,r; P \cup\{+,-\} )
 \label{extend.18} \\
&\hspace{-5cm}+ \sum_{j=1}^{s-1} \big[ 
J^{\te{tree}}_{{\rm YM}+\phi^3}(1,2,\ldots,j,+|-,j{+}1,\ldots,s|s{+}1,\ldots,r;P )
+ {\rm cyc}(1,2,\ldots,s) \big] \notag\\
&\hspace{-5cm}+ \! \! \sum_{j=1}^{r-s-1} \!  \! \big[ 
J^{\te{tree}}_{{\rm YM}+\phi^3}(  1,\ldots,s| s{+}1,\ldots,s{+}j,+|-,s{+}j{+}1,\ldots,r; P )
+ {\rm cyc}(s{+}1,\ldots,r) \big] \, . \notag
\end{align}
While the orders of $g^n  \lambda^r$ and $g^n  \lambda^{r-4}$ follow from straightforward
application of (\ref{colsums1}), the result (\ref{extend.17}) for the intermediate order 
of $g^n  \lambda^{r-2}$ is based on additional simplifications: The color identities
(\ref{colsums1}) in the first place lead to
\begin{align}
J^{\te{1-loop}}_{{\rm YM}+\phi^3}(1,2,\ldots,s|s{+}1,\ldots,r; P ;\ell) \,\big|_{g^n  \lambda^{r-2}} &=
c_2 \bigg\{  \sum_{Q \in {\rm cyc}(1,2,\ldots,s)  } 
 J^{\te{tree}}_{{\rm YM}+\phi^3}(Q,+,-|s{+}1,\ldots,n; P) \notag \\
&\hspace{-2cm}+ \sum_{R \in {\rm cyc}(s+1,\ldots,r)}
 J^{\te{tree}}_{{\rm YM}+\phi^3}(1,2,\ldots,s | R,+,-; P)+ (+\leftrightarrow -)  \bigg\} 
 \notag \\
&\hspace{-3cm}+ c_1 \sum_{j=1}^{s-1}
J^{\te{tree}}_{{\rm YM}+\phi^3}(1,2,\ldots,j,+,j{+}1,\ldots,s,-|s{+}1,\ldots,r;P)
   \label{extend.19} \\
&\hspace{-3cm}+ c_1 \sum_{j=1}^{r-s-1}
J^{\te{tree}}_{{\rm YM}+\phi^3}(1,2,\ldots,s|s{+}1,\ldots,s{+}j,+,s{+}j{+}1,\ldots ,r,- ;P)
\notag \\
&\hspace{-3cm}+ c_1 \! \! \sum_{Q \in {\rm cyc}(1,2,\ldots,s) \atop{ R \in {\rm cyc}(s+1,\ldots,r)} }  \! \! 
\big[ J^{\te{tree}}_{{\rm YM}+\phi^3}(Q,+| R,-; P)
+ J^{\te{tree}}_{{\rm YM}+\phi^3}(Q,-| R,+; P)\big] \, ,
\notag
\end{align}
but the last line cancels by 
$J^{\te{tree}}_{{\rm YM}+\phi^3}(1,2,\ldots,m,+|\ldots ;P) 
+{\rm cyc}(1,2,\ldots,m)=0$, i.e.\ by Kleiss-Kuijf relations. Moreover, the same type of
Kleiss-Kuijf relations implies that the coefficients of $c_1$ in the
third and fourth line of (\ref{extend.19}) conspire to minus the
coefficient of $c_2$ in the first two lines. Based on $c_2-c_1=N$,
one can then confirm the global prefactor in (\ref{extend.17});
the same recombination was already noted in the single-trace context
of (\ref{extend.9}).

As mentioned before, one can anticipate from one-loop Feynman diagrams
that the only $N$-dependence in (\ref{extend.16}) to (\ref{extend.18}) has to
occur via one of $\delta_i^i= N$ and $\delta_a^a= N c_2$.
 
\subsubsection{Any number of traces} 

The multitrace generalizations of the double-trace results (\ref{extend.16}) to (\ref{extend.18})
are given as follows for $r$ scalars in $\Tr_1,\ldots,\Tr_m$
with $m\geq 2$ and $n{-}r$ gluons in $P=\{r{+}1,\ldots,n\}$
\begin{align}
 J^{\te{1-loop}}_{{\rm YM}+\phi^3}(\Tr_1|\Tr_2|\ldots |\Tr_m; P;\ell) \,\big|_{g^n \lambda^{r-2m+4}} &=
\sum_{1\leq i<j}^m \sum_{Q \in {\rm cyc}(\Tr_i) \atop{ R \in {\rm cyc}(\Tr_j)} }
 \label{extend.21}  \\
&\hspace{-3.7cm}
J^{\te{tree}}_{{\rm YM}+\phi^3}(Q,+,R,-|\Tr_1|\ldots | \widehat \Tr_i |\ldots | \widehat \Tr_j |\ldots | \Tr_m;P)+(+\leftrightarrow -)
\notag \\
J^{\te{1-loop}}_{{\rm YM}+\phi^3}(\Tr_1|\Tr_2|\ldots |\Tr_m; P;\ell) \,\big|_{g^n \lambda^{r-2m+2}} &=
N  \sum_{j=1}^m \sum_{Q \in {\rm cyc}(\Tr_j)}   \label{extend.22} \\
&\hspace{-3.7cm}
J^{\te{tree}}_{{\rm YM}+\phi^3}(Q,+,-|\Tr_1|\ldots |  \widehat \Tr_j |\ldots | \Tr_m;P)
+(+\leftrightarrow -)  \notag \\
J^{\te{1-loop}}_{{\rm YM}+\phi^3}(\Tr_1|\Tr_2|\ldots |\Tr_m; P ;\ell) \,\big|_{g^n \lambda^{r-2m}} &= 
N c_2  J^{\te{tree}}_{{\rm YM}+\phi^3}(\Tr_1|\Tr_2|\ldots |\Tr_m|+,-; P ) \notag \\
&\hspace{-4.5cm}+ J^{\te{tree}}_{{\rm YM}+\phi^3}(\Tr_1|\Tr_2|\ldots |\Tr_m; P \cup\{+,-\} )
 \label{extend.23} \\
&\hspace{-4.5cm}+ \sum_{j=1}^{m} \bigg\{ \sum_{\Tr_j = QR \atop{|Q|,|R| \neq 0}}
J^{\te{tree}}_{{\rm YM}+\phi^3}(Q,+|R,-|  \Tr_1|\ldots |  \widehat \Tr_j |\ldots | \Tr_m; P )
+ {\rm cyc}(\Tr_j) \bigg\} \, . \notag
\end{align}
The notation $\widehat \Tr_j $ instructs to omit the respective trace, and
the sum in the last line is over all possibilities to split $\Tr_j = (c_1,c_2,\ldots,c_r)$
into non-empty words $Q=(c_1,\ldots, c_{|Q|})$ and $R=(c_{|Q|+1},\ldots,c_{r})$ with $|Q|=1,2,\ldots,r{-}1$.
The prefactor $N$ of (\ref{extend.22}) again follows from Kleiss-Kuijf relations\footnote{More
specifically, the cancellation of the last line of (\ref{extend.19}) straightforwardly generalizes to
insertions of $+$ and $-$ into any pair of $\Tr_i, \Tr_j$ with $1\leq i<j\leq m$. Similarly, the 
conspiration of $c_1,c_2$ in the first four lines of (\ref{extend.19}) to $c_2-c_1=N$ occurs for the
terms where $+,-$ are inserted into the same $\Tr_j$ with $j=1,2,\ldots,m$.
All of these manipulations are again based on the Kleiss-Kuijf relations (\ref{KKrels})
that hold for any number of traces or gluon insertions}; together with the first
term in (\ref{extend.23}), the complete color dependence lines up with the
factors $\delta_i^i= N$ or $\delta_a^a= N c_2$ expected from Feynman diagrams.
 

\subsection{Parke-Taylor form of the tree-level building blocks}
\label{sec:ptaylor} 
 
The forward-limit representations of $ J^{\te{1-loop}}_{{\rm YM}+\phi^3}$ in 
sections \ref{sec:1trace} and \ref{sec:2trace} are particularly convenient 
if the $\dd \mu^{\rm tree}_{n+2}$ integration over
the punctures in one-loop CHY formula (\ref{looprev.6}) can be performed via (\ref{chysec.5}) in terms of
doubly-partial amplitudes. This is the case when all the contributing $ J^{\te{tree}}_{{\rm YM}+\phi^3}$
are organized in terms of $(n{+}2)$-point Parke-Taylor factors as we assumed in
(\ref{looprev.7}) and in passing to the last line of (\ref{looprev.14}).

In order to make use of the Parke-Taylor decompositions of tree-level half integrands 
in the literature \cite{Nandan:2016pya, Teng:2017tbo, Du:2017gnh}
or the \texttt{Mathematica} package \cite{Edison:2020ehu}, we 
relegate the forward limit $k_{\pm} {\rightarrow} \pm \ell$ to the last step of the computation,
i.e.\ {\it after} performing the $\dd \mu^{\rm tree}_{n+2}$-integral in terms of doubly-partial
amplitudes. This has been implicitly assumed in introducing the partial integrands $a^{\te{1-loop}}_{ \text{YM},\alpha }$ in (\ref{looprev.14}). 


\subsubsection{Examples at leading order in $\lambda$ with external gluons}
\label{sec:ptaylor.1} 

The simplest non-trivial examples of Parke-Taylor decompositions (\ref{looprev.7}) arise
for the $\lambda^r$-order of the single-trace half integrands (\ref{extend.11}) with external gluons.
Their tree-level constituents are given in (\ref{chysec.11}) and in the one-gluon case for instance
reduce to \cite{Stieberger:2016lng, Nandan:2016pya}
\beq
J_{{\rm YM}+\phi^3}^{\rm tree}(1,2,\ldots,r,+,-;\{p\}) = \sum_{j=0}^{r}
\epsilon_p \cdot (k_{-}{+}k_1{+} \ldots{+}k_j) \PT(-,1,2,\ldots,j,p,j{+}1,\ldots,r,+)\, .
 \label{extend.25}
\eeq
The resulting one-loop half integrand in (\ref{extend.11}) then becomes
\begin{align}
&J^{\te{1-loop}}_{{\rm YM}+\phi^3}(1,2,\ldots,r; \{p\} ;\ell)\big|_{g^n\lambda^r}
 = N \Big\{ \epsilon_p {\cdot} \ell \big[ \PT(-,1,\ldots,r,+,p) -  \PT(+,1,\ldots,r,-,p) \big]  \!  \!  \label{extend.26} \\
 &\quad \ \  + \sum_{j=1}^{r-1} \epsilon_p {\cdot} k_{12\ldots j} \big[ \PT(-,1,2,\ldots,j,p,j{+}1,\ldots,r,+)
 + \PT(+,1,2,\ldots,j,p,j{+}1,\ldots,r,-) \big] \Big\}
\notag
\end{align}
and thereby reproduces the expression for $a_{\rm EYM}(+,1,2,\ldots,r,-;p)$ in \cite{He:2016mzd},
averaged over $(+ \leftrightarrow -)$, also see the appendix of the reference for the two-gluon
case. The generalizations of (\ref{extend.25}) to higher numbers of gluons can be found
in \cite{Nandan:2016pya, Teng:2017tbo, Du:2017gnh, Edison:2020ehu} and give the $g^n\lambda^r$-order
of $J^{\te{1-loop}}_{{\rm YM}+\phi^3}(1,\ldots,r; \{r{+}1,\ldots,n\} ;\ell)$ for arbitrary $n,r$ via (\ref{extend.11}).


\subsubsection{Examples at subleading order in $\lambda$ with external scalars}
\label{sec:ptaylor.2} 

The one-loop EYM amplitude relations in \cite{He:2016mzd} are limited to gauge multiplets
in the loop and lowest orders in the gravitational coupling. Already the amplitude relations for the 
subleading orders (\ref{extend.12}) of single-trace $J^{\te{1-loop}}_{{\rm YM}+\phi^3}$ in the 
double copy are a new result of this work. In absence of external gluons, the relevant Parke-Taylor 
decompositions include \cite{Nandan:2016pya}
\begin{align}
J_{{\rm YM}+\phi^3}^{\rm tree}(1,2,\ldots,j,+|-,j{+}1,\ldots ,n;\emptyset) &=
\sum_{i=1}^j \sum_{k=j+1}^n (-1)^{i+k} s_{ik}     \label{extend.27}
\\
&\ \ \ \ \times \! \! \! \sum_{Q \in (1,2,\ldots, i-1) \atop{ \shuffle (j,j-1,\ldots,i+1)}}
\sum_{ R \in (k+1,k+2,\ldots n) \atop{ \shuffle (k-1,\ldots,j+1) }} \! \! \!
J_{{\rm YM}+\phi^3}^{\rm tree}(Q,i,k,R ,-,+;\emptyset) \notag
\\
J_{{\rm YM}+\phi^3}^{\rm tree}(1,2,\ldots,n |+,-;\emptyset)
&= \sum_{j=1}^{n-1} (-1)^{j-1} s_{j,\ell}  \! \! \! \sum_{Q\in (j-1,\ldots,2,1) \atop{ \shuffle (j+1,j+2,\ldots,n-1)}} 
\! \! \! J_{{\rm YM}+\phi^3}^{\rm tree}(  j,Q, n,+,-;\emptyset) 
\, , \notag
\end{align}
where the double traces signal that the EYM amplitudes obtained from double copy via
(\ref{looprev.14}) feature gravity multiplets in the loop.  Note that the right-hand sides of (\ref{extend.27}) admit a variety of alternative Parke-Taylor
representations\footnote{More specifically, the right-hand sides of (\ref{extend.27}) are
attained by rewriting the products $\PT(i_1,i_2,\ldots,i_p)\PT(j_1,j_2,\ldots,j_q)$ entering
the expression (\ref{chysec.12}) for $J_{{\rm YM}+\phi^3}^{\rm tree}( i_1,i_2,\ldots,i_p | 
j_1,j_2,\ldots,j_q; \emptyset)$ in terms
of $(p{+}q)$-point Parke-Taylor factors. In doing so through the identities in section 7 
of \cite{Nandan:2016pya}, the cyclic symmetry in $i_1,i_2,\ldots,i_p$
and $j_1,j_2,\ldots,j_q$ is no longer manifest. Our expression for
$J_{{\rm YM}+\phi^3}^{\rm tree}(1,2,\ldots,j,+|-,j{+}1,\ldots ,n;\emptyset)$ in (\ref{extend.27}) is tailored
to avoid $s_{j,\ell}$ involving loop momenta.}
related by scattering equations. As usual in worldsheet approaches to field-theory amplitudes, modifying the moduli-space integrand via scattering equations or total derivatives might reorganize the cubic-diagram expansion of the loop integrand and amount to {\it generalized gauge transformations} in the lingo of the literature on the color-kinematics duality.
 
The forward limit in a pair
of gluons simplifies the Parke-Taylor decomposition of \cite{Nandan:2016pya} to
\begin{align}
J_{{\rm YM}+\phi^3}^{\rm tree}&(1,2,\ldots,r; \{ +,-\}) = 
- \sum_{j=2}^{r-2} \big[ s_{12\ldots j} \PT(1,2,\ldots,j,+,j{+}1,\ldots,r,-) 
+ {\rm cyc}(1,2,\ldots,r) \big]
\notag \\
&\ + \frac{4{-}D}{2} \sum_{j=1}^{r-1} (-1)^{j-1} s_{j,\ell}  \sum_{Q \in (j-1,\ldots,2,1)
\atop{\shuffle (j+1,j+2,\ldots,r-1)}} \big[ \PT(j,Q,r,-,+) - \PT(j,Q,r,+,-) \big] \, .
\label{extend.28}
\end{align}
This expression is obtained after bringing the Mandelstam invariants $s_{ij}$ with 
$1\leq i<j\leq r$ into an $\frac{r}{2}(r{-}3)$-element basis and exposes that the 
$\ell$-dependent terms in the second line cancel in $D=4$ spacetime dimensions.
We have verified (\ref{extend.28}) up to and including $r=6$ external scalars,
and its validity for higher $r$ is conjectural.

In fact, the second line of (\ref{extend.28}) can be identified with a multiple of
the scalar forward limit $J_{{\rm YM}+\phi^3}^{\rm tree}(1,2,\ldots,r |+,-;\emptyset)$
in (\ref{extend.27}): The latter is symmetric under $(+\leftrightarrow -)$ including
$\ell \rightarrow - \ell$,\footnote{Following our earlier comment, this symmetry is manifest from 
the left-hand side of (\ref{extend.27}), but it requires a substantial amount of scattering 
equations to verify the symmetry on the right-hand side of (\ref{extend.27}).} so the 
two Parke-Taylor factors in the square bracket 
of (\ref{extend.28}) yield the same $J_{{\rm YM}+\phi^3}^{\rm tree}$ upon 
summation over $j$ and $Q$, i.e.
\begin{align}
J_{{\rm YM}+\phi^3}^{\rm tree}(1,2,\ldots,r; \{ +,-\}) &= 
- \sum_{j=2}^{r-2} \big[ s_{12\ldots j} \PT(1,2,\ldots,j,+,j{+}1,\ldots,r,-) 
+ {\rm cyc}(1,2,\ldots,r) \big]
\notag \\
&\quad + (D-4)  J_{{\rm YM}+\phi^3}^{\rm tree}(1,2,\ldots,r |+,-;\emptyset) \, .
\label{extend.29}
\end{align}
On these grounds, the $\lambda^{r-2}$-order
of one-loop half integrands (\ref{extend.12}) simplifies to
\begin{align}
 J^{\te{1-loop}}_{{\rm YM}+\phi^3}(1,2,\ldots,n; \emptyset ;\ell)\big|_{g^n\lambda^{n-2}}
 &= (N c_2 +D - 4)  J^{\te{tree}}_{{\rm YM}+\phi^3}(1,2,\ldots,n|+,-;\emptyset)  \notag \\
 & \! \! \! \! \! \! \! \! \! \! \! \! \! \! \! \! \! \! \! \!  \! \! \! \! \! \! \! \! \! 
 - \sum_{j=2}^{n-2} \big[ s_{12\ldots j} \PT(1,2,\ldots,j,+,j{+}1,\ldots,n,-) 
+ {\rm cyc}(1,2,\ldots,n) \big] \label{extend.30} \\
 & \! \! \! \! \! \! \! \! \! \! \! \! \! \! \! \! \! \! \! \!  \! \! \! \! \! \! \! \! \! 
 +\sum_{j=1}^{n-1} \big[ J^{\te{tree}}_{{\rm YM}+\phi^3}(1,2,\ldots,j,+|-,j{+}1,\ldots,n;\emptyset) 
 + {\rm cyc}(1,2,\ldots,n) \big] \notag
\end{align}
with the $J^{\te{tree}}_{{\rm YM}+\phi^3}$ on the right-hand side given by (\ref{extend.27}).

Based on (\ref{extend.27}) to (\ref{extend.30}), the $\lambda^{n-2}$-order
of one-loop half integrands (\ref{extend.12})
without gluon insertions is available in Parke-Taylor form. Generalizations to additional gluons, 
traces or to different power counting in $g,\lambda$ can be obtained by inserting the tree-level
results of \cite{Nandan:2016pya, Teng:2017tbo, Du:2017gnh, Edison:2020ehu} into (\ref{extend.16}) 
to (\ref{extend.18}) and (\ref{extend.21}) to (\ref{extend.23}).


\section{YM+$\phi^3$ half integrands at one loop: four-point examples}
\label{sec:nexthi}

We shall now specialize the general approach of the previous section to four
external legs and spell out all color-ordered one-loop half integrands in YM+$\phi^3$ theory, 
separately at each order in the couplings $g$ and $\lambda$. Each subsection is
dedicated to a different combination of external scalars and gluons.

\subsection{No external gluons}
\label{sec:nogluons}

We begin with the one-loop four-scalar amplitude in YM+$\phi^3$ theory. In this case, we have a single- and a double trace sector. 


\subsubsection{Single-trace sector} 

\begin{figure}[H]
\centering	\includegraphics[scale=1]{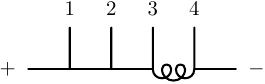}\ \ \ \ \  \ \ \ \ 	\includegraphics[scale=1]{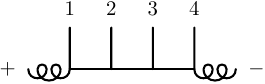}
\caption{The half integrand in (\ref{nogluons.8}) receives contributions from forward limits in both 
scalars (drawn in the left panel) and gluons (drawn in the right panel).}
\label{fig:nogluons.1}
\end{figure}

There are two different orders in the couplings contributing to the single-trace sector by (\ref{extend.5})
\begin{align}
\label{nogluons.1}
& J^{\te{1-loop}}_{{\rm YM}+\phi^3}(1,2,3,4;\emptyset;\ell) =  J^{\te{1-loop}}_{{\rm YM}+{\rm \phi^3}}(1,2,3,4;\emptyset;\ell)\big|_{g^4\lambda^4}+  J^{\te{1-loop}}_{{\rm YM}+\phi^3}(1,2,3,4;\emptyset;\ell)\big|_{g^4\lambda^2} \, .
\end{align}
 At order $g^4\lambda^4$ the color-ordered half integrand comprises cyclic combinations
 of ${\rm PT}(+,\ldots,-)$ in (\ref{extend.7})  known as one-loop Parke-Taylor factors \cite{Geyer:2015bja},
 \begin{align}
 J^{\te{1-loop}}_{{\rm YM}+{\rm \phi^3}}(1,2,3,4;\emptyset;\ell)\big|_{g^4\lambda^4}&= 
N\big[ {\rm PT}(+,1,2,3,4,-) +{\rm PT}(-,1,2,3,4,+) + \te{cyc}(1,2,3,4)\big] 
   \label{nogluons.13} \, .
 \end{align}
According to (\ref{extend.8}), the half integrand at the subleading order $g^4\lambda^2$ receives 
contributions from forward limits in both scalars and gluons,
 \begin{align}
J^{\te{1-loop}}_{{\rm YM}+\phi^3}(1,2,3,4;\emptyset;\ell)\big|_{g^4\lambda^2}
&=  N c_2 J_{{\rm YM}+\phi^3}^{\rm tree}(1,2,3,4|+,-;\emptyset) + J_{{\rm YM}+\phi^3}^{\rm tree}(1,2,3,4;\{+,-\})
\notag \\
&\quad +  \Big[   J_{{\rm YM}+\phi^3}^{\rm tree}(1,2,3,+|-,4;\emptyset)
+ J_{{\rm YM}+\phi^3}^{\rm tree}(1,2,+|-,3,4;\emptyset)    \label{nogluons.8}  \\
&\quad \quad + J_{{\rm YM}+\phi^3}^{\rm tree}(1,+|-,2,3,4;\emptyset) +\text{cyc}(1,2,3,4) \Big] \, .\nonumber
 \end{align}
By the discussion below (\ref{extend.9}), the relative normalization between the
forward limits in scalars and gluons is fixed by requiring a quadratic-propagator representation of
$A^{\te{1-loop}}_{ {\rm EYM}, \alpha}(1,2,3,4)$ resulting from (\ref{looprev.14}).
More specifically, changing the prefactor of $J_{{\rm YM}+\phi^3}^{\rm tree}(1,2,3,4;\{+,-\})$ 
on the right-hand side of (\ref{nogluons.8}) would spoil the recombination of
linearized propagators to quadratic ones in four-gluon one-loop EYM amplitudes
(for any amount of supersymmetry $\alpha$). The two classes of tree-level diagrams
associated with the forward limit in a gluon or scalar are drawn in figure \ref{fig:nogluons.1} (cf. figure \ref{linprops2}).

The tree-level building blocks on the right-hand side of (\ref{nogluons.8}) are given by
the following specializations of (\ref{extend.27})\footnote{Note 
that (\ref{hiexamples.5}) to (\ref{hiexamples.7}) are forward limits of the six-point
tree-level identities (74) and (75) of \cite{Nandan:2016pya} whose right-hand sides obscure 
the cyclic symmetries of the double-trace structures. Cyclic permutations in the underlying 
tree-level expressions lead to alternative representations of (\ref{hiexamples.5}) to (\ref{hiexamples.7})
which may also alter the power-counting of loop momenta, see for instance the factor of $s_{4,\ell}$ in
the following equivalent of (\ref{hiexamples.6}):
\begin{align*}
J_{{\rm YM}+\phi^3}^{\rm tree}(1,2,3,+|-,4;\emptyset)&=- s_{24} \PT(+,3, 2, 4,-,1) -s_{4,\ell} \PT(2,3,+, 4,-,1) \\
&\quad +s_{34}\big[ \PT(2,+, 3, 4,-,1)+  \PT(+,2, 3, 4,-,1)\big]\, .
\end{align*}}
\begin{align}
\label{hiexamples.5}
 J_{{\rm YM}+\phi^3}^{\rm tree}(1,2,3,4|+,-;\emptyset)
&= s_{1,\ell} \PT(4,3,2, 1, -,+)   +s_{3,\ell}\PT(4,1,2, 3, -,+)  \\
&\quad -s_{2,\ell} \big[ \PT(4,1,3, 2, -,+)+\PT(4,3,1, 2, -,+)\big]\notag\\
\label{hiexamples.6}  J_{{\rm YM}+\phi^3}^{\rm tree}(1,2,3,+|-,4;\emptyset)&=- s_{14} \PT(3,2, 1, 4,-,+) -s_{34} \PT(1,2, 3, 4,-,+)  \\
&\quad+s_{24}\big[ \PT(1,3, 2, 4,-,+)+  \PT(3,1, 2, 4,-,+)\big]\notag\\
\label{hiexamples.7}  J_{{\rm YM}+\phi^3}^{\rm tree}(1,2,+|-,3,4;\emptyset)&= s_{13} \PT (2,1,3,4,-,+) - s_{14} \PT(2, 1,4,3,-,+)\\
&\quad -  s_{23}\PT(1, 2,3,4,-,+) +s_{24} \PT(1, 2,4,3,-,+) \notag
\end{align}
and of (\ref{extend.28}),
\begin{align}
J_{{\rm YM}+\phi^3}^{\rm tree}(1,2,3,4;\{+,-\}) &=
 - s_{12} \big[ \PT(1, 2,+,3,4,-)+  \PT(1,2,-,3,4,+)\big] \notag \\
 &\quad  - s_{23} \big[ \PT(2,3,+,4,1,-)+  \PT(2,3,-,4,1,+)\big]
  \notag \\
&\quad + \frac{4{-}D}{2} \Big\{  s_{1,\ell} \PT(1,2,3,4,-,+) + s_{3,\ell} \PT(3,2,1,4,-,+) 
 \label{newgfw.1} \\
  &\quad\quad \  -  s_{2,\ell} 
  \big[ \PT(2,1,3,4,-,+)+ \PT(2,3,1,4,-,+)\big]- (+\leftrightarrow -) 
  \Big\} \, .\notag
\end{align}
Following the general discussion around (\ref{extend.29}), the last two lines of
(\ref{newgfw.1}) are proportional to (\ref{hiexamples.5}) on the support of scattering equations.
Hence, the half integrand (\ref{nogluons.8}) can be simplified to 
 \begin{align}
J^{\te{1-loop}}_{{\rm YM}+\phi^3}(1,2,&3,4;\emptyset;\ell)\big|_{g^4\lambda^2}
=  (N c_2+D-4) J_{{\rm YM}+\phi^3}^{\rm tree}(1,2,3,4|+,-;\emptyset) 
\notag \\
&\quad +  \Big[   J_{{\rm YM}+\phi^3}^{\rm tree}(1,2,3,+|-,4;\emptyset)
+ J_{{\rm YM}+\phi^3}^{\rm tree}(1,2,+|-,3,4;\emptyset)    \label{newgfw.2}  \\
&\quad \quad + J_{{\rm YM}+\phi^3}^{\rm tree}(1,+|-,2,3,4;\emptyset) 
- s_{12} \PT(1,2,+,3,4,-) +\text{cyc}(1,2,3,4) \Big] \, ,\nonumber
 \end{align}
in lines with (\ref{extend.30}). Finally, with the expressions in (\ref{hiexamples.6})
and (\ref{hiexamples.7}), the last two lines of (\ref{newgfw.2}) conspire to a permutation
sum of Parke-Taylor factors,
 \begin{align}
J^{\te{1-loop}}_{{\rm YM}+\phi^3}(1,2,3,4;\emptyset;\ell)\big|_{g^4\lambda^2}
&=  (N c_2+D-4) J_{{\rm YM}+\phi^3}^{\rm tree}(1,2,3,4|+,-;\emptyset) 
\notag \\
&\quad + 2 s_{13} \sum_{\rho \in S_4}  \PT(+,\rho(1,2,3,4),-)   \, . \label{newnewgfw.2}  
\end{align}
 
\subsubsection{Double-Trace Sector} 
\begin{figure}[H]
	\centering	\includegraphics[scale=1]{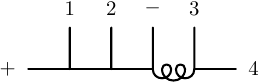}
	\caption{
The one-loop half integrand in (\ref{nogluons.10}) has contributions from forward limits of tree-level half integrands associated to diagrams with six scalars and a double trace.}
	\label{fig:nogluons.2}
\end{figure}

The double-trace sector of the one-loop half integrand with four external scalars 
is compatible with three different powers $\lambda^4,\lambda^2,\lambda^0$, see (\ref{extend.15}),
\begin{align}
\label{nogluons.9}
 J^{\te{1-loop}}_{{\rm YM}+\phi^3}(1,2|3,4;\emptyset;\ell) &= J^{\te{1-loop}}_{{\rm YM}+\phi^3}(1,2|3,4;\emptyset;\ell)\big|_{g^4\lambda^4}\\ 
 &\quad+J^{\te{1-loop}}_{{\rm YM}+\phi^3}(1,2|3,4;\emptyset;\ell)\big|_{g^4\lambda^2}+J^{\te{1-loop}}_{{\rm YM}+\phi^3}(1,2|3,4;\emptyset;\ell)\big|_{g^4}\, .\notag
\end{align}
At leading order $g^4 \lambda^4$ in the couplings, (\ref{extend.16}) specializes to
\begin{align}
J^{\te{1-loop}}_{\phi^3}(1,2|3,4;\emptyset;\ell)\big|_{g^4\lambda^4}
&=2\big[  \PT(1,2,+,3,4,-)
+ \PT(2,1,+,3,4,-)  \label{nogluons.14}  \\
&\ \ \
+ \PT(1,2,+,4,3,-)
+ \PT(2,1,+,4,3,-) \big] \, .\notag
\end{align}
Also the subleading order $g^4 \lambda^2$ entirely stems from tree-level half integrands 
with six scalars (this time distributed over two traces). More specifically, (\ref{extend.17})
with $P = \emptyset$ as well as $s=2$ and $r=4$ yields
	\begin{align}
\label{nogluons.10}  J^{\te{1-loop}}_{{\rm YM}+\phi^3}(1,2|3,4;\emptyset;\ell)\big|_{g^4\lambda^2} &=
2 N \big[ J_{{\rm YM}+\phi^3}^{\rm tree}(1,2,+,-|3,4;\emptyset)
+J_{{\rm YM}+\phi^3}^{\rm tree}(2,1,+,-|3,4;\emptyset) \\
& \quad \quad \quad
+J_{{\rm YM}+\phi^3}^{\rm tree}(1,2|3,4,+,-;\emptyset)
+J_{{\rm YM}+\phi^3}^{\rm tree}(1,2|4,3,+,-;\emptyset) \big]\, ,
\notag
\end{align}
where all terms on the right-hand side are permutations of
\begin{align}
J_{{\rm YM}+\phi^3}^{\rm tree}(1,2,+,-|3,4;\emptyset) &= s_{24} \big[ \PT(1,+,2,4,3,-) + \PT(+,1,2,4,3,-)\big]
\label{newgfw.3} \\
&\quad - s_{14} \PT(+,2,1,4,3,-) - s_{4,\ell} \PT(1,2,+,4,3,-) \, . \notag
\end{align} 
A typical diagram contributing to the forward limit in (\ref{nogluons.10})
is depicted in figure \ref{fig:nogluons.2}.

\begin{figure}[H]
\centering	\includegraphics[scale=1]{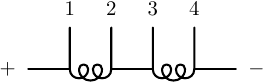}\ \ \ \ \  \ \ \ \ \includegraphics[scale=1]{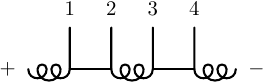}
\caption{The half integrand in (\ref{nogluons.11}) has contributions from forward limits in both
	scalars and gluons. Typical diagrams from $J_{{\rm YM}+\phi^3}^{\rm tree} (+,1|3,4|2,-;\emptyset)$
	and $J_{{\rm YM}+\phi^3}^{\rm tree} (1,2|3,4;\{+,-\})$ are depicted in the left and right panel,
	respectively.}
	\label{fig:nogluons.3}
\end{figure}

At the lowest order in $\lambda$, we also have contributions with gluons in the forward limit
from the second line of (\ref{extend.18}),
\begin{align}
J^{\te{1-loop}}_{{\rm YM}+\phi^3}(1,2|3,4;\emptyset;\ell)\big|_{g^4} &=
Nc_2 J_{{\rm YM}+\phi^3}^{\rm tree} (1,2|3,4|+,-;\emptyset) +
J_{{\rm YM}+\phi^3}^{\rm tree} (1,2|3,4;\{+,-\}) \notag \\
&\quad+ J_{{\rm YM}+\phi^3}^{\rm tree} (+,1|3,4|2,-;\emptyset)
+ J_{{\rm YM}+\phi^3}^{\rm tree} (+,2|3,4|1,-;\emptyset) \label{nogluons.11} \\
&\quad + J_{{\rm YM}+\phi^3}^{\rm tree} (+,3|1,2|4,-;\emptyset)
+ J_{{\rm YM}+\phi^3}^{\rm tree} (+,4|1,2|3,-;\emptyset) \, , \notag
\end{align}
see figure \ref{fig:nogluons.3} for typical diagrams that contribute. It would be interesting
to investigate multitrace generalizations of (\ref{newgfw.2}), e.g.\ whether the
expression for $J_{{\rm YM}+\phi^3}^{\rm tree} (1,2|3,4;\{+,-\}) $ simplifies
after peeling off $(D{-}4) J_{{\rm YM}+\phi^3}^{\rm tree} (1,2|3,4|+,-;\emptyset)$.

\subsection{One external gluon}
\label{sec:onegluon}

\begin{figure}[H]
	\centering	\includegraphics[scale=1]{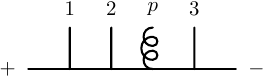}
	\caption{The forward limits contributing to the integrand in (\ref{onegluon.2}) are performed in scalars as illustrated in this figure. }
	\label{fig:onegluon.1}
\end{figure}

For one-loop half integrands of ${\rm YM}+\phi^3$ with three external scalars and
one external gluon~$p$, (\ref{extend.5}) admits two different powers of the coupling
$\lambda$ from the forward limit of trees,
\begin{align}
\label{onegluon.1}
& J^{\te{1-loop}}_{{\rm YM}+\phi^3}(1,2,3;\{p\};\ell) =  J^{\te{1-loop}}_{{\rm YM}+\phi^3}(1,2,3;\{p\};\ell)\big|_{g^4\lambda^3}+ J^{\te{1-loop}}_{{\rm YM}+\phi^3}(1,2,3;\{p\};\ell)\big|_{g^4\lambda}\, .
\end{align}
Following the lines of \cite{He:2016mzd}, at order $g^4\lambda^3$ 
we use (\ref{extend.11}) to obtain
\begin{align}
 J^{\te{1-loop}}_{{\rm YM}+\phi^3}(1,2,3;\{p\};\ell)\big|_{g^4\lambda^3}&= N \, \big[ 
J_{{\rm YM}+\phi^3}^{\rm tree}(+,1,2,3,-;\{p\})
+J_{{\rm YM}+\phi^3}^{\rm tree}(-,1,2,3,+;\{p\})  \big]  \notag \\
&\quad +\text{cyc}(1,2,3)\, ,
\label{onegluon.2} 
\end{align}
where the Parke-Taylor decomposition (\ref{extend.26}) for the 
$J_{{\rm YM}+\phi^3}^{\rm tree}$ on the right-hand side yields
\begin{align}
J_{{\rm YM}+\phi^3}^{\rm tree}(+,1,2,3,-;\{p\})
& = -(\epsilon_p\cdot \ell){\rm PT}(+,1,2,3,-,p)+  (\epsilon_p\cdot k_1){\rm PT}(+,1,p,2,3,-) 
\notag \\
&\quad +  (\epsilon_p\cdot k_{12}){\rm PT}(+,1,2,p,3,-) \, . \label{hiexamples.1}
\end{align}
A typical diagram at the order of $g^4\lambda^3$ is depicted
in figure \ref{fig:onegluon.1}.

\begin{figure}[H]
	\centering	\includegraphics[scale=1]{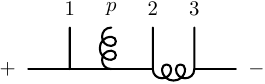}\ \ \ \ \ \ \ 	\includegraphics[scale=1]{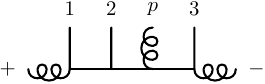}
	\caption{The half integrand in (\ref{onegluon.3}) receives contributions from 
	forward limits in both scalars (left panel) and gluons (right panel).}
	\label{fig:onegluon.2}
\end{figure}

The half integrand at the order of $g^4\lambda$ receives contributions from
forward limits in both scalars and gluons (see figure \ref{fig:onegluon.2} for
typical diagrams in both cases), and (\ref{extend.12}) specializes to
 \begin{align}
J^{\te{1-loop}}_{{\rm YM}+\phi^3}(1,2,3;\{p\};\ell)\big|_{g^4\lambda}
 &=   J_{{\rm YM}+\phi^3}^{\rm tree}(1,2,3;\{p,+,-\})+
 N c_2\, J_{{\rm YM}+\phi^3}^{\rm tree}(1,2,3|+,-;\{p\}) \notag\\
&\quad \hspace{-3.2cm} +  \big[ J_{{\rm YM}+\phi^3}^{\rm tree}(1,2,+|-,3;\{p\})+   J_{{\rm YM}+\phi^3}^{\rm tree}(1,+|-,2,3;\{p\})+ \text{cyc}(1,2,3) \big]  \, .  \label{onegluon.3} 
\end{align}
The tree-level half integrands on the right-hand side have both been brought into
Parke-Taylor form in sections 5 and 8 of \cite{Nandan:2016pya}.

%


\subsection{Two external gluons}
\label{sec:twogluons}
\begin{figure}[H]
	\centering	\includegraphics[scale=1]{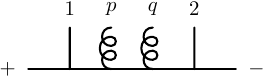}
	\caption{The forward limits contributing to the integrand in (\ref{twogluons.2}) are performed in scalars of tree-level half integrands with two gluons and four scalars. }
	\label{fig:twogluons.1}
\end{figure}

Also for two external scalars and two external gluons $p,q$,
one-loop half integrands of ${\rm YM}+\phi^3$ exhibit two different powers of the coupling
$\lambda$ from the forward limit of trees, 
\begin{align}
\label{twogluons.1}
& J^{\te{1-loop}}_{{\rm YM}+\phi^3}(1,2;\{p,q\};\ell) =  J^{\te{1-loop}}_{{\rm YM}+\phi^3}(1,2;\{p,q\};\ell)\big|_{g^4\lambda^2}+ J^{\te{1-loop}}_{{\rm YM}+\phi^3}(1,2;\{p,q\};\ell)\big|_{g^4}\, .
\end{align}
As in the previous case, at the leading order in $\lambda$ we use (\ref{extend.11}) to obtain
\begin{align}
 J^{\te{1-loop}}_{{\rm YM}+\phi^3}(1,2;\{p,q\};\ell)\big|_{g^4\lambda^2} &=2 N \big[ J_{{\rm YM}+\phi^3}^{\rm tree}(1,2,+,-;\{p,q\})+J_{{\rm YM}+\phi^3}^{\rm tree}(2,1,+,-;\{p,q\}) \big] \, ,\label{twogluons.2}
\end{align}
see figure \ref{fig:twogluons.1} for typical diagrams that contribute. The $J_{{\rm YM}+\phi^3}^{\rm tree}$ on the right-hand side are of the following form,
\begin{align}
J_{{\rm YM}+\phi^3}^{\rm tree}(2,1,+,-;\{p,q\})
& =\PT(1,2,p,q,-,+)\big((\epsilon_p\cdot \ell_{12})(\epsilon_q\cdot \ell)-\tfrac{1}{2}(\epsilon_p\cdot \epsilon_q)(s_{p,\ell}-s_{pq})\big) \notag \\
&\quad+\PT(1,p,2,q,-,+)\big((\epsilon_p\cdot \ell_1)(\epsilon_q\cdot\ell)-\tfrac{1}{2}(\epsilon_p\cdot \epsilon_q) (s_{1p}+s_{p,\ell})\big)
\notag\\
&\quad+\PT(1,p,q,2,-,+)\big((\epsilon_p\cdot \ell_1) (\epsilon_q\cdot \ell_{1p}) -\tfrac{1}{2}(\epsilon_p\cdot \epsilon_q) (s_{1p}+s_{p,\ell})\big)
\notag\\
&\quad+\PT(p,1,2,q,-,+)\big((\epsilon_p\cdot \ell)(\epsilon_q\cdot \ell)-\tfrac{1}{2}(\epsilon_p\cdot \epsilon_q) s_{p,\ell}\big)
\notag\\
&\quad+ \PT(p,1,q,2,-,+)\big((\epsilon_p\cdot \ell) (\epsilon_q\cdot \ell_{1p})-\tfrac{1}{2}(\epsilon_p\cdot \epsilon_q)s_{p,\ell}\big)\notag\\
&\quad+\PT(p,q,1,2,-,+)\big((\epsilon_p\cdot \ell) (\epsilon_q\cdot \ell_p)-\tfrac{1}{2}(\epsilon_p\cdot \epsilon_q) s_{p,\ell}\big)
\notag\\
& \quad+ ( p\leftrightarrow q)\, , \label{hiexamples.3}
\end{align}
which is equivalent to the $n=2$ instance of (28) in \cite{He:2016mzd}.

\begin{figure}[H]
	\centering	\includegraphics[scale=1]{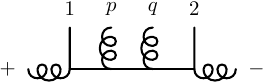}\ \ \ \ \ \ \ \ \ \ 	\includegraphics[scale=1]{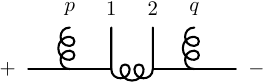}
	\caption{Forward limits in gluons and scalars that contribute to the half integrand (\ref{twogluons.3}) are illustrated in the left and right panel, respectively.}
	\label{fig:twogluons.2}
\end{figure}

At the subleading order in $\lambda$, (\ref{extend.12}) specializes to
\begin{align}
J^{\te{1-loop}}_{{\rm YM}+\phi^3}(1,2;\{p,q\};\ell)\big|_{g^4}
 &=   J_{{\rm YM}+\phi^3}^{\rm tree}(1,2;\{p,q,+,-\})
+ N c_2 J_{{\rm YM}+\phi^3}^{\rm tree}(1,2|+,-;\{p,q\}) \notag \\
&\quad + J_{{\rm YM}+\phi^3}^{\rm tree}(1,+|2,-;\{p,q\})+   J_{{\rm YM}+\phi^3}^{\rm tree}(2,+|1,-;\{p,q\}) \, ,
 \label{twogluons.3}
\end{align}
and typical diagrams associated with the gluonic and scalar forward limit 
are depicted in figure \ref{fig:twogluons.2}.

\subsection{Four external gluons}
\label{sec:fourgluons}

\begin{figure}[H]
	\centering	\includegraphics[scale=1]{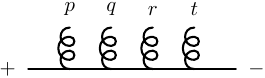}\ \ \ \ \ \ \ \ \ \ 	\includegraphics[scale=1]{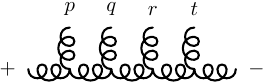}
	\caption{The four-gluon one-loop half integrand has a contribution with a scalar loop (left panel) and
	a gluon loop (right panel).}
	\label{fig:fourgluons.1}
\end{figure}

In case of four external gluons $p,q,r,t$, half integrands of ${\rm YM}+\phi^3$ only receive
contributions at the order of $g^4$ according to (\ref{trunc.12}). This time, scalars and gluons in the loop 
occur at the same powers in the couplings,
\begin{align}
 J^{\te{1-loop}}_{{\rm YM}+\phi^3}(\{p,q,r,t\};\ell) &= 
N c_2\,  J_{{\rm YM}+\phi^3}^{\rm tree}(+,-; \{p,q,r,t\})
+ J^{\te{tree}}_{{\rm YM}}(\emptyset; \{p,q,r,t,+,-\}) \, ,
 \label{fourgluons.2}
\end{align}
but with an extra factor of $Nc_2$ in case of the scalars in the loop. The master numerators
in the Parke-Taylor decomposition of $N c_2\,  J_{{\rm YM}+\phi^3}^{\rm tree}(+,-; \{p,q,r,t\})$
can be found in the ancillary file. The master numerators associated with
$J^{\te{tree}}_{{\rm YM}}(\emptyset; \{p,q,r,t,+,-\})$ in turn are described in section 3
of \cite{Geyer:2017ela}. The associated master diagrams are depicted in figure \ref{fig:fourgluons.1}. 

\section{Four-point EYM amplitudes at one loop with maximal supersymmetry}
\label{sec:maxsusy}

In this and the following section, we apply the method of this work to obtain expressions 
for one-loop integrands of four-point amplitudes in EYM theories that expose the simplifications
due to supersymmetry. The calculations are driven by the results of section \ref{sec:nexthi} for the four-point
one-loop half integrands of YM+$\phi^3$ which apply to EYM with any number of supercharges.
We perform separate 
calculations for the color-ordered EYM amplitudes (\ref{looprev.14}) at different orders 
in the couplings $g$ and $\kappa$
which are related to the orders of the couplings $g$ and $\lambda$ of YM+$\phi^3$
via (\ref{review.10}), see table~\ref{tab:couplings}.

Following the last line of (\ref{looprev.14}), we first obtain the loop integrands of EYM in terms
of one-loop half integrands of YM+$\phi^3$ on the nodal Riemann sphere and partial integrands 
$a(\ldots)$ in YM theory. While the partial integrands are defined in terms
of linearized propagators, we combine different terms in the permutation sum of 
(\ref{looprev.14}) to attain the conventional form of Feynman integrals with 
quadratic propagators. The maximal supersymmetry of the partial integrands in this section
leads to extra simplifications in comparison to the half-maximally supersymmetric
case in section \ref{sec:halfmax}.

The recombination of the loop integrands in (\ref{looprev.14}) to quadratic propagators is
a first consistency check of our method and the expressions for the half integrands 
$J^{\te{1-loop}}_{{\rm YM}+\phi^3}$ in the previous section. Moreover, we have verified 
all EYM amplitudes with external gravitons
to respect linearized diffeomorphism invariance. This is non-trivial for the polarization vectors
from $J^{\te{1-loop}}_{{\rm YM}+\phi^3}$ that we denote by $\bar{\epsilon}_i$ (in contradistinction to
the polarizations $\epsilon_i$ of the YM half integrands). All the explicit results for four-point
EYM amplitudes in this section are checked to vanish under the linearized gauge transformation
$\bar{\epsilon}_p\rightarrow p$ that double copies to linearized diffeomorphisms.

The results of this section exemplify that the no-triangle property \cite{Bern:1994zx} of 
maximally supersymmetric YM and supergravity does not apply to EYM with 16 supercharges.
In spite of their maximal supersymmetry, the one-loop integrands of EYM amplitudes in
(\ref{nogravmax.2}), (\ref{nogravmax.4}) and later equations feature triangle and bubble 
diagrams.\footnote{We refrain from re-interpreting diagrams by introducing spurious propagators
$1=  \frac{ (\ell+K)^2}{( \ell+K)^2}$ (as one may need to manifest the color-kinematics duality in the results of this section). For instance, the triangle with propagators $\frac{1}{\ell^2 \ell_1^2\ell_{12}^2} = \frac{\ell^2_{123}}{\ell^2 \ell_1^2\ell_{12}^2 \ell^2_{123}}$ is not counted as a box with the inverse propagator $\ell^2_{123}$ in the numerator.} The loop integrands presented in this section 
can also be found in the ancillary file.
	
\subsection{Partial integrands with maximal supersymmetry}
\label{sec:pimax}

Maximally supersymmetric EYM theory has 16 supercharges and can be defined in spacetime 
dimensions $D\leq 10$. Its four-dimensional incarnation is said to have $\mathcal{N}=4$ supersymmetry. 
In the CHY construction of EYM one-loop amplitudes, all the supersymmetries arise from the YM 
half integrand which takes the simple form (\ref{looprev.3}) at four points. The simplicity of the 
underlying correlation function on a torus was first revealed in \cite{Green:1982sw}, and the associated 
partial integrands in terms of linearized propagators are 
given by \cite{He:2016mzd}\footnote{Actually, only the first line of (\ref{pimax.3}) is independent under
the Kleiss-Kuijf relations while the second and the last two lines
can be obtained from $-a^{\te{1-loop}}_{ \text{YM},{\rm max}}((4\shuffle 1,2,3),-,+)$
and $a^{\te{1-loop}}_{ \text{YM},{\rm max}}((1,2\shuffle 4,3),-,+)$, respectively.}
\begin{align}
a^{\te{1-loop}}_{ \text{YM},{\rm max}}(1,2,3,4,-,+) &= \frac{t_8(1,2,3,4)}{s_{1,\ell}s_{12,\ell}s_{123,\ell}}
\notag\\ 
a^{\te{1-loop}}_{ \text{YM},{\rm max}}(1,2,3,-,4,+) &= \frac{t_8(1,2,3,4)}{s_{1,\ell}s_{12,\ell}s_{4,\ell}}+\frac{t_8(1,2,3,4)}{s_{1,\ell}s_{12,\ell}s_{3,\ell}}+\frac{t_8(1,2,3,4)}{s_{1,\ell}s_{14,\ell}s_{3,\ell}}+\frac{t_8(1,2,3,4)}{s_{4,\ell}s_{14,\ell}s_{3,\ell}} \notag \\ \label{pimax.3}
a^{\te{1-loop}}_{ \text{YM},{\rm max}}(1,2,-,3,4,+) &= \frac{t_8(1,2,3,4)}{s_{1,\ell}s_{12,\ell}s_{124,\ell}}+ \frac{t_8(1,2,3,4)}{s_{1,\ell}s_{14,\ell}s_{124,\ell}}+ \frac{t_8(1,2,3,4)}{s_{4,\ell}s_{14,\ell}s_{124,\ell}}\\
&\quad+ \frac{t_8(1,2,3,4)}{s_{4,\ell}s_{34,\ell}s_{134,\ell}}
+ \frac{t_8(1,2,3,4)}{s_{1,\ell}s_{14,\ell}s_{134,\ell}}
+ \frac{t_8(1,2,3,4)}{s_{4,\ell}s_{14,\ell}s_{134,\ell}} \, .\notag
\end{align}
The $t_8$-tensor defined in (\ref{looprev.4}) prescribes a dimension-agnostic contraction of
$D$-dimensional polarization vectors and momenta. Given that also the YM$+\phi^3$ ingredients
of the previous sections are dimension agnostic, the results of this section apply to any
dimensional reduction of ten-dimensional EYM with maximal supersymmetry to $D\leq 10$.

\subsection{No external gravitons}
\label{sec:nogravmax}

The amplitude with four external gluons has a single- and a double-trace sector.

\subsubsection{Single-trace sector} 

By the two contributions (\ref{nogluons.1}) to the YM$+\phi^3$ half integrand, there are
two different combinations of couplings in the single-trace sector,
\begin{align}
\label{nogravmax.6} 
A^{\te{1-loop}}_{\text{EYM}, \text{max}}(1,2,3,4;\emptyset)&= A^{\te{1-loop}}_{\text{EYM}, \text{max}}(1,2,3,4;\emptyset)\big|_{g^4}+A^{\te{1-loop}}_{ \text{EYM}, \text{max}}(1,2,3,4;\emptyset)\big|_{\kappa^2g^2}\, .
\end{align}
The first term describing a four-gluon amplitude with the maximally supersymmetric gauge multiplet
in the loop coincides with the one-loop amplitude in SYM \cite{Green:1982sw} ($g^4$ does not leave any room for
gravitational exchange). The YM+$\phi^3$ half integrand is the one-loop Parke-Taylor factor given in (\ref{nogluons.13}). In combination with the partial integrands (\ref{pimax.3}), we find
\begin{align}
\label{nogravmax.9}
A^{\te{1-loop}}_{ \text{EYM}, \text{max}}(1,2,3,4;\emptyset)\big|_{g^4}&= A^{\te{1-loop}}_{ \text{YM}, \text{max}}(1,2,3,4) \notag \\
&\hspace{-1.5cm}=   N\, \int \frac{ \dd^D \ell}{\ell^2} \lim_{k_{\pm} \rightarrow \pm \ell} \int \dd \mu_{6}^{\te{tree}}\,  I^{\te{1-loop}}_{\rm YM,\te{max}}(\{1,2,3,4\};\ell) \notag \\
& \times \big[ \PT(+,1,2,3,4,-) +\PT(-,1,2,3,4,+) + {\rm cyc}(1,2,3,4)\big]\, \notag \\
&\hspace{-1.5cm}= N\, t_8(1,2,3,4) \int \frac{ \dd^D \ell}{\ell^2} \left(\left( \frac{1}{s_{1,\ell}s_{12,\ell}s_{123,\ell}} +\frac{1}{s_{1,\ell}s_{14,\ell}s_{143,\ell}}\right)+ \te{cyc}(1,2,3,4)\right) \notag\\
&\hspace{-1.5cm}= 8\, N\, t_8(1,2,3,4)\int \frac{ \dd^D \ell}{\ell^2}\left( \frac{1}{\ell_1^2\ell_{12}^2\ell_{123}^2}+\frac{1}{\ell_1^2\ell_{14}^2\ell_{143}^2}\right) \notag \\
&\hspace{-1.5cm}= 16\, N\, t_8(1,2,3,4)\int \frac{ \dd^D \ell}{\ell^2 \ell_1^2\ell_{12}^2\ell_{123}^2}
\, ,
\end{align}
where we used (\ref{looprev.11}) to obtain the quadratic propagators
$\ell^{-2}_{12\ldots p} =( \ell {+}k_{12\ldots p})^{-2}$. The last step is
based on the reflection property $k_2\leftrightarrow k_4$ of the scalar box with propagators
$\ell^{-2} \ell_1^{-2}\ell_{12}^{-2}\ell_{123}^{-2}$.

\begin{figure}
	\centering
	\includegraphics[scale=0.8]{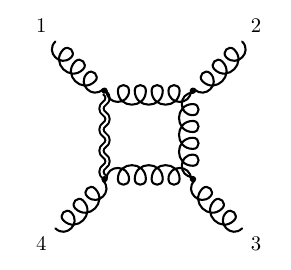}\ \ \ \ \ \ \ \ \ \ \ \ \includegraphics[scale=0.7]{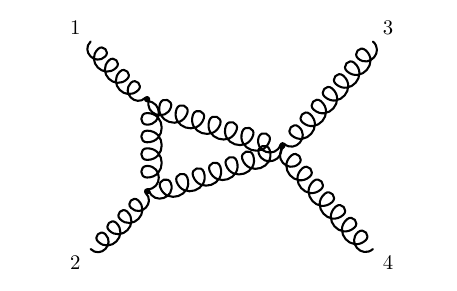}
\caption{At order $\kappa^2g^2$ the four-gluon EYM amplitude at one loop has a single- and a double-trace sector. The single-trace sector in (\ref{nogravmax.1}) is exclusively furnished by the box diagrams
in the left panel with one propagator of the gravity multiplet (double wavy line). The tree diagrams in its partial-fraction decomposition correspond
to forward limits in different particles, see figure \ref{fig:nogluons.1} for their YM$+\phi^3$ analogue. The
double-trace sector in (\ref{nogravmax.2}) of the maximally supersymmetric four-gluon amplitude is built
from the triangle diagrams in the right panel. All graphs in this figure and later ones do not
represent Feynman diagrams but instead illustrate the propagator structure of the maximally
supersymmetric loop integrands.}
	\label{fig:nogravmax.1}
\end{figure}

At order $\kappa^2g^2$, the relevant half integrand is given by (\ref{newnewgfw.2}) and
leads to the following single-trace contribution to the four-gluon amplitude in EYM theory: 
\begin{align}\label{nogravmax.1}
&A^{\te{1-loop}}_{ \text{EYM}, \text{max}}(1,2,3,4;\emptyset)\big|_{\kappa^2g^2}\\&= \frac{1}{16}\int \frac{ \dd^D \ell}{\ell^2} \lim_{k_{\pm} \rightarrow \pm \ell} \int \dd \mu_{6}^{\te{tree}} I^{\te{1-loop}}_{\rm YM,\te{max}}(\{1,2,3,4\};\ell)J^{\te{1-loop}}_{{\rm YM}+\phi^3}(1,2,3,4;\emptyset;\ell)\big|_{g^4\lambda^2}\notag\\
&= s_{13}\, t_8(1,2,3,4) \int\frac{ \dd^D \ell}{\ell^2} ~
\left[\, \frac{1}{\ell_1^2\ell_{12}^2\ell_{123}^2 } +\text{perm}(2,3,4)\right]\notag
\\
&=2 s_{13}\, t_8(1,2,3,4) \int\frac{ \dd^D \ell}{\ell^2} ~
\left[\, \frac{1}{\ell_1^2\ell_{12}^2\ell_{123}^2 } +\text{cyc}(2,3,4)\right]\notag
\end{align}
As illustrated in figure \ref{fig:nogravmax.1}, the result stems from box graphs with
a graviton propagator, and the Mandelstam invariant $s_{13}$ can be attributed to 
its two gravitational vertices $\sim \kappa^2$. Note that the $\kappa^2g^2$ order
does not share the prefactor of $N$ in the $g^4$ order in (\ref{nogravmax.9}), 
and the first line $\sim (Nc_2{+}D{-}4)$ of the half integrand (\ref{newnewgfw.2})
does not contribute in the maximally supersymmetric case.

By the permutation-symmetric combination of boxes in (\ref{nogravmax.1}) 
and $s_{13}+{\rm cyc}(1,2,3)=0$, our result obeys Kleiss-Kuijf relations 
$A^{\te{1-loop}}_{ \text{EYM}, \text{max}}(1,2,3,4;\emptyset)\big|_{\kappa^2g^2}
+{\rm cyc}(1,2,3) = 0$. This ensures that the accompanying color factors combine
to permutations of contracted structure constants $\hat f^{A_1 A_2 B } \hat f^{B A_3 A_4}$ 
as expected from the left panel of figure \ref{fig:nogravmax.1}.

\subsubsection{Double-trace sector}

Based on the half integrand (\ref{nogluons.9}) of YM$+\phi^3$, the double-trace sector of the 
four-gluon amplitude can come with three different powers of the couplings
\begin{align}
\label{nogravmax.7} 
A^{\te{1-loop}}_{\text{EYM}, \text{max}}(1,2| 3,4;\emptyset)&= A^{\te{1-loop}}_{ \text{YM}, \text{max}}(1,2  |3,4;\emptyset)\big|_{g^4}\\
&\ \ \ \ +A^{\te{1-loop}}_{\text{EYM}, \text{max}}(1,2  | 3,4;\emptyset)\big|_{\kappa^2g^2}+A^{\te{1-loop}}_{ \text{EYM}, \text{max}}(1,2  | 3,4;\emptyset)\big|_{\kappa^4}\, .\notag
\end{align}
At order $g^4$, we recover the double-trace amplitude of SYM which is
determined by the half integrand in (\ref{nogluons.14})
\begin{align}
\label{nogravmax.10}
A^{\te{1-loop}}_{ \text{YM}, \text{max}}(1,2  | 3,4;\emptyset)\big|_{g^4} =  32\, t_8(1,2,3,4)\int \frac{\dd^D\ell}{\ell^2}\left(\frac{1}{ \ell_1^2\ell_{12}^2\ell_{123}^2}+\te{cyc}(2,3,4)\right) \, .
\end{align}
Since this is proportional to a permutation sum of the single-trace amplitude (\ref{nogravmax.9}), 
we can verify the relation of \cite{Bern:1994zx} between planar and
non-planar one-loop gauge-theory amplitudes which only holds for the
$\kappa \rightarrow 0$ limit of EYM amplitudes.

At the order of $\kappa^2 g^2$ in the double-trace sector, the net effect of the gravitational
vertices is to cancel one of the propagators of the box diagrams:
The YM+$\phi^3$ half integrand constructed in (\ref{nogluons.10}) leads to
triangle diagrams 
\begin{align}
&A^{\te{1-loop}}_{\text{EYM}, \text{max}}(1,2|3,4;\emptyset)\big|_{\kappa^2 g^2} \notag\\&= \frac{1}{16}\int \frac{ \dd^D \ell}{\ell^2} \lim_{k_{\pm} \rightarrow \pm \ell} \int \dd \mu_{6}^{\te{tree}} I^{\te{1-loop}}_{\rm YM,\te{max}}(\{1,2,3,4\};\ell)J^{\te{1-loop}}_{{\rm YM}+\phi^3}(1,2|3,4;\emptyset;\ell)\big|_{g^4\lambda^2}\notag\\
&=-\frac{N}{2}\, t_8(1,2,3,4)\int \frac{\dd^D\ell}{\ell^2}\left[\left( \frac{1}{\ell_1^2\ell_{12}^2}+(1\leftrightarrow 2)\right)+(1,2\leftrightarrow 3,4)\right] \label{nogravmax.2} \\
&= - 2 \, N\, t_8(1,2,3,4)\int \frac{\dd^D\ell}{\ell^2 \ell_1^2\ell_{12}^2} \, .  \notag
\end{align}
These triangular contributions are illustrated diagrammatically in the right panel of 
figure \ref{fig:nogravmax.1}. The recombination in terms of quadratic propagators is based on the identity 
 \begin{align}
\label{nogravmax.3}&4 \int \frac{\dd^D \ell}{\ell^2} \frac{f(\ell)}{\ell_P^2\ell_{PQ}^2}=\int \frac{\dd^D \ell }{\ell^2} \bigg[ \frac{f(\ell)}{s_{P,\ell}s_{PQ,\ell}} +\frac{f(\ell-k_P)}{s_{Q,\ell}s_{QR,\ell} }
+\frac{f(\ell-k_{P}-k_{Q})}{s_{R,\ell}s_{RP,\ell}}\bigg]
\end{align}
for multiparticle momenta subject to $k_P+k_Q+k_R=0$ 
which can be derived from partial-fraction identities and shifts of the loop 
momenta similar to those employed to obtain (\ref{looprev.11}). In passing 
to the last line of (\ref{nogravmax.2}), we have used elementary properties of
the scalar triangle integrals which allow to identify the four terms in the square bracket of the third line.

The triangles in (\ref{nogravmax.2}) are the first example where one-loop EYM 
amplitudes with 16 supercharges violate the no-triangle property of maximally supersymmetric
YM and supergravity \cite{Bern:1994zx}. As we will see in (\ref{nogravmax.4}) and below, generic one-loop
integrands of maximally supersymmetric EYM additionally involve bubble diagrams which go even
further beyond the no-triangle property.

\begin{figure}
	\centering	\includegraphics[scale=0.55]{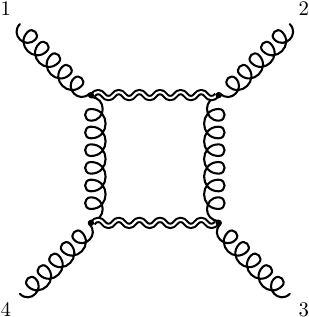}\ \ \ \ \ \ \ \ \ \ \ \ \includegraphics[scale=0.55]{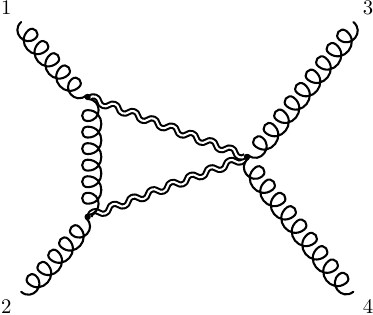}\ \ \ \ \ \ \ \ \ \ \ \ \includegraphics[scale=0.55]{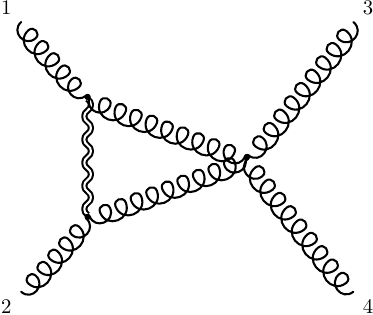}\\ \includegraphics[scale=0.55]{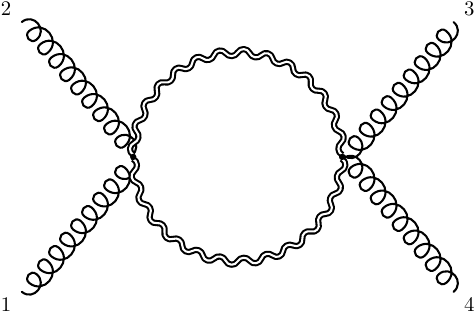}\ \ \ \ \ \ \ \ \ \ \ \ \includegraphics[scale=0.55]{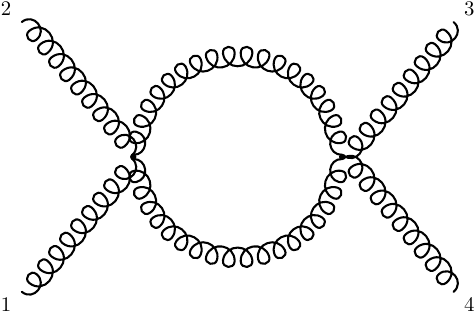}
	\caption{The expression (\ref{nogravmax.4}) for the $\kappa^4$ order of the four-gluon EYM 
	amplitude at one loop mixes box integrals with triangles and bubbles.}
	\label{fig:nogravmax.2}
\end{figure}

 At order $\kappa^4$ we use the YM+$\phi^3$ half integrand constructed in (\ref{nogluons.11}) to obtain: 
 \begin{align}
&A^{\te{1-loop}}_{ \text{EYM}, \text{max}}(1,2|3,4;\emptyset)\big|_{\kappa^4} \notag \\&= \frac{1}{256}\int \frac{ \dd^D \ell}{\ell^2} \lim_{k_{\pm} \rightarrow \pm \ell} \int \dd \mu_{6}^{\te{tree}} I^{\te{1-loop}}_{\rm YM,\te{max}}(\{1,2,3,4\};\ell)J^{\te{1-loop}}_{{\rm YM}+\phi^3}(1,2|3,4;\emptyset;\ell)\big|_{g^4}\label{nogravmax.4}\\
&=\frac{t_8(1,2,3,4)}{256}\int \frac{\dd^D\ell}{\ell^2}   \left\{\left[ \frac{8s_{14}^2}{\ell_{1}^2\ell_{12}^2\ell_{123}^2}+\frac{4s_{12}}{\ell_1^2\ell_{12}^2} +\frac{4s_{12}}{\ell_{3}^2\ell_{34}^2}
+\frac{(D{-}3{+}Nc_2)}{2\ell_{12}^2}+(3\leftrightarrow 4)\right]+(1\leftrightarrow 2)\right\}  \notag\\
&=\frac{t_8(1,2,3,4)}{16}\int \frac{\dd^D\ell}{\ell^2}   \left\{ 
\frac{s_{14}^2}{\ell_{1}^2\ell_{12}^2\ell_{123}^2}
+\frac{s_{13}^2}{\ell_{1}^2\ell_{12}^2\ell_{124}^2}
+\frac{2s_{12}}{\ell_1^2\ell_{12}^2} 
+\frac{(D{-}3{+}Nc_2)}{8\ell_{12}^2}
\right\}
\, .\nonumber
\end{align}
In addition to the partial-fraction manipulations (\ref{looprev.11}) and (\ref{nogravmax.3}) we use the identity
\begin{align}
\label{nogravmax.5}   2\int \frac{\dd^D \ell }{\ell^2} \frac{1}{\ell_{P}^2}=\int \frac{~\dd^D\ell }{\ell^2}\left[ \frac{1}{s_{P,\ell}}+\frac{1}{s_{Q,\ell}}\right]
\end{align}
for multiparticle momenta $k_P+k_Q=0$
to obtain the quadratic propagators of the bubble integral in (\ref{nogravmax.4}). 
Again, we have used relabelling symmetries of the boxes, triangles and bubbles
in passing to the last line of (\ref{nogravmax.4}).
The contributions to the one-loop amplitude with four external gluons at the $\kappa^4$ 
order are depicted diagrammatically in figure \ref{fig:nogravmax.2}. 

%

\subsection{One external graviton}
\label{sec:onegravmax}

\begin{figure}
\centering	\includegraphics[scale=0.8]{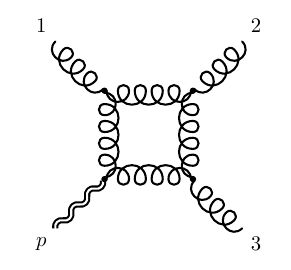}
\caption{The box graphs in the three-gluon amplitude at order  $\kappa g^3$ given by (\ref{onegravmax.2})}
\label{fig:onegravmax.1}
\end{figure}

The structure of the YM$+\phi^3$ half integrand in (\ref{onegluon.1}) admits
two contributions to the EYM four-point amplitude with one external graviton and three gluons,
\begin{align}
\label{onegravmax.3}
A^{\te{1-loop}}_{\text{EYM}, \text{max}}(1,2,3;\{p\})= A^{\te{1-loop}}_{\text{EYM}, \text{max}}(1,2,3;\{p\})\big|_{\kappa g^3}+ A^{\te{1-loop}}_{\text{EYM}, \text{max}}(1,2,3;\{p\})\big|_{\kappa^3 g} \, .
\end{align}
At order $\kappa g^3$, the YM+$\phi^3$ half integrand in (\ref{onegluon.2}) leads to
\begin{align}\label{onegravmax.2}
&A^{\te{1-loop}}_{\text{EYM}, \text{max}}(1,2,3;\{p\})\big|_{\kappa g^3}\\&= \frac{1}{4}\int \frac{ \dd^D \ell}{\ell^2} \lim_{k_{\pm} \rightarrow \pm \ell} \int \dd \mu_{6}^{\te{tree}} I^{\te{1-loop}}_{\rm YM,\te{max}}(\{1,2,3,p\};\ell) J^{\te{1-loop}}_{{\rm YM}+\phi^3}(1,2,3;\{p\};\ell)\big|_{g^4\lambda^3}\notag\\
&=4\, N\, t_8(1,2,3,p) \int \frac{\dd^D \ell}{\ell^2}  ~\left[\frac{ (\bar{\epsilon}_p\cdot \ell) }{\ell_{1}^2\ell_{12}^2\ell_{123}^2}  + \text{cyc}(1,2,3) \right] \, .\notag
\end{align}
As already derived in \cite{He:2016mzd}, maximal supersymmetry leads to a cyclic orbit
of box integrals as depicted in figure \ref{fig:onegravmax.1}. 
The recombination to quadratic propagators is done by supplementing
(\ref{looprev.11}) with a shift $\ell \rightarrow \ell - k_{12\ldots i}$ of the loop momentum
in the numerator in the $i^{\rm th}$ term of the second line.
Linearized gauge invariance of (\ref{onegravmax.2}) can be shown 
via shifts of loop momenta in the cyclic orbit of
\beq
\delta_{ \bar{\epsilon}_p \rightarrow p} 
\bigg( \frac{2 (\bar{\epsilon}_p\cdot \ell) }{\ell^2 \ell_{1}^2\ell_{12}^2\ell_{123}^2} \bigg) = 
 \frac{ \ell^2 - \ell^2_{123} }{\ell^2 \ell_{1}^2\ell_{12}^2\ell_{123}^2} 
 = \frac{ 1 }{ \ell_{1}^2\ell_{12}^2\ell_{123}^2}  - \frac{ 1 }{\ell^2 \ell_{1}^2\ell_{12}^2} \, .
\eeq

\begin{figure}[H]
\centering	
\includegraphics[scale=0.8]{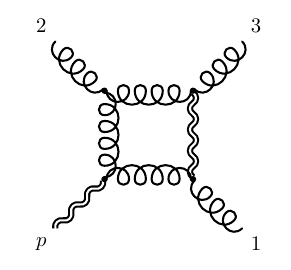}\ \ \ \ \ \ \ \ \ \ \ \ \includegraphics[scale=0.8]{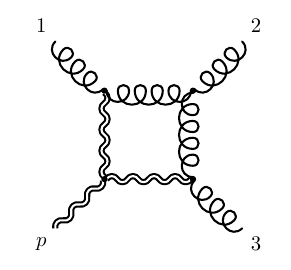}
\caption{The box graphs in the three-gluon amplitude at  order  $\kappa^3 g$ given by (\ref{onegravmax.1})}
\label{fig:onegravmax.2}
\end{figure}

For the  order $\kappa^3 g$ we use the half integrand (\ref{onegluon.3}) to get, 
\begin{align}\label{onegravmax.1}
&A^{\te{1-loop}}_{ \text{EYM}, \text{max}}(1,2,3;\{p\})\big|_{\kappa^3 g}\\
&=\frac{1}{64}\int \frac{ \dd^D \ell}{\ell^2} \lim_{k_{\pm} \rightarrow \pm \ell} \int \dd \mu_{6}^{\te{tree}} I^{\te{1-loop}}_{\rm YM,\te{max}}(\{1,2,3,p\};\ell) J^{\te{1-loop}}_{{\rm YM}+\phi^3}(1,2,3;\{p\};\ell)\big|_{g^4\lambda}\notag\\
&= \frac{1}{4} \, t_8(1,2,3,p)\left(k_1\cdot \bar f_p\cdot k_2\right) 
\int \frac{\dd^D\ell}{\ell^2}\left[ \frac{1}{\ell_{1}^2\ell_{12}^2\ell_{123}^2}+\text{cyc}(1,2,3)\right]\, ,\notag
\end{align}
where we introduce the linearized field strength in (\ref{looprev.4}) to manifest gauge invariance of 
\beq
k_1\cdot \bar f_p\cdot k_2 = s_{1p} (\bar{\epsilon}_p \cdot k_2) - s_{2p} (\bar{\epsilon}_p \cdot k_1)  \, .
\label{kinfactor3}
\eeq
The configurations of vertices $\sim \kappa^3 g$ in the box diagrams of
(\ref{onegravmax.1}) are depicted in figure~\ref{fig:onegravmax.2}.
Moreover, the kinematic factor (\ref{kinfactor3}) manifests the permutation antisymmetry of
$A^{\te{1-loop}}_{ \text{EYM}, \text{max}}(1,2,3;\{p\})\big|_{\kappa^3 g}$ in $1,2,3$,
consistent with the color structure $\hat f^{A_1 A_2 A_3}$ expected by the figure.

%

\subsection{Two external gravitons}
\label{sec:twogravmax}

\begin{figure}[H]
\centering	
\includegraphics[scale=0.8]{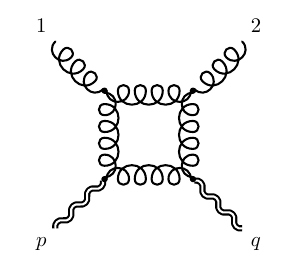}\ \ \ \ \ \ \ \ \ \ \ \ \includegraphics[scale=0.7]{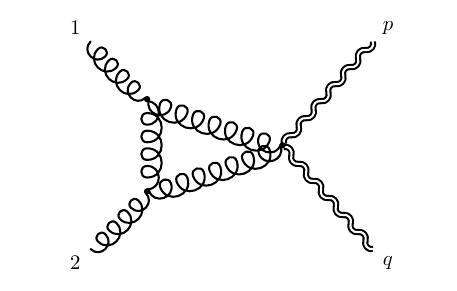}
\caption{The $\kappa^2g^2$ order of the two-gluon-two-graviton amplitude in (\ref{twogravmax.1}) 
is given by the depicted boxes and triangles.}
	\label{fig:twogravmax.1}
\end{figure}

With two external gravitons, the YM$+\phi^3$ half integrand
(\ref{twogluons.1}) introduces the following coupling dependence:
\begin{align}
\label{twogravmax.3}
A^{\te{1-loop}}_{\text{EYM}, \text{max}}(1,2;\{p,q\})= A^{\te{1-loop}}_{\text{EYM}, \text{max}}(1,2;\{p,q\})\big|_{\kappa^2 g^2}+A^{\te{1-loop}}_{ \text{EYM}, \text{max}}(1,2;\{p,q\})\big|_{\kappa^4}
\end{align}
Using the YM+$\phi^3$ half integrand (\ref{twogluons.2}) we obtain 
the following combination of boxes and triangles at order  $\kappa^2 g^2$,
see figure \ref{fig:twogravmax.1} for an illustration of both contributions.
\begin{align}\label{twogravmax.1}
&A^{\te{1-loop}}_{\text{EYM}, \text{max}}(1,2;\{p,q\})\big|_{\kappa^2 g^2}\\&=\frac{1}{8} \int \frac{ \dd^D \ell}{\ell^2} \lim_{k_{\pm} \rightarrow \pm \ell} \int \dd \mu_{6}^{\te{tree}} I^{\te{1-loop}}_{\rm YM,\te{max}}(\{1,2,p,q\};\ell) J^{\te{1-loop}}_{{\rm YM}+\phi^3}(1,2;\{p,q\};\ell)\big|_{g^4\lambda^2}\notag\\
&=2N \, t_8(1,2,p,q) \int \frac{\dd^D \ell}{\ell^2}
\bigg\{ \bigg[\bigg( 
\frac{ (\bar{\epsilon}_p\cdot \ell)(\bar{\epsilon}_q\cdot \ell)}{\ell_p^2\ell_{p1}^2\ell_{p12}^2}
+\frac{(\bar{\epsilon}_p\cdot (\ell+k_1))(\bar{\epsilon}_q\cdot\ell)}{2\, \ell_1^2\ell_{1p}^2\ell_{1p2}^2}
- \frac{(\bar{\epsilon}_p\cdot\bar{\epsilon}_q)}{4 \, \ell_1^2\ell_{12}^2}\bigg)
 \notag \\
&\quad \quad\quad \quad \quad \quad
 \quad \quad \quad \quad  \quad \quad  \quad \quad
  + (p\leftrightarrow q) \bigg]+(1\leftrightarrow 2)\bigg\} \, .\notag
\end{align}

\begin{figure}[H]
\centering
\includegraphics[scale=0.55]{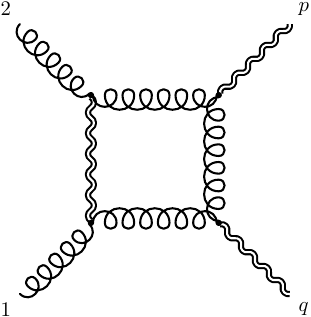}\ \ \ \ \ \ \ \ \ \ \ \ \includegraphics[scale=0.55]{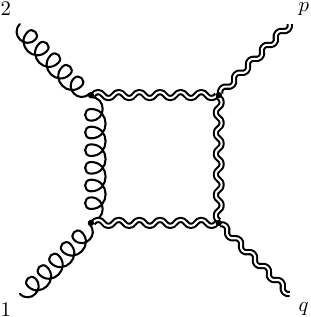}\ \ \ \ \ \ \ \ \ \ \ \ \includegraphics[scale=0.55]{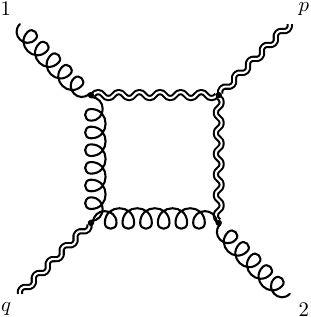} \\	\includegraphics[scale=0.55]{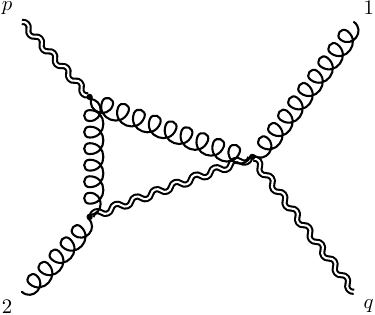}\ \ \ \ \ \ \ \ \ \ \ \ \includegraphics[scale=0.55]{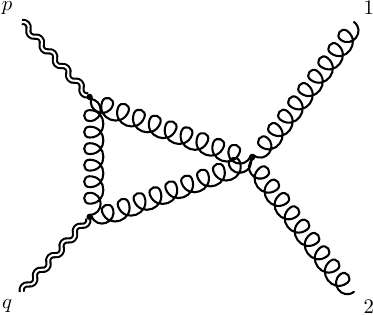}\ \ \ \ \ \ \ \ \ \ \ \ \includegraphics[scale=0.55]{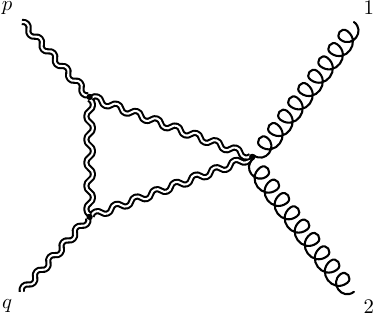}\\
\includegraphics[scale=0.55]{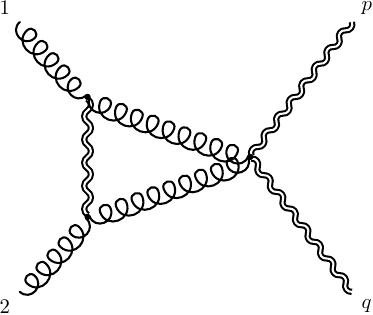}\ \ \ \ \ \ \ \ \ \ \ \ \includegraphics[scale=0.55]{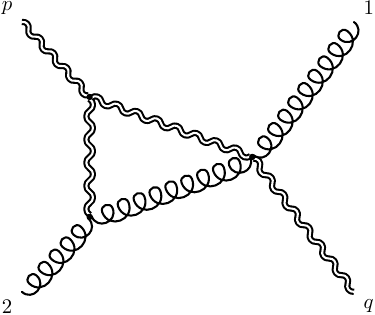}\\ \includegraphics[scale=0.55]{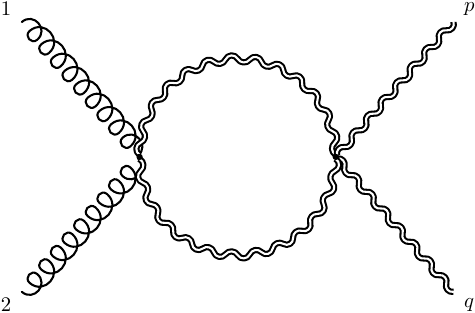} \includegraphics[scale=0.55]{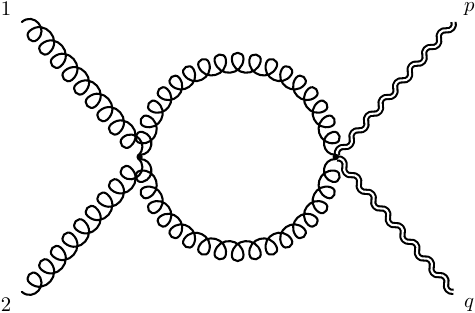}
\caption{At the $\kappa^4$ order of the two-gluon and two-graviton amplitude of EYM, 
various boxes, triangles and bubbles contribute to the integrand~(\ref{twogravmax.2}).}
\label{fig:twogravmax.2}
\end{figure}

For the order $\kappa^4$ contribution we use (\ref{twogluons.3}) to get
\begin{align}\label{twogravmax.2}
&A^{\te{1-loop}}_{ \text{EYM}, \text{max}}(1,2;\{p,q\})\big|_{\kappa^4}\\&= \int \frac{ \dd^D \ell}{\ell^2} \lim_{k_{\pm} \rightarrow \pm \ell} \int \dd \mu_{6}^{\te{tree}} I^{\te{1-loop}}_{\rm YM,\te{max}}(\{1,2,p,q\};\ell) J^{\te{1-loop}}_{{\rm YM}+\phi^3}(1,2;\{p,q\};\ell)\big|_{g^4}\notag\\
&=\frac{1}{256} \, t_8(1,2,p,q)\int \frac{\dd^D\ell}{\ell^2}\bigg\{\Big[\Big[(N c_2{+}D{-}3)\left(\frac{1}{2}\frac{(\bar{\epsilon}_p\cdot \bar{\epsilon}_q)}{\ell_{pq}^2}-\frac{2(\bar{\epsilon}_p\cdot \ell)( \bar{\epsilon}_q\cdot \ell_{p})}{\ell_{p}^2\ell_{pq}^2}\right)\notag\\
&\quad+\frac{4s_{12}(\bar{\epsilon}_p\cdot \bar{\epsilon}_q)}{\ell_{1}^2\ell_{12}^2}+\frac{4s_{12}(\bar{\epsilon}_p\cdot \bar{\epsilon}_q)+4(\bar{\epsilon}_p\cdot q)( \bar{\epsilon}_q\cdot p)}{\ell_{p}^2\ell_{pq}^2}\notag\\
&\quad- 8s_{12}\frac{2(\bar{\epsilon}_p \cdot \ell_{1})(\bar{\epsilon}_q\cdot \ell)+(\bar{\epsilon}_p\cdot \ell_{1})(\bar{\epsilon}_q\cdot k_1)+ (\bar{\epsilon}_p\cdot k_2)(\bar{\epsilon}_q\cdot \ell)}{\ell_{1}^2\ell_{12}^2\ell_{12p}^2}\notag\\
&\quad+8s_{1q}\frac{(\bar{\epsilon}_p\cdot \bar{\epsilon}_q)s_{1q}+ (\bar{\epsilon}_p\cdot \ell)(\bar{\epsilon}_q\cdot p)+(\bar{\epsilon}_p\cdot k_2)(\bar{\epsilon}_q\cdot k_1)- (\bar{\epsilon}_p\cdot q)(\bar{\epsilon}_q\cdot \ell)-(\bar{\epsilon}_p\cdot k_1)(\bar{\epsilon}_q\cdot k_2)}{\ell_{1}^2\ell_{12}^2\ell_{12p}^2}
\notag\\
&\quad-4s_{12}\frac{2(\bar{\epsilon}_p\cdot \ell_{1})(\bar{\epsilon}_q\cdot \ell)+(\bar{\epsilon}_p\cdot \ell_{1})(\bar{\epsilon}_q\cdot k_1)+(\bar{\epsilon}_p\cdot k_2)(\bar{\epsilon}_q\cdot \ell)}{\ell_{1}^2\ell_{1p}^2\ell_{1p2}^2}\notag\\
&\quad+4s_{1q}\frac{(\bar{\epsilon}_p\cdot \ell)(\bar{\epsilon}_q\cdot p)+(\bar{\epsilon}_p\cdot q)(\bar{\epsilon}_q\cdot \ell)+(\bar{\epsilon}_p\cdot k_1)(\bar{\epsilon}_q\cdot p) }{\ell_{1}^2\ell_{1p}^2\ell_{1p2}^2}
\Big]+(1\leftrightarrow 2)\Big]+ (p\leftrightarrow q) \bigg\} \notag\, .
\end{align}
The contributing diagrams are illustrated in figure \ref{fig:twogravmax.2}.
For both of (\ref{twogravmax.1}) and (\ref{twogravmax.2}), we have verified
gauge invariance in both $\bar \epsilon_p$ and $\bar \epsilon_q$. 

%
\subsection{Four external gravitons}
\label{sec:fourgravmax}

\begin{figure}[H]
	\centering	
	\includegraphics[scale=0.7]{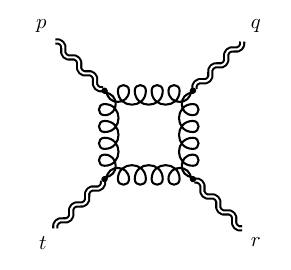}\ \ \ \ \ \ \ \ \includegraphics[scale=0.6]{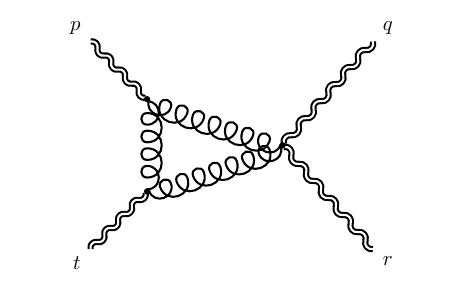}\ \ \ \includegraphics[scale=0.66]{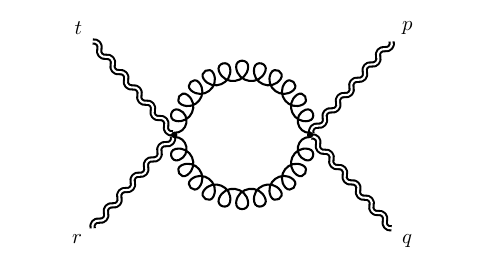}
\caption{The contribution (\ref{fourgravmax.2}) of a gauge-multiplet loop to the four-graviton amplitude
contains box, triangle and bubble graphs.}
	\label{fig:fourgravmax.1}
\end{figure}

Finally, the entire one-loop four-graviton amplitude is proportional to $\kappa^4$,
but it can be organized into contributions from gauge and gravity multiplets in the loop,
\begin{align}
\label{fourgravmax.2}
A^{\te{1-loop}}_{ \text{EYM}, \text{max}}(\{p,q,r,t\})&= A^{\te{1-loop}}_{ \text{EYM}, \text{max}}(\{p,q,r,t\})\big|_{\kappa^4}\\
&= A^{\te{1-loop}}_{ \text{EYM}, \text{max}}(\{p,q,r,t\})\big|_{\text{graviton loop}}+A^{\te{1-loop}}_{\text{EYM}, \text{max}}(\{p,q,r,t\})\big|_{\text{gluon loop}}\, ,\notag
\end{align}
see (\ref{fourgluons.2}) for the analogous organization of the YM$+\phi^3$ half integrand.
The contribution to (\ref{fourgravmax.2}) from a gluon loop follows from inserting 
the Parke-Taylor form of $J_{{\rm YM}+\phi^3}^{\rm tree}(+,-; \{p,q,r,t\})$ in the ancillary file
into (\ref{looprev.14}):
\begin{align}\label{fourgravmax.1}
&A^{\te{1-loop}}_{\text{EYM}, \text{max}}(\{p,q,r,t\})\big|_{\text{gluon loop}}\\&= \frac{1}{256}\int \frac{ \dd^D \ell}{\ell^2} \lim_{k_{\pm} \rightarrow \pm \ell} \int \dd \mu_{6}^{\te{tree}}\, I^{\te{1-loop}}_{\rm YM,\te{max}}(\{p,q,r,t\};\ell)  N c_2\,  J_{{\rm YM}+\phi^3}^{\rm tree}(+,-; \{p,q,r,t\})
\notag\\
&= \frac{N c_2}{64}\, t_8(p,q,r,t) \int \frac{\dd^D \ell}{\ell^2} \bigg\{ 
\frac{1}{2} \bigg(\frac{(\bar{\epsilon}_p\cdot \bar{\epsilon}_q)(\bar{\epsilon}_r\cdot \bar{\epsilon}_t)}{\ell_{pq}^2}+\frac{(\bar{\epsilon}_p\cdot \bar{\epsilon}_r)(\bar{\epsilon}_q\cdot \bar{\epsilon}_t)}{\ell_{pr}^2}+\frac{(\bar{\epsilon}_p\cdot \bar{\epsilon}_t)(\bar{\epsilon}_q\cdot \bar{\epsilon}_r)}{\ell_{pt}^2}\bigg)\notag\\
&\quad\quad\quad+ {}\bigg(\left[2\frac{(\bar{\epsilon}_p\cdot \ell)( \bar{\epsilon}_q\cdot \ell_{p})(\bar{\epsilon}_r \cdot \ell_{pq})( \bar{\epsilon}_t\cdot \ell) }{\ell_{p}^2\ell_{pq}^2\ell_{pqr}^2}-(\bar{\epsilon}_p\cdot \bar{\epsilon}_q) \left(\frac{(\bar{\epsilon}_r\cdot \ell )(\bar{\epsilon}_t\cdot \ell_{r})}{\ell_{r}^2\ell_{tr}^2}+\frac{(\bar{\epsilon}_r\cdot \ell_{t} )(\bar{\epsilon}_t\cdot \ell)}{\ell_{t}^2\ell_{tr}^2}\right)\right] \notag \\
&\quad\quad\quad\quad\quad\quad +\text{perm}(p,q,r,t)\bigg)\bigg\}\, . \notag
\end{align}
The different diagrams with a gluon loop are depicted in figure \ref{fig:fourgravmax.1},
and their interplay gives rise to a gauge-invariant amplitude under
$\bar{\epsilon}_p\rightarrow p$.

One can similarly obtain the contributions from a graviton in the loop via 
\begin{align}\label{altfourgrav}
&A^{\te{1-loop}}_{\text{EYM}, \text{max}}(\{p,q,r,t\})\big|_{\text{graviton loop}}\\&= \frac{1}{256}\int \frac{ \dd^D \ell}{\ell^2} \lim_{k_{\pm} \rightarrow \pm \ell} \int \dd \mu_{6}^{\te{tree}}\, I^{\te{1-loop}}_{\rm YM,\te{max}}(\{p,q,r,t\};\ell)   \,  J^{\te{tree}}_{{\rm YM}+\phi^3}(\emptyset; \{p,q,r,t,+,-\}) \, ,
\notag
\end{align}
where the master-numerator decomposition of 
$J^{\te{tree}}_{{\rm YM}+\phi^3}(\emptyset; \{p,q,r,t,+,-\})$ -- the forward limit
of a six-point Pfaffian (\ref{chysec.6}) -- can be found in section 3 of \cite{Geyer:2017ela}.

\section{Four-point one-loop EYM amplitudes with half-maximal supersymmetry}
\label{sec:halfmax}

In this section, we investigate four-point one-loop amplitudes in EYM theory with half-maximal supersymmetry,
i.e.\ 8 supercharges instead of 16. This kind of half-maximal supersymmetry can be realized in spacetime 
dimensions $D \leq 6$ and is referred to as $\mathcal{N}=2$ supersymmetry in four dimensions. 
Our procedure to construct the loop integrands of EYM is very similar to the one presented for the 
maximally supersymmetric case in section \ref{sec:maxsusy}: We start by introducing half integrands 
$I^{\te{1-loop}}_{\rm YM,\frac{1}{2}-\text{max}}$ tailored to a chiral hypermultiplet 
in the loop\footnote{Maximally supersymmetric 
gauge multiplets can be decomposed into a vector multiplet of 
half-maximal supersymmetry (one vector and two Weyl fermions in six dimensions) and 
two hypermultiplets (two scalars and a single Weyl fermion in 
six dimensions). Accordingly, one can extract the contribution of a vector multiplet in the loop
from the linear combinations (\ref{linearcomb}).} and spell out the associated half-maximally 
supersymmetric partial integrands of YM. Based on the CHY formula (\ref{looprev.14}),
one-loop amplitudes of EYM are constructed from the double copy of these YM partial integrands 
with the YM$+\phi^3$ building blocks of sections \ref{sec:hi} and~\ref{sec:nexthi}.
In all cases, we convert the linearized propagators in the loop integrand of (\ref{looprev.14})
to conventional quadratic ones.

All supercharges in our EYM results are from the half-maximally supersymmetric YM theory
which is double-copied with the universal, non-supersymmetric YM+$\phi^3$ theory. 
By double copy of the hypermultiplets of YM with the scalars and gluons of YM+$\phi^3$, the 
internal states for the EYM results in this
section are hypermultiplets in the adjoint representation of $U(N)$ and gravitino supermultiplets.
In six dimensions, the gravitino multiplet has 8 bosonic and 8 fermionic on-shell degrees of freedom
from the double copy of a hypermultiplet with the 4 physical polarizations of a $D=6$ gluon.
The contributions to the EYM loop integrand from adjoint vector multiplets and graviton multiplets 
with 8 supercharges each can be reconstructed from combining the results of this section with the maximally supersymmetric loop integrands of section \ref{sec:maxsusy}.

Massless gravitino multiplets generically conflict with local supersymmetry unless they are embedded into a larger gravity multiplet or rendered massive via compactification, see for instance \cite{Chiodaroli:2013upa} or section 2.6 of \cite{Bern:2019prr}. Hence, one can view the expressions of this section for gravitino multiplets in the loop as formal building blocks that compactly encode the difference between supergravity multiplets with 16 or 8 supercharges in the loop. 

The half-maximally supersymmetric EYM amplitudes in this section are once more expressed
in terms of dimension-agnostic gluon polarization vectors. Hence, the loop integrands in this
section apply to both the maximal dimension $D=6$ for 8 supercharges and dimensional reductions thereof.
We also track parity-odd contributions from the chiral fermions in the loop
in terms of the six-dimensional Levi-Civita tensor. The running of chiral fermions in 
fact leads to gauge and diffeomorphism anomalies in some of the four-point amplitudes. 
We perform the loop integrals for these six-dimensional anomalies and obtain rational 
functions of the momenta as expected.

The reduction from 16 to 8 supercharges leads to longer expressions for the EYM loop integrands
in this section as compared to those in section \ref{sec:maxsusy}. Hence, we
only present a subset of the possible four-point amplitudes with external gluons or gravitons
in the main text and relegate some cases (or certain orders in $g,\kappa$) to the ancillary files.

\subsection{Partial integrands with half-maximal supersymmetry}
\label{sec:pihalfmax} 

The four-point one-loop half integrand of YM with half-maximal supersymmetry
will again be used in Parke-Taylor form \cite{He:2017spx}
\begin{align}
\label{pihalfmax.1} 
&I^{\te{1-loop}}_{\rm YM,\frac{1}{2}-\text{max}}(\{\hat1,2,3,4\};\ell)=  \frac{1}{2}\sum_{\rho\in S_4} {\rm PT}(+, \rho(1,2,3,4),-) 
\\
&\ \ \ \ \times  \bigg\{ \ell_\mu \ell_\nu C_{1|2,3,4}^{\mu \nu} + \ell_\mu \left[ {\rm sgn}_{23}^{\rho} s_{23} C_{1|23,4}^{\mu}
+{\rm sgn}_{24}^{\rho}  s_{24} C_{1|24,3}^{\mu}
+ {\rm sgn}_{34}^{\rho} s_{34} C_{1|34,2}^{\mu} \right] \bigg\} \, ,\notag
\end{align}
where the coefficients depend on the permutation $\rho$ in the Parke-Taylor
ordering via
\beq
{\rm sgn}_{ij}^{\rho} = \left\{ \begin{array}{rl}
+1 &: \ \te{$i$ is on the right of $j$ in $\rho(1,2,3,4)$} \\
-1 &: \ \te{$i$ is on the left of $j$ in $\rho(1,2,3,4)$} \, . \\
\end{array} \right.
\eeq
The vectorial and tensorial kinematic factors $ C_{1|23,4}^{\mu}$ and $C_{1|2,3,4}^{\mu \nu}$ 
introduced in \cite{Berg:2016wux, Berg:2016fui} are multilinear in $D$-dimensional gluon 
polarization vectors, see appendix \ref{app:cs} for a brief review. In contrast to the 
scalar $t_8(1,2,3,4)$ that we encountered in the maximally supersymmetric case, they are 
not permutation invariant and obey
\beq
C_{1|2,3,4}^{\mu \nu} = C_{1|3,2,4}^{\mu \nu} = C_{1|2,4,3}^{\mu \nu} \, , \ \ \ \ \ \
C_{1|23,4}^{\mu}= - C_{1|32,4}^{\mu}
\eeq
as well as more complicated identities under exchange of leg 1 in front of the vertical bar \cite{Berg:2016fui}
and contractions with external momenta, see for instance (\ref{moreids}). 
Moreover, they exhibit simple poles in $s_{ij}$ from their expansion in terms of polarization vectors
and momenta in the ancillary file.

The analogous half integrand associated with a half-maximally supersymmetric
vector multiplet in the loop (instead of a hypermultiplet) is a linear combination 
of (\ref{pihalfmax.1}) with the maximally supersymmetric one (\ref{looprev.3}) \cite{Geyer:2015jch, He:2017spx}
\beq
I^{\te{1-loop}}_{\rm YM,\frac{1}{2}-\text{max}}(\{\hat 1,2,3,4\};\ell) \big|_{\rm vector} 
= I^{\te{1-loop}}_{\rm YM,\text{max}}(\{1,2,3,4\};\ell)
-2 I^{\te{1-loop}}_{\rm YM,\frac{1}{2}-\text{max}}(\{\hat 1,2,3,4\};\ell) \, .
\label{linearcomb}
\eeq
Their relative coefficients follow from the fact that the maximally supersymmetric 
gauge multiplet with $8+8$ on-shell degrees of freedom decomposes into one vector multiplet 
and two hypermultiplets associated with 8 supercharges.

\subsubsection{Singling out an anomaly leg}

While the vector $ C_{1|23,4}^{\mu}$ is gauge invariant with respect to
$\epsilon_j^{\mu} \rightarrow k_j^{\mu} \ \forall  \ j=1,2,3,4$, the parity-odd part
of the tensor $C_{1|2,3,4}^{\mu \nu}$ has an 
anomalous gauge variation in the first leg,
\begin{align}
\label{pihalfmax.3}
\delta_{\epsilon_1\rightarrow k_1 }C_{1|2,3,4}^{\mu \nu}  = 2 i \eta^{\mu \nu}\varepsilon_6(k_2,\epsilon_2,k_3,\epsilon_3, k_4,\epsilon_4)\, , \ \ \ \
\delta_{\epsilon_j\rightarrow k_j }C_{1|2,3,4}^{\mu \nu}
 = 0 \ \te{for}\ j= 2,3,4\, ,
\end{align}
where $\varepsilon_6$ is the six-dimensional Levi-Civita tensor contracting the
six vectors in the brackets.

By the asymmetric gauge variation (\ref{pihalfmax.3}) of the tensor, the half integrand
(\ref{pihalfmax.1}) is gauge invariant in external legs $2,3,4$ but anomalous in the first leg
\cite{He:2017spx, Edison:2020uzf}
\begin{align}
\label{anomHint}
\delta_{\epsilon_1\rightarrow k_1 }
I^{\te{1-loop}}_{\rm YM,\frac{1}{2}-\text{max}}(\{\hat1,2,3,4\};\ell)
= i \ell^2  \varepsilon_6(k_2,\ep_2,k_3,\ep_3,k_4,\ep_4)
\sum_{\rho\in S_4} {\rm PT}(+, \rho(1,2,3,4),-)\, .
\end{align}
The hat above leg 1 in the notation on the left-hand side of (\ref{pihalfmax.1}) keeps track of 
the external leg that carries the anomaly. While generic kinematic half integrand 
$I^{\te{1-loop}}_{\rm YM,\alpha}$
in (\ref{looprev.14}) are supposed to be permutation symmetric, the anomaly introduces a mild 
asymmetry \cite{He:2017spx, Edison:2020uzf}\footnote{The asymmetry of the half integrand can be traced back to the assignment of superghost pictures to the vertex operators of the RNS ambitwistor string \cite{Adamo:2013tsa} that determine
$I^{\te{1-loop}}_{\rm YM,\alpha}$ through their genus-one correlators. More precisely, the prescription for the 
parity-odd part of
genus-one amplitudes in the reference requires one of the vertex operators to appear in the superghost picture $-1$.
In contrast to the remaining vertex operators in the zero picture, the gauge variation in the $-1$ picture yields a derivative in moduli space and ultimately leads to the factor of $\ell^2$ in (\ref{asymmHI}).}
\begin{align}
&I^{\te{1-loop}}_{\rm YM,\frac{1}{2}-\text{max}}(\{\hat 1,2,3,4\};\ell) - I^{\te{1-loop}}_{\rm YM,\frac{1}{2}-\text{max}}(\{\hat 2,1,3,4\};\ell) 
\label{asymmHI} \\
& \ \ = - i \ell^2 \varepsilon_6(\ep_1,\ep_2,k_3,\ep_3,k_4,\ep_4) \sum_{\rho\in S_4} {\rm PT}(+, \rho(1,2,3,4),-) \notag
\end{align}
while maintaining permutation invariance in the unhatted legs $2,3,4$ in (\ref{pihalfmax.1}).
As we will see in section \ref{sec:anomaly}, the factor of $\ell^2$ in the asymmetry (\ref{asymmHI}) and the anomalous
gauge variation $\delta_{\ep_1 \rightarrow k_1}$ of (\ref{pihalfmax.1}) implies that all the Feynman integrals
obtained from integration over the $\sigma_j$ evaluate to rational functions of the momenta, 
i.e.\ no logarithms in the Mandelstam invariants.

\subsubsection{Partial integrands}

A Kleiss-Kuijf basis of partial integrands (\ref{defpartint}) for an internal hypermultiplet
resulting from (\ref{pihalfmax.1}) can be assembled from permutations in $\{2,3,4\}$ of
\begin{align}
&\label{pihalfmax.4}a^{\text{1-loop}}_{ \te{YM},\frac{1}{2}-\text{max}}(\hat 1,2,3,4,-,+) = \frac{\ell_\mu C_{1|23,4}^\mu}{s_{1,\ell}s_{4,\ell} }-\frac{\ell_\mu C_{1|34,2}^\mu}{s_{1,\ell}s_{12,\ell}} \\
&\hspace{1.2cm}-\frac{\ell_\mu\ell_\nu C_{1|2,3,4}^{\mu\nu}
	-(\ell_\mu C_{1|24,3}^\mu s_{24}+\ell_\mu C_{1|23,4}^\mu s_{23}+\ell_\mu C_{1|34,2}^\mu s_{34})
}{2s_{1,\ell}s_{4,\ell}s_{12,\ell}}
\notag\\
&\label{pihalfmax.5}a^{\text{1-loop}}_{ \te{YM},\frac{1}{2}-\text{max}}(4,\hat1,2,3,-,+) = -\frac{\ell_\mu C_{1|23,4}^\mu}{s_{4,\ell}s_{14,\ell}}\\
&\hspace{1.2cm}-\frac{\ell_\mu \ell_\nu C_{1|2,3,4}^{\mu\nu}
	+\ell_\mu C_{1|24,3}^\mu s_{24}-\ell_\mu C_{1|23,4}^\mu s_{23}+\ell_\mu C_{1|34,2}^\mu s_{34}
}{2s_{4,\ell}s_{3,\ell}s_{14,\ell}}\notag\\ 
&\label{pihalfmax.6}a^{\text{1-loop}}_{ \te{YM},\frac{1}{2}-\text{max}}(3,4,\hat1,2,-,+) = -\frac{\ell_\mu C_{1|34,2}^\mu}{s_{2,\ell}s_{34,\ell}} \\
&\hspace{1.2cm}-\frac{\ell_\mu\ell_\nu C_{1|2,3,4}^{\mu\nu}
	+\ell_\mu C_{1|24,3}^\mu s_{24}+\ell_\mu  C_{1|23,4}^\mu s_{23}-\ell_\mu C_{1|34,2}^\mu s_{34}
}{2s_{2,\ell}s_{3,\ell}s_{34,\ell}} \notag\\
&\label{pihalfmax.7}a^{\text{1-loop}}_{ \te{YM},\frac{1}{2}-\text{max}}(2,3,4,\hat1,-,+) =
\frac{\ell_\mu C_{1|34,2}^\mu}{s_{1,\ell}s_{2,\ell} }
-\frac{\ell_\mu C_{1|23,4}^\mu}{s_{1,\ell}s_{23,\ell}}
\\
&\hspace{1.2cm}-\frac{\ell_\mu \ell_\nu C_{1|2,3,4}^{\mu \nu}
	-(\ell_\mu C_{1|24,3}^\mu s_{24}+\ell_\mu C_{1|23,4}^\mu s_{23}+\ell_\mu C_{1|34,2}^\mu s_{34})
}{2s_{1,\ell}s_{2,\ell}s_{23,\ell}} \, .
\notag
\end{align} 
Vector multiplets in the loop can be accommodated by taking linear combinations with
the maximally supersymmetric partial integrands in (\ref{pimax.3}) according to (\ref{linearcomb}).
In (\ref{pihalfmax.4}) to (\ref{pihalfmax.7}) and similar equations below, the hat indicates the leg that
carries the anomaly i.e.\ singles out the variant of the underlying half integrand (\ref{pihalfmax.1}).

\subsubsection{The anomalous kinematic factor $P_{1|2|3,4}$}
\label{sec:kinPs}

As we will see below in (\ref{nogravhalfmax.3}) and (\ref{nogravhalfmax.1}), we will also encounter
scalar kinematic factors $P_{1|a|b,c} = P_{1|a|c,b}$ introduced in \cite{Berg:2016wux, Berg:2016fui},
\begin{align}
s_{12}P_{1|2|3,4} &= \frac{1}{4} \big[ {\rm tr}(f_1 f_3) {\rm tr}(f_2 f_4)
+ {\rm tr}(f_1 f_4) {\rm tr}(f_2 f_3) - {\rm tr}(f_1 f_2) {\rm tr}(f_3 f_4) \big] - {\rm tr}(f_1f_3f_2f_4)
\label{defPs} \\
&\ \ \ \ + i \big[ (\epsilon_1 \cdot k_2) \varepsilon_6(k_2,\ep_2,k_3,\ep_3,k_4,\ep_4) + (1\leftrightarrow 2)\big]
- i s_{12} \varepsilon_6(\ep_1,\ep_2,k_3,\ep_3,k_4,\ep_4) \, ,\notag
\end{align}
where the traces refer to the Lorentz indices of the linearized field
strengths, e.g.\ ${\rm tr}(f_1 f_2) = (f_1)_\mu{}^\nu (f_2)_\nu{}^\mu$.
They are symmetric in the last two legs, $P_{1|2|3,4}=P_{1|2|4,3}$, and 
related to the vectors and tensors in (\ref{pihalfmax.1}) via identities like
\begin{align}
(k_1)_{\mu }C^\mu_{1|23,4} &= P_{1|3|2,4} - P_{1|2|3,4} \notag \\
(k_1)_{\mu}  C^{\mu \nu}_{1|2,3,4} &=-k_2^\nu  P_{1|2|3,4}-k_3^\nu P_{1|3|2,4} - k_4^\nu P_{1|4|2,3}
\label{moreids} \\
\eta_{\mu\nu}C^{\mu\nu}_{1|2,3,4} &= 2(P_{1|2|3,4} + P_{1|3|2,4} + P_{1|4|2,3} )\, .
\notag
\end{align}
Similar to (\ref{pihalfmax.3}), the parity-odd terms in (\ref{defPs}) exhibit an anomalous
gauge variation in the first leg,
\begin{align}
\label{pihalfmax.P}
\delta_{\epsilon_1\rightarrow k_1 }P_{1|2|3,4} = 2 i  \varepsilon_6(k_2,\epsilon_2,k_3,\epsilon_3, k_4,\epsilon_4), \ \ \ 
\delta_{\epsilon_j\rightarrow k_j }P_{1|2|3,4} = 0 \ \te{for}\ j= 2,3,4\, .
\end{align}
The anomalies of the six-dimensional loop integrands due to (\ref{pihalfmax.3}) and 
(\ref{pihalfmax.P}) are discussed in more detail in section \ref{sec:anom.1}. 

Note that spinor-helicity components of $ C_{1|23,4}^{\mu},C_{1|2,3,4}^{\mu \nu}$ 
and $P_{1|2|3,4}$ upon dimensional reduction to $D=4$ vanish outside the MHV 
sector as required by supersymmetry, see section 5.1 of \cite{Berg:2016fui} for the non-zero MHV expressions.


\subsection{No external gravitons}
\label{sec:nogravhalfmax}

We shall now determine the loop integrand of the four-gluon amplitude
from the half-maximally supersymmetric ingredients (\ref{pihalfmax.1})
and (\ref{pihalfmax.4}) to (\ref{pihalfmax.7}). As before, separate calculations 
are performed for the single- and double-trace sector.

\subsubsection{Single-trace sector}

As in the maximally supersymmetric case, the two contributions (\ref{nogluons.1}) to the YM$+\phi^3$ 
half integrand give two different dependences on the couplings in the single-trace sector,
\begin{align}
\label{nogravhalfmax.6} 
A^{\te{1-loop}}_{\text{EYM}, \frac{1}{2}-\text{max}}(\hat 1,2,3,4;\emptyset)&= A^{\te{1-loop}}_{\text{EYM}, \frac{1}{2}-\text{max}}(\hat 1,2,3,4;\emptyset)\big|_{g^4}+A^{\te{1-loop}}_{ \text{EYM}, \frac{1}{2}-\text{max}}( \hat 1,2,3,4;\emptyset)\big|_{\kappa^2g^2}\, ,
\end{align}
where the hat above the first leg tracks the anomalous leg in the underlying half integrand (\ref{pihalfmax.1}).
Using the half integrand in (\ref{nogluons.13}) and the half-maximally supersymmetric partial integrands in (\ref{pihalfmax.4}) to (\ref{pihalfmax.7}), we obtain the single-trace sector of a half-maximally supersymmetric four-gluon amplitude in EYM theory at order $g^4$: 
\begin{align}\label{nogravhalfmax.3}
&A^{\te{1-loop}}_{\text{EYM}, \frac{1}{2}-\text{max}}(\hat 1,2,3,4;\emptyset)\big|_{g^4}\\&=\int \frac{ \dd^D \ell}{\ell^2} \lim_{k_{\pm} \rightarrow \pm \ell} \int \dd \mu_{6}^{\te{tree}} I^{\te{1-loop}}_{\rm YM,\frac{1}{2}-\text{max}}(\{\hat 1,2,3,4\};\ell)J^{\te{1-loop}}_{{\rm YM}+\phi^3}(1,2,3,4;\emptyset;\ell)\big|_{g^4\lambda^4}\notag\\
&=4N \int \frac{ \dd^D \ell}{\ell^2}\bigg\{ \frac{\ell_{\mu}C^{\mu}_{1|23,4}}{\ell_1^2\ell_{14}^2} - \frac{\ell_{\mu}C^{\mu}_{1|23,4}}{\ell_4^2\ell_{41}^2} + \frac{\ell_{\mu}C^{\mu}_{1|34,2}}{\ell_2^2\ell_{21}^2}- \frac{\ell_{\mu}C^{\mu}_{1|34,2}}{\ell_1^2\ell_{12}^2} - \frac{P_{1|2|3,4}}{\ell_4^2\ell_{43}^2}  - \frac{P_{1|4|2,3}}{\ell_2^2\ell_{23}^2} \notag  \\
&\quad + \frac{\ell_{\mu}\ell_{\nu} C^{\mu \nu}_{1|2,3,4}- \ell_{\mu}  (C_{1|34,2}^\mu s_{34} + C^{\mu}_{1|24,3} s_{24}+ C_{1|23,4}^\mu s_{23} )}{\ell_1^2\ell_{12}^2\ell_{123}^2} \notag\\
&\quad + \frac{\ell_{\mu}\ell_{\nu} C^{\mu \nu}_{1|2,3,4}+ \ell_{\mu}  (C_{1|34,2}^\mu s_{34} + C^{\mu}_{1|24,3} s_{24}+ C_{1|23,4}^\mu s_{23}  )}{\ell_1^2\ell_{14}^2\ell_{143}^2}  \bigg\} \, .\notag
\end{align} 
This expression matches the representation of the four-gluon amplitude in half-maximally supersymmetric
YM obtained in \cite{Berg:2016fui} after symmetrizing the loop integrand in the reference 
w.r.t.\ $2\leftrightarrow 4$. The scalar kinematic factors $P_{1|a|b,c}$ have been introduced
in section \ref{sec:kinPs}.

Similarly, we use the YM$+\phi^3$ half integrand (\ref{newnewgfw.2}) at subleading order in $\lambda$ to get:\footnote{We have
discarded a tadpole diagram proportional to $A_{\rm YM}^{\rm tree}(1,2,3,4)  \int \frac{ \dd^D \ell}{\ell^2}$ in 
deriving the result (\ref{nogravhalfmax.1}) which integrates to zero in dimensional regularization.}
\begin{align}\label{nogravhalfmax.1}
&A^{\te{1-loop}}_{\text{EYM}, \frac{1}{2}-\text{max}}(\hat 1,2,3,4;\emptyset)\big|_{\kappa^2g^2}\\&=\frac{1}{16} \int \frac{ \dd^D \ell}{\ell^2} \lim_{k_{\pm} \rightarrow \pm \ell} \int \dd \mu_{6}^{\te{tree}} I^{\te{1-loop}}_{\rm YM,\frac{1}{2}-\text{max}}(\{\hat 1,2,3,4\};\ell)J^{\te{1-loop}}_{{\rm YM}+\phi^3}(1,2,3,4;\emptyset;\ell)\big|_{g^4\lambda^2}\notag\\
&=\frac{1}{2} s_{13}\int \frac{ \dd^D \ell}{\ell^2}\bigg\{
{-} t_8(1,2,3,4)  \left(\frac{1}{\ell_{2}^2\ell_{23}^2}
+\frac{1}{\ell_{3}^2\ell_{32}^2}\right) \notag \\
&\ \  + \frac{\ell_{\mu} \ell_{\nu} C_{1|2,3,4}^{\mu \nu}-\ell_{\mu}(s_{23}C_{1|23,4}^{\mu}+s_{24}C_{1|24,3}^{\mu}+s_{34}C_{1|34,2}^{\mu})}{\ell_{1}^2\ell_{12}^2\ell_{123}^2}\notag\\
&\ \ +  \frac{  \ell_{\mu} \ell_{\nu} C_{1|2,3,4}^{\mu \nu}-\ell_{\mu} (s_{23} C_{1|23,4}^\mu+s_{24}C_{1|24,3}^{\mu}-s_{34}C_{1|34,2}^{\mu} ) }{\ell_{1}^2\ell_{12}^2\ell_{124}^2}\notag\\	
&\ \ +  \frac{ \ell_{\mu} \ell_{\nu} C_{1|2,3,4}^{\mu \nu}-\ell_{\mu} (-s_{23} C_{1|23,4}^\mu+s_{24}C_{1|24,3}^{\mu}+s_{34}C_{1|34,2}^{\mu} ) }{\ell_{1}^2\ell_{13}^2\ell_{132}^2}\notag\\	
&\ \ +  \frac{ \ell_{\mu} \ell_{\nu} C_{1|2,3,4}^{\mu \nu}-\ell_{\mu} (-s_{23} C_{1|23,4}^\mu-s_{24}C_{1|24,3}^{\mu}+s_{34}C_{1|34,2}^{\mu} ) }{\ell_{1}^2\ell_{13}^2\ell_{134}^2}\notag\\		
&\ \ +  \frac{  \ell_{\mu} \ell_{\nu} C_{1|2,3,4}^{\mu \nu}-\ell_{\mu} (s_{23} C_{1|23,4}^\mu-s_{24}C_{1|24,3}^{\mu}-s_{34}C_{1|34,2}^{\mu} ) }{\ell_{1}^2\ell_{14}^2\ell_{142}^2}\notag\\	
&\ \ +  \frac{ \ell_{\mu} \ell_{\nu} C_{1|2,3,4}^{\mu \nu}+\ell_{\mu} (s_{23} C_{1|23,4}^\mu+s_{24}C_{1|24,3}^{\mu}+s_{34}C_{1|34,2}^{\mu} )}{\ell_{1}^2\ell_{14}^2\ell_{143}^2}\notag\\	
&\ \ -P_{1|4|2,3}\left(\frac{1}{\ell_{3}^2\ell_{32}^2}+\frac{1}{\ell_{2}^2\ell_{23}^2}\right)-P_{1|2|3,4}\left(\frac{1}{\ell_{3}^2\ell_{34}^2}+ \frac{1}{\ell_{4}^2\ell_{43}^2}\right)-P_{1|3|2,4}\left(\frac{1}{\ell_{4}^2\ell_{42}^2}+ \frac{1}{\ell_{2}^2\ell_{24}^2}\right)  \bigg\} \notag\, .
\end{align}
Since the loop integral over the expression in the curly brackets is permutation invariant
w.r.t.\ $2,3,4$, the color-ordered amplitude (\ref{nogravhalfmax.1}) obeys Kleiss-Kuijf
relations just like its maximally supersymmetric counterpart in (\ref{nogravmax.1}).

\subsubsection{Double-trace sector}

The double-trace sector of the half-maximally supersymmetric
four-gluon amplitude, based on the half integrand (\ref{nogluons.9}) of YM$+\phi^3$, introduces 
three different powers of the couplings: 
\begin{align}
\label{nogravhalfmax.7} 
A^{\te{1-loop}}_{\text{EYM}, \frac{1}{2}-\text{max}}(\hat 1,2 | 3,4;\emptyset)&= A^{\te{1-loop}}_{ \text{YM},\frac{1}{2}-\text{max}}(\hat 1,2 |3,4;\emptyset)\big|_{g^4}\\
&\ \ \ \ +A^{\te{1-loop}}_{\text{EYM},\frac{1}{2}-\text{max}}(\hat 1,2 |3,4;\emptyset)\big|_{\kappa^2g^2}+A^{\te{1-loop}}_{ \text{EYM},\frac{1}{2}-\text{max}}(\hat 1,2 | 3,4;\emptyset)\big|_{\kappa^4}\, .\notag
\end{align}
For the order $g^4$ of the double-trace sector we get 
\begin{align}
\label{nogravhalfmax.5} 
&A^{\te{1-loop}}_{\text{EYM}, \frac{1}{2}-\text{max}}(\hat 1,2|3,4;\emptyset)\big|_{g^4} \\
&=\int \frac{ \dd^D \ell}{\ell^2}  \lim_{k_{\pm} \rightarrow \pm \ell} \int \dd \mu_{6}^{\te{tree}} I^{\te{1-loop}}_{\rm YM,\frac{1}{2}-\text{max}}(\{\hat 1,2,3,4\};\ell)J^{\te{1-loop}}_{{\rm YM}+\phi^3}(1,2|3,4;\{\emptyset\};\ell)\big|_{g^4\lambda^4}  \notag \\
&=8\int \frac{ \dd^D \ell}{\ell^2}\bigg\{  \frac{\ell_{\mu}\ell_{\nu} C^{\mu \nu}_{1|2,3,4} - \ell_{\mu}  (C^{\mu}_{1|24,3} s_{24} + C^{\mu}_{1|34,2} s_{34}+ C^{\mu}_{1|23,4}s_{23}  )}{\ell_1^2\ell_{12}^2\ell_{123}^2}\notag \\
&\quad\quad\quad +\frac{\ell_{\mu}\ell_{\nu} C^{\mu \nu}_{1|2,3,4} - \ell_{\mu}  (C^{\mu}_{1|24,3} s_{24}- C^{\mu}_{1|34,2} s_{34}+ C^{\mu}_{1|23,4}s_{23}  )}{\ell_1^2\ell_{12}^2\ell_{124}^2} \notag\\
&\quad\quad\quad +\frac{\ell_{\mu}\ell_{\nu} C^{\mu \nu}_{1|2,3,4} - \ell_{\mu}  (C^{\mu}_{1|24,3} s_{24}+ C^{\mu}_{1|34,2} s_{34}-  C^{\mu}_{1|23,4}s_{23}   )}{\ell_1^2\ell_{13}^2\ell_{132}^2} \notag\\
&\quad\quad\quad +\frac{\ell_{\mu}\ell_{\nu} C^{\mu \nu}_{1|2,3,4}+ \ell_{\mu}  (C^{\mu}_{1|24,3} s_{24}- C^{\mu}_{1|34,2} s_{34}+  C^{\mu}_{1|23,4}s_{23}  )}{\ell_1^2\ell_{13}^2\ell_{134}^2}\notag \\
&\quad\quad\quad +\frac{\ell_{\mu}\ell_{\nu} C^{\mu \nu}_{1|2,3,4}+ \ell_{\mu}  (C^{\mu}_{1|24,3} s_{24}+ C^{\mu}_{1|34,2} s_{34}-  C^{\mu}_{1|23,4}s_{23} )}{\ell_1^2\ell_{14}^2\ell_{142}^2}\notag \\
&\quad\quad\quad +\frac{\ell_{\mu}\ell_{\nu} C^{\mu \nu}_{1|2,3,4}+ \ell_{\mu}  (C^{\mu}_{1|24,3} s_{24}+ C^{\mu}_{1|34,2} s_{34}+  C^{\mu}_{1|23,4}s_{23}  )}{\ell_1^2\ell_{14}^2\ell_{143}^2}\notag \\
&\quad\quad\quad 
- P_{1|4|2,3} \bigg( \frac{1}{\ell_2^2\ell_{23}^2}+ \frac{1}{\ell_3^2\ell_{32}^2} \bigg)
-  P_{1|2|3,4} \bigg(\frac{1}{\ell_3^2\ell_{34}^2}+ \frac{1}{\ell_4^2\ell_{43}^2} \bigg)
- P_{1|3|2,4} \bigg( \frac{1}{\ell_2^2\ell_{24}^2}+ \frac{1}{\ell_4^2\ell_{42}^2} \bigg)\bigg\}\, . \notag
\end{align}
This is again proportional to a permutation sum of the single-trace amplitude (\ref{nogravhalfmax.3}), 
so the results of this section respect the supersymmetry-agnostic relations of \cite{Bern:1994zx} between
planar and non-planar one-loop amplitudes at zeroth order in $\kappa$.

Similarly, for the order of $\kappa^2g^2$ in the double-trace sector, we obtain
\begin{align}
\label{nogravhalfmax.2}
&A^{\te{1-loop}}_{ \text{EYM},\frac{1}{2}-\text{max}}(\hat 1,2|3,4;\emptyset)\big|_{\kappa^2g^2}\\&=  \frac{1}{16}\int \frac{ \dd^D \ell}{\ell^2} \lim_{k_{\pm} \rightarrow \pm \ell} \int \dd \mu_{6}^{\te{tree}} I^{\te{1-loop}}_{\rm YM,\frac{1}{2}-\text{max}}(\{\hat 1,2,3,4\};\ell)J^{\te{1-loop}}_{{\rm YM}+\phi^3}(1,2|3,4;\emptyset;\ell)\big|_{g^4\lambda^2}\notag\\
&=  \frac{N}{4} \int \frac{ \dd^D \ell}{\ell^2}	\bigg\{ 2s_{3,\ell} 
\ell_{\mu} C_{1|34,2}^\mu\left[\frac{1}{\ell_{1}^2\ell_{12}^2}+ \frac{1}{\ell_{2}^2\ell_{21}^2}\right]\notag\\
& \ \ \ \ \ \ \ \ \ \ \ \ \ \ + \frac{\ell_{\mu} (s_{34} C_{1|34,2}^{\mu}-s_{23} C_{1|23,4}^{\mu}-s_{24} C_{1|24,3}^{\mu} )-\ell_{\mu} \ell_{\nu} C_{1|2,3,4}^{\mu \nu}}{\ell_{3}^2\ell_{34}^2} \notag\\
& \ \ \ \ \ \ \ \ \ \ \ \ \ \ + \frac{\ell_{\mu} ({-}s_{34} C_{1|34,2}^{\mu}-s_{23} C_{1|23,4}^{\mu}-s_{24} C_{1|24,3}^{\mu} )-\ell_{\mu} \ell_{\nu} C_{1|2,3,4}^{\mu \nu}}{\ell_{4}^2\ell_{43}^2} \notag\\
& \ \ \ \ \ \ \ \ \ \ \ \ \ \ + \frac{\ell_{\mu} (s_{23} C_{1|23,4}^{\mu}+s_{24} C_{1|24,3}^{\mu} -s_{34} C_{1|34,2}^{\mu})-\ell_{\mu} \ell_{\nu} C_{1|2,3,4}^{\mu \nu}}{\ell_{1}^2\ell_{12}^2} \notag\\
& \ \ \ \ \ \ \ \ \ \ \ \ \ \ + \frac{\ell_{\mu} (s_{23} C_{1|23,4}^{\mu}+s_{24} C_{1|24,3}^{\mu} -s_{34} C_{1|34,2}^{\mu} )-\ell_{\mu} \ell_{\nu} C_{1|2,3,4}^{\mu \nu}}{\ell_{2}^2\ell_{21}^2} 
\bigg\}\, . \notag
\end{align}
The $\kappa^4$ order of the double-trace sector (\ref{nogravhalfmax.7}) is more lengthy
and therefore relegated to the ancillary file of the arXiv submission.


\subsection{One external graviton}
\label{sec:onegravhalfmax}

For all results with external gravitons and half-maximal supersymmetry, we chose to have a graviton
carry the anomaly of $I^{\te{1-loop}}_{\rm YM,\frac{1}{2}-\text{max}}$ and therefore write $\hat p$ in 
the place of $p$. The analogous expressions with a gluon in the anomaly leg can be found in 
the ancillary file. 

The amplitude with one external graviton is based 
on the half integrand in (\ref{onegluon.1}) and therefore contains two different combinations of couplings,  
\begin{align}
\label{onegrahalfvmax.3}
A^{\te{1-loop}}_{\text{EYM},\frac{1}{2}-\text{max}}(1,2,3;\{\hat p\})= 
A^{\te{1-loop}}_{\text{EYM}, \frac{1}{2}-\text{max}}(1,2,3;\{\hat p\})\big|_{\kappa g^3}+ 
A^{\te{1-loop}}_{\text{EYM}, \frac{1}{2}-\text{max}}(1,2,3;\{\hat p\})\big|_{\kappa^3 g} \, .
\end{align}
At order $\kappa g^3$ we find
\begin{align}\label{onegravhalfmax.1}
&A^{\te{1-loop}}_{\text{EYM}, \frac{1}{2}-\text{max}}(1,2,3;\{\hat p\})\big|_{\kappa g^3}\\&= \frac{1}{4}\int \frac{ \dd^D \ell}{\ell^2} \lim_{k_{\pm} \rightarrow \pm \ell} \int \dd \mu_{6}^{\te{tree}}  I^{\te{1-loop}}_{\rm YM,\frac{1}{2}-\text{max}}(\{1,2,3,\hat p\};\ell) J^{\te{1-loop}}_{{\rm YM}+\phi^3}(1,2,3;\{p\};\ell)\big|_{g^4\lambda^3}\notag\\
&=2N\, \int \frac{ \dd^D \ell}{\ell^2}\bigg\{\bigg[-\frac{ (\bar{\epsilon}_p\cdot \ell)\ell_{\mu} C_{p|12,3}^\mu}{\ell_{p}^2\ell_{p3}^2} -\frac{ (\bar{\epsilon}_p\cdot \ell_{3})\ell_{\mu} C_{p|12,3}^\mu}{\ell_{3}^2\ell_{3p}^2}-\frac{ (\bar{\epsilon}_p\cdot \ell)P_{p|3|1,2}}{\ell_{1}^2\ell_{12}^2}\notag \\
& \ \ \ \ +	\frac{ (\bar{\epsilon}_p\cdot \ell) \ell_{\mu}\ell_{\nu} C_{p|1,2,3}^{\mu \nu}-(\bar{\epsilon}_p\cdot \ell) (s_{23}\ell_{\mu} C_{p|23,1}^\mu{+}s_{13}\ell_{\mu} C_{p|13,2}^\mu{+}s_{12}\ell_{\mu} C_{p|12,3}^\mu )}{\ell_{p}^2\ell_{p1}^2\ell_{p12}^2}\, \bigg]+ \te{cyc}(1,2,3)\bigg\}\, ,\notag
\end{align}
which we confirmed to be invariant under linearized gauge transformations
$\bar{\epsilon}_p\rightarrow p$ in the YM$+\phi^3$ half integrand. The
analogous integrand at order $\kappa^3 g$ can be found in the ancillary file. 
At both orders $\kappa g^3$ and $\kappa^3 g$, the anomalous variations 
$\epsilon_j\rightarrow k_j$ on the supersymmetric side can be found 
in section \ref{sec:anom.2}.
%

\subsection{Two external gravitons}
\label{sec:twogravhalfmax}

The amplitude with two external gravitons, based on the half integrand (\ref{twogluons.1}) of YM$+\phi^3$, can come with two different powers of the couplings: 
\begin{align}
\label{twogravhalfmax.3}
A^{\te{1-loop}}_{\text{EYM},\frac{1}{2}-\text{max}}(1,2;\{\hat p,q\})= 
A^{\te{1-loop}}_{\text{EYM}, \frac{1}{2}-\text{max}}(1,2;\{\hat p,q\})\big|_{\kappa^2 g^2}+A^{\te{1-loop}}_{ \text{EYM}, \frac{1}{2}-\text{max}}(1,2;\{\hat p,q\})\big|_{\kappa^4}
\end{align}
The half-maximally supersymmetric amplitude at order $\kappa^2 g^2$ is found to be
\begin{align}\label{twogravhalfmax.1}
&A^{\te{1-loop}}_{\text{EYM},\frac{1}{2}-\text{max}}(1,2;\{\hat p,q\})\big|_{\kappa^2 g^2}\\&= \frac{1}{8}\int \frac{ \dd^D \ell}{\ell^2} \lim_{k_{\pm} \rightarrow \pm \ell} \int \dd \mu_{6}^{\te{tree}} I^{\te{1-loop}}_{\rm YM,\frac{1}{2}-\text{max}}(\{1,2,\hat p,q\};\ell) J^{\te{1-loop}}_{{\rm YM}+\phi^3}(1,2;\{p,q\};\ell)\big|_{g^4\lambda^2}\notag\\
&=\frac{N}{2} \int \frac{ \dd^D \ell}{\ell^2} \bigg\{    
\frac{(\bar{\epsilon}_p\cdot \ell) (\bar{\epsilon}_q\cdot k_2) \ell_{\mu} C_{p|q2,1}^\mu}{\ell_{p}^2\ell_{p1}^2}
+\frac{(\bar{\epsilon}_p\cdot \ell_{1}) (\bar{\epsilon}_q\cdot k_2 ) \ell_{\mu} C_{p|q2,1}^\mu}{\ell_{1}^2\ell_{p1}^2} \notag \\
&\quad +\frac{ (\bar{\epsilon}_p\cdot \bar{\epsilon}_q)\left[s_{pq}\ell_{\mu} (C_{p|q2,1}^{\mu}+C_{p|12,q}^\mu )+s_{p2} \ell_{\mu} (C_{p|q2,1}^\mu - C_{p|q1,2}^\mu )-\ell_{\mu}\ell_{\nu}  C_{p|q,1,2}^{\mu \nu}\right] }{2\ell_{1}^2\ell_{12}^2}  \notag\\
&\quad +\frac{(\bar{\epsilon}_p\cdot \ell)(\bar{\epsilon}_q\cdot \ell_{p})\left[ \ell_{\mu}\ell_{\nu} C_{p|q,1,2}^{\mu \nu}-(s_{p2} \ell_{\mu} C_{p|q1,2}^{\mu}+s_{pq}\ell_{\mu} C_{p|12,q}^{\mu}+ s_{p1} \ell_{\mu} C_{p|q2,1}^{\mu})\right]}{\ell_{p}^2\ell_{pq}^2\ell_{pq1}^2} \notag\\
&\quad +\frac{(\bar{\epsilon}_p\cdot \ell)(\bar{\epsilon}_q\cdot (\ell-k_2)) \left[\ell_{\mu}\ell_{\nu} C_{p|q,1,2}^{\mu \nu}-(s_{p1}\ell_{\mu} C_{p|q2,1}^{\mu}-s_{p2}\ell_{\mu} C_{p|q1,2}^{\mu}+s_{pq} \ell_{\mu} C_{p|12,q}^{\mu})\right]}{\ell_{p}^2\ell_{p1}^2\ell_{p1q}^2} \notag\\	
&\quad +\frac{(\bar{\epsilon}_p\cdot \ell)(\bar{\epsilon}_q\cdot \ell)\left[ \ell_{\mu} \ell_{\nu}  C_{p|q,1,2}^{\mu \nu}+ (s_{p1} \ell_{\mu}  C_{p|q2,1}^{\mu}+s_{p2} \ell_{\mu}  C_{p|q1,2}^{\mu}-s_{pq} \ell_{\mu}  C_{p|12,q}^{\mu} )\right]}{\ell_{p}^2\ell_{p1}^2\ell_{p12}^2} \notag\\
&\quad - \frac{(\bar{\epsilon}_p\cdot \ell)(\bar{\epsilon}_q\cdot (\ell{-}p)) P_{p|q|1,2}}{\ell_{1}^2\ell_{12}^2}	- \frac{(\bar{\epsilon}_p\cdot \ell)(\bar{\epsilon}_q\cdot \ell) P_{p|2|q,1}}{\ell_{q}^2\ell_{q1}^2}	- \frac{(\bar{\epsilon}_p\cdot \ell)(\bar{\epsilon}_q\cdot \ell_{1}) P_{p|1|q,2}}{\ell_{1}^2\ell_{1q}^2} +(1\leftrightarrow 2)\bigg\}\, .\notag
\end{align}
Our result for the integrand at order $\kappa^4$ as well as the one for the four-graviton amplitude can be found in the ancillary file. 

\subsection{Gauge anomalies in six dimensions}
\label{sec:anomaly} 

In this section, we integrate the anomalous gauge variations of the six-dimensional versions of 
the amplitudes in this section. These
anomalies result from the variations (\ref{pihalfmax.3}) and (\ref{pihalfmax.P}) of the tensor $C_{1|2,3,4}^{\mu \nu}$
and scalar $P_{1|2|3,4}$ which yield rational functions of the momenta upon loop integration.

\subsubsection{Amplitudes with no external gravitons}
\label{sec:anom.1}

First, we consider the variation $\epsilon_1 \rightarrow k_1$ of the four-gluon amplitude at order $g^4$. In the single-trace sector, the anomaly due to the expression (\ref{nogravhalfmax.3}) is given by
\begin{align}
&\delta_{\epsilon_1\rightarrow k_1 } A^{\te{1-loop}}_{\text{EYM}, \frac{1}{2}-\text{max}}(\hat 1,2,3,4;\emptyset)\big|_{g^4} \label{anomhalfmax.3} \\
&=8iN \varepsilon_6(k_2,\epsilon_2, k_3, \epsilon_3, k_4,\epsilon_4) \int  \dd^D \ell \,  \bigg\{
 \frac{\eta_{\mu \nu} \ell^\mu \ell^\nu}{\ell^2 \ell_1^2\ell_{12}^2\ell_{123}^2}
+ \frac{\eta_{\mu \nu} \ell^\mu \ell^\nu }{\ell^2 \ell_1^2\ell_{14}^2\ell_{143}^2} 
- \frac{1}{\ell^2 \ell_4^2\ell_{43}^2}  - \frac{1}{\ell^2 \ell_2^2\ell_{23}^2} 
  \bigg\}  \notag \\
&= 8iN \varepsilon_6(k_2,\epsilon_2, k_3, \epsilon_3, k_4,\epsilon_4) \int  \dd^D \ell \,  \bigg\{
 \frac{ \eta_{\mu \nu} \ell^\mu \ell^\nu }{\ell^2 \ell_1^2\ell_{12}^2\ell_{123}^2}
  - \frac{1}{\ell_1^2 \ell_{12}^2\ell_{123}^2} 
+(2\leftrightarrow 4)  \bigg\} \, ,
\notag
\end{align} 
where a naive contraction $ \eta_{\mu \nu} \ell^\mu \ell^\nu \rightarrow \ell^2$ would give rise to a
vanishing loop integrand. However, all the integrals with measure $\dd^D\ell$ in this work are
understood in dimensional regularization where $D=2m-2 \varepsilon$ is displaced from
integer values $2m\in \mathbb N$ by some infinitesimal parameter $ \varepsilon$.
In an expansion around six dimensions (with $m=3$), the inverse propagators $\ell_{12\ldots}^2$
are $(6{-}2 \varepsilon)$-dimensional contractions while the anomalous contributions
in the numerator are lacking the formal $({-}2 \varepsilon)$-dimensional components of $\ell$,
\beq
 \eta_{\mu \nu} \ell^\mu \ell^\nu = \ell^2 - \ell^2_{({-}2 \varepsilon)} \, .
 \label{ellsquares}
\eeq
In this way, the integrand of (\ref{anomhalfmax.3}) is found to be non-vanishing and 
proportional to $\ell^2_{({-}2 \varepsilon)}$,
\begin{align}
&\delta_{\epsilon_1\rightarrow k_1 } A^{\te{1-loop}}_{\text{EYM}, \frac{1}{2}-\text{max}}(\hat 1,2,3,4;\emptyset)\big|_{g^4} \label{anomhalfmax.4} \\
&= - 8iN \varepsilon_6(k_2,\epsilon_2, k_3, \epsilon_3, k_4,\epsilon_4) \int  \dd^{6-2 \varepsilon} \ell \,  \bigg\{
 \frac{ \ell^2_{({-}2 \varepsilon)}}{\ell^2 \ell_1^2\ell_{12}^2\ell_{123}^2}
+(2\leftrightarrow 4)  \bigg\} \, ,
\notag
\end{align} 
The key formulae for the evaluation of $\ell^2_{({-}2 \varepsilon)}$ integrals
in $D=6-2 \varepsilon$ dimensions are reviewed in appendix \ref{app:gaugemeth}.
Specifically, the identity (\ref{integrals.3}) for the integral in (\ref{anomhalfmax.4}) yields 
\begin{align}
\label{anomaly.5}
\delta_{\epsilon_1\rightarrow k_1 }A^{\te{1-loop}}_{\text{EYM}, \frac{1}{2}-\text{max}}(\hat 1,2,3,4;\emptyset)\big|_{g^4} = \frac{8}{3} \,  \pi^3 \, N \, \varepsilon_6(k_2,\epsilon_2, k_3, \epsilon_3, k_4,\epsilon_4)\, ,
\end{align}
consistent with the anomaly of the four-gluon amplitude in \cite{Berg:2016fui}. Here
and below, we discard the ${\cal O}(\varepsilon)$ contributions that vanish in $D=6$.

Based on the permutation sum of the computations in
(\ref{anomhalfmax.3}) to (\ref{anomaly.5}), the double-trace
sector (\ref{nogravhalfmax.5}) of the one-loop four-gluon amplitude is given by
\begin{align}
\label{anomaly.6}
\delta_{\epsilon_1\rightarrow k_1 }A^{\te{1-loop}}_{\text{EYM}, \frac{1}{2}-\text{max}}(\hat 1,2|3,4;\emptyset)\big|_{g^4} =16\,  \pi^3\, \varepsilon_6(k_2,\epsilon_2, k_3, \epsilon_3, k_4,\epsilon_4)\, .
\end{align}
At the order $\kappa^2 g^2$ of the single-trace amplitude in (\ref{nogravhalfmax.1}),
the same mechanism leads to
\beq
\label{anomaly.1}
\delta_{\epsilon_1\rightarrow k_1 }A^{\te{1-loop}}_{\text{EYM}, \frac{1}{2}-\text{max}}(\hat 1,2,3,4;\emptyset)\big|_{\kappa^2 g^2} =  \pi^3\,  s_{13}\, \varepsilon_6(k_2,\epsilon_2, k_3, \epsilon_3, k_4,\epsilon_4)\, .
\eeq
For the anomaly at the $\kappa^2 g^2$ order of the double-trace sector in turn,
the variation of (\ref{nogravhalfmax.2}) introduces
\begin{align}
\label{newhalfmax.2}
&\delta_{\epsilon_1\rightarrow k_1 }A^{\te{1-loop}}_{ \text{EYM},\frac{1}{2}-\text{max}}(\hat 1,2|3,4;\emptyset)\big|_{\kappa^2g^2}\\
&=  -  \frac{iN}{2}  \varepsilon_6(k_2,\epsilon_2, k_3, \epsilon_3, k_4,\epsilon_4) \int \frac{ \dd^{6-2 \varepsilon} \ell}{\ell^2} \, \eta_{\mu \nu} \ell^\mu \ell^\nu \, 
\bigg\{  \frac{1}{\ell_{3}^2\ell_{34}^2} 
+ \frac{1}{\ell_{4}^2\ell_{43}^2} 
+ \frac{1}{\ell_{1}^2\ell_{12}^2}
+ \frac{1}{\ell_{2}^2\ell_{21}^2} 
\bigg\}\, . \notag
\end{align}
With the rewriting (\ref{ellsquares}) of the six-dimensional $\eta_{\mu \nu} \ell^\mu \ell^\nu$,
we obtain one-mass bubble integrals that vanish in dimensional regularization such as
\beq
\int  \frac{ \dd^{6-2 \varepsilon} \ell }{\ell^2_3 \ell^2_{34}} = 0 \, .
\eeq
The $({-}2 \varepsilon)$-dimensional parts of $\eta_{\mu \nu} \ell^\mu \ell^\nu$ in (\ref{newhalfmax.2}) 
introduce rational versions (\ref{integrals.4}) of scalar triangles which yield the anomaly
\begin{align}
&\delta_{\epsilon_1\rightarrow k_1 }A^{\te{1-loop}}_{ \text{EYM},\frac{1}{2}-\text{max}}(\hat 1,2|3,4;\emptyset)\big|_{\kappa^2g^2} \label{newhalfmax.2a} \\
&= \frac{iN}{2}  \varepsilon_6(k_2,\epsilon_2, k_3, \epsilon_3, k_4,\epsilon_4) \int  \dd^{6-2 \varepsilon} \, 
\bigg\{ \bigg( \frac{\ell^2_{(-2\varepsilon)}}{\ell^2 \ell_{3}^2\ell_{34}^2} 
+ (3\leftrightarrow 4) \bigg) + 
(1,2\leftrightarrow 3,4)
\bigg\} \notag \\
%
&=  \frac{ \pi^3}{6} \, N \, s_{12}\, \varepsilon_6(k_2,\epsilon_2, k_3, \epsilon_3, k_4,\epsilon_4)  \, .
\notag
\end{align}
Finally, the expression for the order $\kappa^4$ of the double-trace sector in the ancillary 
file and the integrals in appendix \ref{app:gaugemeth} give rise to the following anomaly
\begin{align}
\delta_{\epsilon_1\rightarrow k_1 }A^{\te{1-loop}}_{\text{EYM}, \frac{1}{2}-\text{max}}(\hat 1,2| 3,4;\emptyset)\big|_{\kappa^4}&= \frac{\pi^3}{256}\varepsilon_6(k_2,\epsilon_2, k_3, \epsilon_3, k_4,\epsilon_4) \label{anomaly.3} \\
& \ \ \times
\left(\frac{2(Nc_2{+}D{-}3)}{15}s_{12}^2-\frac{16 }{3} s_{14}s_{13}\right) \, . 
\notag
\end{align}
 \subsubsection{Amplitudes with one external graviton} 
\label{sec:anom.2}

For the amplitude with one external graviton at order $\kappa g^3$ that was given in (\ref{onegravhalfmax.1}), the sum of the anomalous variations from the tensor and scalar building blocks conspire to
\begin{align}\label{anohalfmax.1}
&\delta_{\epsilon_p\rightarrow p }A^{\te{1-loop}}_{\text{EYM}, \frac{1}{2}-\text{max}}(1,2,3;\{\hat p\})\big|_{\kappa g^3} \\
&=-2iN\,  \varepsilon_6(k_1,\epsilon_1,k_2,\epsilon_2, k_3, \epsilon_3) \int   \dd^{6-2\varepsilon} \ell \, (\bar{\epsilon}_p\cdot \ell)
\bigg\{ \frac{\ell^2_{(-2\varepsilon)}}{\ell^2 \ell_{p}^2\ell_{p1}^2\ell_{p12}^2}
+ \te{cyc}(1,2,3)\bigg\}\, .\notag
\end{align}
By the expression (\ref{integrals.6}) for the vector integral, the anomaly is proportional 
to the cyclic sum over $\bar{\epsilon}_p\cdot (k_1{+}2k_2{+}3k_3)$ which vanishes
by $\bar{\epsilon}_p\cdot p=0$,
\beq
\label{anomaly.4}
\delta_{\epsilon_p\rightarrow p}A^{\te{1-loop}}_{ \text{EYM},\frac{1}{2}-\text{max}}(1,2,3;\{\hat p\})\big|_{\kappa g^3} = 0\, .
\eeq
The same conclusion can be reached for a gluon in the anomalous leg,
\beq
\label{anomaly.9}
\delta_{\epsilon_1\rightarrow k_1}A^{\te{1-loop}}_{ \text{EYM},\frac{1}{2}-\text{max}}(\hat 1,2,3;\{ p\})\big|_{\kappa g^3} = 0\, .
\eeq
The integrand for the amplitude with one external graviton at order $\kappa^3 g$ can be found in the ancillary file. In six dimensions  a gauge transformation in the leg $p$ in the half-maximally supersymmetric half integrand together with (\ref{integrals.3}) results in the anomaly
\begin{align}
\label{anomaly.10}
&\delta_{\epsilon_p\rightarrow p}A^{\te{1-loop}}_{ \text{EYM},\frac{1}{2}-\text{max}}(1,2,3;\{\hat p\})\big|_{\kappa^3 g} \\
&=\frac{i}{8}\,  \varepsilon_6(k_1,\epsilon_1, k_2, \epsilon_2, k_3,\epsilon_3) \int \frac{ \dd^{6-2\varepsilon} \ell}{\ell^2} 
\notag\\
&\hspace{0.65cm}
\times \bigg[  (k_1\cdot\bar  f_p \cdot k_3) \,\frac{ \ell^2_{(-2\varepsilon)} }{\ell_{p}^2\ell_{p1}^2\ell_{p13}^2}
- (k_1\cdot\bar  f_p \cdot k_2) \, \frac{ \ell^2_{(-2\varepsilon)} }{\ell_{p}^2\ell_{p1}^2\ell_{p12}^2}  +\te{cyc}(1,2,3)\bigg]
\notag \\
&= \frac{\pi^3}{8}  \,(k_1\cdot \bar f_p \cdot k_2) \varepsilon_6(k_1,\epsilon_1, k_2, \epsilon_2, k_3,\epsilon_3)
\, ,\notag
\end{align}
(see (\ref{looprev.4}) for the linearized field strength $\bar  f_p $)
and similarly 
\begin{align}
\label{anomaly.7}
&\delta_{\epsilon_1\rightarrow k_1}A^{\te{1-loop}}_{ \text{EYM},\frac{1}{2}-\text{max}}(\hat 1,2,3;\{ p\})\big|_{\kappa^3 g} = \frac{\pi^3}{8}  \,(k_1\cdot\bar  f_p \cdot k_2) \varepsilon_6(k_2,\epsilon_2, k_3, \epsilon_3, p,\epsilon_p)
\, .
\end{align}

\subsubsection{Amplitudes with two external gravitons} 
\label{sec:anom.3} 
 
For the amplitude with two external gravitons at order $\kappa^2 g^2$ in half-maximal supersymmetry the anomaly in six dimensions is:
 \begin{align}\label{anomaly.8}
&\delta_{\epsilon_p\rightarrow p} A^{\te{1-loop}}_{4, \text{EYM}, \frac{1}{2}-\text{max}}(1,2;\{\hat p,q\})\big|_{\kappa^2g^2}\\
&= iN\varepsilon_6(k_q,\epsilon_q,k_1,\epsilon_1, k_2,\epsilon_2) \int \frac{ \dd^{6-2\varepsilon} \ell}{\ell^2}  \, 
 \ell^2_{(-2\varepsilon)} \,\bigg\{
  \frac{ (\bar{\epsilon}_p\cdot \bar{\epsilon}_q)  }{2\ell_1^2\ell_{12}^2} 
 +  \frac{(\bar{\epsilon}_p\cdot \bar{\epsilon}_q) }{2\ell_2^2\ell_{21}^2}   \notag \\
 &\hspace{3cm}-   \frac{(\bar{\epsilon}_p\cdot \ell)(\bar{\epsilon}_q\cdot (\ell+p+k_1)) }{\ell_p^2\ell_{p1}^2\ell_{p1q}^2}
 - \frac{(\bar{\epsilon}_p\cdot \ell)(\bar{\epsilon}_q\cdot( \ell+p))}{\ell_p^2\ell_{pq}^2\ell_{pq1}^2} 
 -   \frac{(\bar{\epsilon}_p\cdot \ell)(\bar{\epsilon}_q\cdot (\ell+p))  }{\ell_p^2\ell_{pq}^2\ell_{pq2}^2}
 \notag  \\
&\hspace{3cm}  -   \frac{(\bar{\epsilon}_p\cdot \ell)(\bar{\epsilon}_q\cdot \ell)}{\ell_p^2\ell_{p1}^2 \ell_{p12}^2}
- \frac{(\bar{\epsilon}_p\cdot \ell)(\bar{\epsilon}_q\cdot (\ell-k_1)) }{\ell_p^2\ell_{p2}^2\ell_{p2q}^2}
- \frac{(\bar{\epsilon}_p\cdot \ell)(\bar{\epsilon}_q\cdot \ell) }{\ell_p^2\ell_{p2}^2\ell_{p21}^2}
\bigg\} \notag\\
&=- \frac{N}{24}\,  \pi^3 \varepsilon_6(k_q,\epsilon_q,k_1,\epsilon_1, k_2,\epsilon_2) (\bar f_p)_{\mu \nu} (\bar f_q)^{\mu \nu}\, . \notag \end{align}
 The integrals were calculated as indicated in appendix \ref{app:gaugemeth}, and
 we have rewritten the kinematic factor in terms of linearized field strengths via
 $(\bar f_p)_{\mu \nu} (\bar f_q)^{\mu \nu} =
2 (\bar{\epsilon}_p\cdot \bar{\epsilon}_q) s_{pq} -
2 (\bar{\epsilon}_p\cdot q)(\bar{\epsilon}_q\cdot p) $.

\section{Conclusion}
\label{sec:conclusion}

We have introduced a method to determine one-loop integrands of EYM theories
with any number of external gauge and gravity multiplets to all orders in the couplings
$\kappa$ and $g$. Our construction is based on the double copy of (possibly supersymmetric)
gauge theories with YM$+\phi^3$ theory, implemented via one-loop CHY formulae involving forward limits
of tree-level integrands on the Riemann sphere. More specifically, the forward limits of YM$+\phi^3$ 
building blocks yield new relations between loop integrands of EYM and those of YM theories, 
see (\ref{fwlimits}) and (\ref{looprev.14}) for the main formulae. These 
relations take a universal form for any number of supersymmetries and spacetime dimensions.

We have worked out the composition rules for tree-level building blocks
in color-ordered EYM loop integrands with any number of traces.
At four points, we have applied our method to determine one-loop EYM 
amplitudes with 8 and 16 supercharges and exposed their supersymmetry
cancellations. In particular, the linearized Feynman propagators resulting
from the CHY integrals are recombined to conventional quadratic ones. Moreover, we have
evaluated the rational expressions for six-dimensional gauge and diffeomorphism anomalies 
in the half-maximally supersymmetric case due to chiral hypermultiplets in the loop.

This methods and results of this work suggest a variety of follow-up research directions:
\begin{itemize}
\item {\it higher multiplicity}: With the availability of supersymmetric YM loop integrands
at $n\geq 5$ points \cite{Mafra:2014gja, He:2016mzd, He:2017spx, Edison:2020uzf, Bridges:2021ebs}, 
there is no obstruction to constructing higher-point one-loop EYM amplitudes from our method. 
It is conceivable that the half integrands of YM$+\phi^3$ in section
\ref{sec:hi} admit further all-multiplicity simplifications as exemplified in (\ref{extend.30}) for $n$
external scalars at subleading order in the coupling $\lambda$.
\item {\it higher loops}: Based on ambitwistor-string methods, the integrands of two-loop amplitudes
in gauge theories and (super-)gravity can be derived from double-forward limits of
tree-level building blocks \cite{Geyer:2016wjx, Geyer:2018xwu, Geyer:2019hnn}. It would
be interesting to perform the same double-forward limits in the half-integrands of YM$+\phi^3$
and deduce two-loop EYM amplitudes from the setup in the references.
\item {\it more general supergravities:} The double-copy structure of EYM
theories generalizes to magical, general homogeneous or gauged ${\cal N}=2$
supergravities \cite{Chiodaroli:2015wal, Chiodaroli:2018dbu}.
The non-supersymmetric double-copy constituents in the references augment
YM$+\phi^3$ by fundamental fermions and mass
terms that preserve the color-kinematics duality. It would be a rewarding line
of follow-up research to investigate worldsheet descriptions of the double copies
and one-loop integrands of these ${\cal N}=2$ supergravities.
\item {\it comparison with conventional string theories}: It would give a valuable crosscheck of our results to 
match the one-loop EYM amplitudes in this work with the point-particle limit 
$\alpha' \rightarrow0$ of genus-one amplitudes of heterotic and type-I superstrings.
In particular, the EYM amplitude relations in this work call for comparison with the $\alpha' \rightarrow0$ limit 
of the relations for mixed open- and closed-string type-I amplitudes at genus one in 
\cite{Stieberger:2021daa}.
\item {\it uplift to higher-mass-dimension operators}: It would be rewarding to incorporate
$\alpha'$-corrections into our forward-limit approach to EYM one-loop amplitudes as
done for loop integrands of pure SYM and supergravity in \cite{Edison:2021ebi}. 
Adapting the methods of the reference to genus-one correlators mixing
gauge and gravity multiplets should yield one-loop matrix elements of higher-mass-dimension
operators\footnote{The shorthand $D^{2k} F^m R^n$ refers to effective operators 
involving $m$ powers of the non-abelian gluon field strength $F$, $n$ powers of the Riemann 
tensor $R$ and $2k$ gauge- and diffeomorphism-covariant derivatives.} $D^{2k} F^m R^n$ and the non-analytic contributions to the $\alpha'$-expansion
of the respective genus-one string amplitudes.
\end{itemize}


\acknowledgments

We are grateful to Marco Chiodaroli, Alex Edison, Henrik Johansson and Fei Teng for combinations of inspiring discussions, collaboration on related topics and valuable comments on a draft version of this work. 
Moreover, we thank Jan Plefka for initiating this collaboration, participating in early stages 
of this project and a variety of valuable discussions. The research of FP was supported in part by a Humboldt Research Track Scholarship. OS is supported by the European Research Council under ERC-STG-804286 UNISCAMP.


\appendix


\section{YM+$\phi^3$ half integrands at tree-level}
\label{app:ymphi3tree}

In (\ref{chysec.8}), (\ref{chysec.11}), (\ref{chysec.12}) and (\ref{chysec.13}) we gave some specific examples for the half integrands of YM+$\phi^3$ in the CHY representation (\ref{chysec.0}) of EYM tree-level amplitudes. These can be deduced from a general expression for multiple traces given in \cite{Cachazo:2014xea} that we review here. We denote the cyclically ordered set of scalars in the $i^{\rm th}$ trace as $\Tr_i$. The color-decomposed YM+$\phi^3$ half integrand with $r$ gluons $\{p_1,p_2,\ldots ,p_r\}$ and scalars in $m$ traces $\Tr_1,\Tr_2,\cdots \Tr_m$ is given by
\begin{align}
\label{appymphi3.1}
J_{{\rm YM}+\phi^3}^{\rm tree}(\Tr_1|\dots|\Tr_m; \{p_1,\dots,p_r\}) 
 = \PT({\Tr_1}) \dots \PT({\Tr_m}) \text{Pf}'\Pi(\text{Tr}_1,\dots,\text{Tr}_m, \{p_1,\dots,p_r\}) \, .
\end{align}
The antisymmetric $2(r{+}m)\times 2(r{+}m)$-matrix $\Pi(\text{Tr}_1,\dots,\text{Tr}_m, \{p_1,\dots,p_r\})$ is constructed from the  $(2r\times 2r)$-matrix $\Psi(\{p_1,\dots,p_r\})$ in the YM half integrand (\ref{chysec.7}) by adding rows and columns for each of the traces: 
\begin{align}
\label{appymphi3.2}\Pi:=
\begin{blockarray}{ccccc}
b{\in} \{p_1,\dots,p_r\} & j{\in} \{1,\dots,m\} & b{\in} \{p_1,\dots,p_r\}&j' {\in}\{1,\dots,m\} &\\&&&&\\
\begin{block}{(cccc)c}
A_{ab} &\Pi_{a,j} & (-C)^T_{ab} &\Pi_{a,j'}& a\in  \{p_1,\dots,p_r\}\\&&&&\\
\Pi_{i,b}& \Pi_{i,j} & \tilde{\Pi}_{i,b}& \Pi_{i,j'}&i\in \{1,\dots,m\}
\\&&&&\\
C_{ab} & \tilde{\Pi}_{a,j}& B_{ab} & \tilde{\Pi}_{a,j'}& a\in \{p_1,\dots,p_r\}
\\&&&&\\
\Pi_{i',b} &\Pi_{i',j}&
\tilde{\Pi}_{i',b}& \Pi_{i',j'}&i'\in \{1,\dots,m\}\\
\end{block}
\end{blockarray}
\end{align}
The submatrices are defined as
\begin{align}
\label{appymphi3.3}
&  \Pi_{i,b} = \sum_{c\in \text{Tr}_i} \frac{k_c\cdot k_b}{\sigma_{c,b}} ,\ \   \tilde{\Pi}_{i,b}=\sum_{c\in\text{Tr}_i} \frac{k_c\cdot \epsilon_b}{\sigma_{c,b}} , \ \ \Pi_{i',b} = \sum_{c\in \text{Tr}_i} \frac{\sigma_c k_c\cdot k_b}{\sigma_{c,b}} , \ \ \tilde{\Pi}_{i',b} = \sum_{c\in \text{Tr}_i} \frac{\sigma_ck_c\cdot \epsilon_b}{\sigma_{c,b}} \\
& \Pi_{i,j} = \sum_{c\in \text{Tr}_i, d\in \text{Tr}_j} \frac{k_c\cdot k_d}{\sigma_{c,d}},\ \ \Pi_{i',j}=\sum_{c\in\text{Tr}_i,d\in \text{Tr}_j} \frac{\sigma_ck_c\cdot k_d}{\sigma_{c,d}}, \ \ \Pi_{i',j'} = \sum_{c\in \text{Tr}_i,d\in \text{Tr}_{j} } \frac{\sigma_ck_c\cdot k_d\notag \sigma_d}{\sigma_{c,d}}\, .
\end{align}
The modified Pfaffian $\Pf'$ of $\Pi$ can be evaluated in several equivalent ways
\begin{align}
\label{appymphi3.4}
\Pf'\Pi := \Pf \big[\Pi\big]_{ij'}^{ij'}= \frac{(-1)^a}{\sigma_a} \Pf \big[\Pi\big]^{j'a}_{{j'a}}= \frac{(-1)^a}{\sigma_a} \Pf\big[\Pi\big]_{ia}^{ia}= \frac{(-1)^{a+b}}{\sigma_{a,b}} \Pf \big[\Pi\big]_{ab}^{ab} \, ,
\end{align}
where $[\ldots]^{ab}_{ab}$ once more instructs to remove the $a^{\rm th}$ and $b^{\rm th}$
rows and columns. Note that in the absence of scalars, the matrix $\Pi$ reduces to the matrix $\Psi$ 
in (\ref{chysec.7}) and thus the YM+$\phi^3$ half integrand reduces to a YM half integrand. Similarly, 
(\ref{appymphi3.2}) smoothly reduces to the $2m\times 2m$ matrix $\big( \begin{smallmatrix}  \Pi_{i,j} 
& \Pi_{i,j'} \\ \Pi_{i',j} & \Pi_{i',j'}\end{smallmatrix} \big)$ with $i,j,i',j' \in \{1,2,\ldots, m\}$ in absence of gluons.


\section{Building blocks of the half-maximally supersymmetric integrand} 
\label{app:cs}

The tensor $C^{\mu \nu}_{1|2,3,4}$ and the vectors $C^{\mu}_{1|ab,c}$ are the central building blocks of the 
half-maximally supersymmetric integrands in section \ref{sec:halfmax}. Together with a closely related
scalar $C_{1|abc}$, they are defined in terms of 
more elementary tensors~$t$ \cite{Berg:2016wux, Berg:2016fui}
\begin{align}
C_{1|234} &= t_{1,234}+t_{12,34}+t_{123,4} - t_{124,3} - t_{14,23} -t_{142,3} +t_{143,2} \notag \\
C_{1|23,4}^\mu &= t_{1,23,4}^\mu + t_{12,3,4}^\mu -t_{13,2,4}^\mu + k_3^\mu t_{123,4} -k_2^\mu t_{132,4} +k_4^\mu \left[t_{14,23}-t_{214,3}+t_{314,2} \right] \label{appceq.1}\\
C_{1|2,3,4}^{\mu \nu} & = t_{1,2,3,4}^{\mu \nu} +2 \left[k_2^{(\mu}t_{12,3,4}^{\nu)}+ (2\leftrightarrow 3,4) \right]- 2 \left[k_2^{(\mu}k_3^{\nu)} t_{213,4}+ (2,3|2,3,4)\right] \, , \notag
\end{align}
which in turn are defined in terms of Berends-Giele currents $\mathfrak{e}$ and $\mathfrak{f}$ as 
\begin{align}
t_{A,B}& = -\frac{1}{2} (\mathfrak{f}_A)_{\mu \nu} \mathfrak{f}_B^{\mu \nu} \notag \\
t_{A,B,C}^\mu&= \left[\mathfrak{e}_A^\mu t_{B,C} + (A\leftrightarrow B,C) \right] + \frac{i}{4}\varepsilon_6^\mu (\mathfrak{e}_A, \mathfrak{f}_B, \mathfrak{f}_C) \label{appceq.2} \\
t_{A,B,C,D}^{\mu \nu} &= 2\left[\mathfrak{e}_A^{(\mu}\mathfrak{e}_B^{\nu)} t_{C,D} + (A,B|A,B,C,D) \right] +\frac{i}{2} \left[\mathfrak{e}_B^{(\mu}\varepsilon_6^{\nu)}(\mathfrak{e}_A,\mathfrak{f}_C, \mathfrak{f}_D) + (B\leftrightarrow C,D)\right]\, .  \notag
\end{align}
These currents are labelled by words $P=(p_1,p_2,\ldots,p_{|P|})$ and recursively defined by
\begin{align}
\mathfrak{e}_P^\mu&= \frac{1}{2s_P} \sum_{XY=P} \left[\mathfrak{e}_{Y}^\mu (k_Y \cdot \mathfrak{e}_Y) + (\mathfrak{e}_Y)_{\nu} \mathfrak{f}_X^{\mu \nu} -(X\leftrightarrow Y)\right] \notag \\
\mathfrak{f}_P^{\mu\nu} &= k_P^\mu \mathfrak{e}_P^\nu - k_P^\nu \mathfrak{e}_P^\mu - \sum_{XY=P} \left(\mathfrak{e}_X^\mu\mathfrak{e}_Y^\nu -\mathfrak{e}_Y^\mu\mathfrak{e}_X^\nu \right)\, , \label{appceq.3}
\end{align}
starting with the single-particle cases $ \mathfrak{e}_j^\mu=\epsilon_j^\mu$ and
$\mathfrak{f}_j^{\mu\nu}= k_j^\mu\epsilon_j^\nu- k_j^\nu \epsilon_j^\mu$. Vector
indices are symmetrized according to the normalization convention
$2k_2^{(\mu}k_3^{\nu)}=k_2^{\mu}k_3^{\nu}+k_2^{\nu}k_3^{\mu}$, and
the sums over deconcatenations $P=XY$ exclude the empty words $X=\emptyset$
and $Y=\emptyset$.

The propagators $s_P^{-1}$ in the recursion (\ref{appceq.3}) for $\mathfrak{e}_P^\mu$ expose
simple poles $t_{12,3,4}^\mu \sim s_{12}^{-1}$, and one might naively expect
a pole structure of $(s_{12}s_{123})^{-1}$ and $(s_{23}s_{123})^{-1}$ for $t_{123,4}$.
The propagators $s_{123}^{-1}$ diverge in the momentum phase-space of four
massless particles and are in fact absent from $t_{123,4}$ based on the 
Minahaning procedure \cite{Minahan:1987ha, Berg:2016wux, Berg:2016fui}.\footnote{Following
J.\ Minahan's prescription to relax momentum conservation in intermediate steps \cite{Minahan:1987ha}, 
factors of $s_{123}$ are temporarily taken to be non-zero and cancel 
between numerators and denominators
of $t_{123,4}$ \cite{Berg:2016wux, Berg:2016fui}.} 
Similarly, the poles $s^{-1}_{12}$ and $s^{-1}_{34}$ of
$(\mathfrak{f}_{12})_{\mu \nu}$ and $\mathfrak{f}_{34}^{\mu \nu}$ do
not occur simultaneously in $t_{12,34}$. On these grounds, all 
of $C_{1|234}, C_{1|23,4}^\mu$ and $C_{1|2,3,4}^{\mu \nu}$ only have simple
poles in $s_{ij}$.

More details on the symmetries and relations of these 
building blocks can be found in \cite{Berg:2016fui, He:2017spx}.

\section{Feynman integrals in six-dimensional anomalies}
\label{app:gaugemeth}

In this appendix, we review the expressions for the Feynman integrals in
the anomalous gauge variations of half-maximally supersymmetric EYM amplitudes
presented in section \ref{sec:anomaly}. As a common feature of the Feynman integrals in 
such anomalies, their loop integrand is proportional to $\ell_{(-2\varepsilon)}^2$, the
formal $(-2\varepsilon)$-dimensional part of the $(6{-}2\varepsilon)$-dimensional square $\ell^2$.
These factors of $\ell_{(-2\varepsilon)}^2$ arise along with the propagators of box and triangle 
integrals via
\begin{align}
\label{integrals.8} 
\int\dd^{6-2\varepsilon} \ell\, \bigg\{ \frac{f(\ell) \ell^\mu \ell^\nu \eta_{\mu \nu} }{\ell^2 \ell_1^2\ell_{12}^2\ell_{123}^2} - \frac{f(\ell{-}k_1) }{\ell^2\ell_{2}^2 \ell_{23}^2} \bigg\}= 
- \int\dd^{6-2\varepsilon} \ell\,  \frac{f(\ell) \ell_{(-2\varepsilon)}^2 }{\ell^2 \ell_1^2\ell_{12}^2\ell_{123}^2} \\
\label{integrals.9}
\int\dd^{6-2\varepsilon} \ell\,  \bigg\{ \frac{f(\ell)  \ell^\mu \ell^\nu \eta_{\mu \nu} }{\ell^2 \ell_1^2\ell_{12}^2} - \frac{f(\ell{-}k_1) }{\ell^2\ell_{2}^2 } \bigg\} = 
- \int\dd^{6-2\varepsilon} \ell\,  \frac{f(\ell) \ell_{(-2\varepsilon)}^2 }{\ell^2 \ell_1^2\ell_{12}^2} \, ,
\end{align}
where $f(\ell)$ is a polynomial in the loop momentum.
Such integrals can be related to the poles in dimensionally regulated integrals in 
higher dimensions with the relation
\cite{Bern:1995db, Bern_1997, Weinzierl:2006qs}
\begin{align}
\label{integrals.1}
\int \frac{\dd^{2m-2\varepsilon}\ell}{i \pi^{m-\varepsilon}}(-\ell^2_{(-2\varepsilon)})^r \, 
F(\ell_{(2m)}, \ell^2_{(-2\varepsilon)}) = \frac{\Gamma(r-\varepsilon)}{\Gamma(-\varepsilon)} \int \frac{\dd^{2m+2r-2\varepsilon} \ell}{i\pi^{m+r-\varepsilon}} \, F(\ell_{(2m)}, \ell^2_{(-2\varepsilon)})\, ,
\end{align}
valid for $m,r \in \NN$ and arbitrary functions $F(\ell_{(2m)}, \ell^2_{(-2\varepsilon)})$ of 
the $2m$- and the $(-2\varepsilon)$-dimensional components of the loop momentum.
Specifically, we use this relation with $r=1$ and $m = 3$,
\begin{align}
\label{integrals.2}
\int  \dd^{6-2\varepsilon}\ell\, \ell^2_{(-2\varepsilon)} F(\ell_{(6)}, \ell^2_{(-2\varepsilon)})
=\frac{\varepsilon}{\pi} \int \dd^{8-2\varepsilon}\ell\,  F(\ell_{(6)}, \ell^2_{(-2\varepsilon)})\, , 
\end{align}
where the resulting scalar boxes, triangles and bubbles in $8{-}2\varepsilon$ dimensions can 
be obtained with standard methods \cite{Weinzierl:2006qs}, for instance
\begin{align}
\label{integrals.3}
\int \dd^{6-2\varepsilon} \ell\,  \frac{\ell^2_{(-2\varepsilon)}}{\ell^2\ell_1^2\ell_{12}^2\ell_{123}^{2}}&=\frac{\varepsilon}{\pi} \int \dd^{8-2\varepsilon}\ell\,  \frac{1}{\ell^2\ell_1^2\ell_{12}^2\ell_{123}^{2}}=  \frac{i\pi^3}{6} +{\cal O}(\varepsilon)\\
\label{integrals.4} 
\int \dd^{6-2\varepsilon} \ell\,  \frac{\ell^2_{(-2\varepsilon)}}{\ell^2\ell_1^2\ell_{12}^2}&=\frac{\varepsilon}{\pi} \int \dd^{8-2\varepsilon}\ell\,  \frac{1}{\ell^2\ell_1^2\ell_{12}^2}=  -\frac{i\pi^3s_{12}}{12} +{\cal O}(\varepsilon)\\
\label{integrals.5} 
\int \dd^{6-2\varepsilon} \ell\,  \frac{\ell^2_{(-2\varepsilon)}}{\ell^2\ell_{12}^2}&= \frac{\varepsilon}{\pi} \int \dd^{8-2\varepsilon}\ell\,  \frac{1}{\ell^2\ell_{12}^2}=  \frac{i\pi^3s_{12}^2}{15} + {\cal O}(\varepsilon) \, ,
\end{align}
where the ${\cal O}(\varepsilon)$ terms on the right-hand sides are not tracked
in the six-dimensional anomalies of section \ref{sec:anomaly}. Additionally, the anomalies in half-maximally supersymmetric EYM amplitudes with external
gravitons feature vector and tensor integrals with $\ell^2_{(-2\varepsilon)}$ insertions. We 
employ Passarino-Veltman reduction \cite{PASSARINO1979151}, in particular its
implementation in the \texttt{Mathematica} package \texttt{FeynCalc} \cite{Shtabovenko_2016},
to obtain these vector and tensor integrals from scalar ones. Upon inserting the expressions
(\ref{integrals.3}) to (\ref{integrals.5}) for scalar integrals, we arrived at
\begin{align}
\label{integrals.6}
\int \dd^{6-2\varepsilon} \ell\,  \frac{\ell^2_{(-2\varepsilon)}\ell^\mu_{(6)}}{\ell^2\ell_1^2\ell_{12}^2\ell_{123}^{2}}&= \frac{\varepsilon}{\pi} \int \dd^{8-2\varepsilon}\ell\,  \frac{\ell^\mu_{(6)}}{\ell^2\ell_1^2\ell_{12}^2\ell_{123}^{2}}= \frac{i\pi^3}{24} \left[k_2^\mu + 2k_3^\mu+3k_4^\mu\right] + {\cal O}(\varepsilon)\\
\label{integrals.7}
\int \dd^{6-2\varepsilon} \ell\,  \frac{\ell^2_{(-2\varepsilon)}\ell^\mu_{(6)}\ell^\nu_{(6)}}{\ell^2\ell_1^2\ell_{12}^2\ell_{123}^{2}}&= \frac{\varepsilon}{\pi} \int \dd^{8-2\varepsilon}\ell\,  \frac{\ell^\mu_{(6)}\ell^\nu_{(6)}}{\ell^2\ell_1^2\ell_{12}^2\ell_{123}^{2}}\\
&= \frac{i\pi^3}{120} \, \big[2k_2^\mu k_2^\nu + 3(k_2^\mu k_3^\nu + k_2^\nu k_3^\mu)  +4(k_2^\mu k_4^\nu + k_2^\nu k_4^\mu) \notag \\
& \ \ \ \ \ \ \   +6 k_3^\mu k_3^\nu +8(k_3^\mu k_4^\nu + k_3^\nu k_4^\mu) +12 k_4^\mu k_4^\nu+ \eta^{\mu\nu}s_{13} \big]+ {\cal O}(\varepsilon)\, . \notag
\end{align}

\bibliographystyle{JHEP}

\providecommand{\href}[2]{#2}\begingroup\raggedright\endgroup


\end{document}